\documentclass[twocolumn,showpacs,preprintnumbers,superscriptaddress]{revtex4-1}                
\usepackage[T1]{fontenc}    
\usepackage{amsfonts}     
\usepackage{amsmath,amsbsy,amssymb,graphicx}                              
\usepackage{times}
\usepackage{color}

\begin{document}                 
\title{Spin-Pumping-Induced Non-Linear Electric Current on the Surface of a Ferromagnetic Topological Insulator  }          
\author{Yusuke~Hama}  
 \altaffiliation[current address:]{ Quemix Inc., 2-11-2 Nihombashi, Chuo-ku, Tokyo 103-0027, Japan}
 \affiliation{National Institute of Informatics, 2-1-2 Hitotsubashi, Chiyoda-ku, Tokyo 101-8430, Japan} 
\author{Kentaro~Nomura}
\affiliation{Institute of Material Research, Tohoku University 2-1-1 Katahira, Aoba-ku, Sendai,  980-8577, Japan} 
\affiliation{Center for Spintronics Research Network, Tohoku University 2-1-1 Katahira, Aoba-ku, Sendai,  980-8577, Japan }
\date{\today}
\begin{abstract}{We investigate the spin-pumping-induced electric current on the surface of a three-dimensional topological insulator hybridized with a ferromagnet, namely, ferromagnetic topological insulator. 
In order to do this, we establish the microscopic formalism and construct the perturbation theory using a Keldysh Green's function approach.
We analyze how this electric current is generated by an exchange interaction and an external ac magnetic field, which is the driving force of ferromagnetic resonance as well as the spin pumping.
The mechanism is as follows. 
First, the ferromagnetic resonance is driven and a zero-momentum magnon emerges. It is the fluctuation from the saturation magnetization pointing parallel to the precession axis of the ferromagnetic resonance.
After then, the spin pumping is generated with the zero-momentum magnon being the carrier of spin. 
The zero-momentum magnon and the topological insulator surface state couples through the exchange interaction and the spin carried by the magnon is transferred to it.   
Owing to the spin-momentum locking, the transferred spin is converted into the momentum of topological insulator surface state leading to the generation of electric current flowing perpendicular to the precession axis of the ferromagnetic resonance.  
It is quadratic in the amplitude of external ac magnetic field whereas it is linear to the strength of the exchange interaction. 
The associated electric voltage is described by the spectrum of zero-momentum magnon.
The non-linearity of spin-pumping-induced electric current in the ac magnetic field  as well as the linearity in the exchange-interaction strength
 reflects that the surface of ferromagnetic topological insulator has a high-performing functionality of generating the electric charge current by  magnetic controlling. }  
  \end{abstract}
 
\maketitle 

\section{Introduction}\label{intro}      

Quantum technologies for hybridizing two or more sub quantum systems have been advancing rapidly with many types of elements ranging from solid-state systems to atomic-molecular and optical systems
having been used, for example, electrons and nuclei in GaAs semiconductors, nitrogen-vacancy centers in diamonds, superconducting qubits, and
atoms and  cavities composing cavity quantum electrodynamic systems \cite{hybrid1,hybrid2,hybrid3,quantumdotreview1,electronnuclear1,electronnuclear2,electronnuclear3,nvcenterreview1,nvcenterreview2,cavityqedreview1,cavityqedreview2,SCQRPP2017,SCQNISQ2019}. 
The functionalities of these hybrid quantum systems are superior to or richer than those of any individual sub quantum systems and are characterized in the way they are composed of.  
By selecting sets of sub quantum systems which are the best choices to engineer the hybrid quantum system which has the functionality to perform the task you are aiming,  
it becomes a powerful tool to execute quantum-state controlling, quantum information processing, and spintronics.  

The key issue for spintronics is to perform a high-efficient conversion of electric charge and spin degrees of freedom or the coherent controlling of electricity and magnetism
with lowering sufficiently an energy consumption (Joule heating). 
In order to accomplish these tasks, we have to search for materials having potentials to create physical processes which can be utilized for them and use these materials to engineer quantum devices. 
 Examples include the non-magnetic heavy metals with strong spin-orbit interaction which exhibits (inverse) spin Hall effect like Pt
 and materials composed of metal and oxide possessing Rashba interfaces    \cite{spintronicsRMP2004,spintronicsRMP2005,spintronicsPR2008,spintronicsannrev2010,spintronicsnatmat2012nb1,yohnumaetalsspinpumpingPRB2014,SHERMP2015,spintronicsarticle2016,interfacemagnetismRMP2017,antiferrospintronicsRMP2018,spintronicsreviewnpj2018}. 
Recently, topological insulator (TI) is considered to be a good candidate for a component of spintronics devices because TI exhibits bulk state with strong spin-orbit coupling as well as
surface state whose spin and momentum are strongly coupled which is called the spin-momentum locking (high-efficient convertibility of spin and electric charge current) \cite{HansanKaneRMP2010,QiZhangTITSCRMP2011,AndoTIreview,TIbook}.
In addition, the hybrid quantum system of magnetic materials and TI, namely, the magnetic TI, has been intensively investigated from both theoretical and experimental points of view \cite{Yokoyamaetal,Nomuraetal,Garateetal,Tseetal1,Tserkovnyaketal1,mahfouzietal2014,ISHETIPRB2014,ShiomiTISP,Sakaietal,Taguchietal,SPTISR2015,TISOTPRB20162017,Checkelskyetal1,Henketal,Changetal,expFermiEdependencemagneticTI1,Mooderagroup1,MellniketalTISTT,Garateetal,Fanetal,Leeetal,Kouetal,Mogietal,ChangandLi,Vdoping1,Vdoping2,SPISHETINanoLett2015,magneticTISPprl2016,Mogietal2,SHETINatMat2018,ChiTangetal2,expFermiEdependencemagneticTI2}.   
Including the quantum anomalous Hall effect, the magnetic TI exhibits rich quantum phenomena owing to the composition of magnetism and spin-momentum locking (multifunctionality).  
Because of the multifunctionality and the high-efficient convertibility of spin and electric charge current, the magnetic TI is considered to be one of the promising candidate for spintronics devices and a large number of investigations have been made toward this goal \cite{Yokoyamaetal,mahfouzietal2014,ISHETIPRB2014,ShiomiTISP,Sakaietal,Taguchietal,SPTISR2015,SPISHETINanoLett2015,magneticTISPprl2016,TISOTPRB20162017,MellniketalTISTT,Fanetal,Kouetal,expFermiEdependencemagneticTI1,SHETINatMat2018,ChiTangetal2,expFermiEdependencemagneticTI2}. 
Although great efforts have been made for this, we still have not satisfactorily achieved the microscopic understanding of physics at the interface between the magnet and TI.  
For instance, we have not understand satisfactorily the way and how efficiently the spin transferred from the magnet can be converted into the electric current  and/or voltage (spin pumping and the associated phenomena; 
inverse spin Hall effect and inverse Edelstein effect) whereas the electric current of TI  being converted into the magnetization dynamics and/or a spin current (spin-orbit torque, spin Hall effect, and Edelstein effect).    
Such complexities are arising from the fact that the spin current is not a conserved current in the macroscopic systems and the difficulties to distinguish 
whether the contribution to the electric charge current under observation is coming from the surface state or the bulk state. 
It is important and an urgent issue to challenge analyzing these problems in order to achieve a deeper understanding of
the conversion between the electric current (orbital degrees of freedom) and the magnetization dynamics (spin polarization as well as the spin current) in the magnetic TI,
the physics at the interface between magnets and TI surface state
both theoretically  and experimentally,  and further, to realize the coherent controlling of TI surface state and magnetization toward spintronics application.

In this paper, we will focus on the physics of TI surface state and construct the microscopic theory for the quantum transport phenomena at the interface between ferromagnet and TI. 
In order to do this, we use a Keldysh (non-equilibrium or contour-time) Green's function approach.  
We investigate the electric current of TI surface state as well as the associated electric voltage induced by the spin pumping originating in the ferromagnetic resonance (FMR) driven by an external ac magnetic field. 
We analyze in detail how this electric current is created by the ac magnetic field and the exchange interaction between the localized spin in the ferromagnet and the TI surface state.
We show that when the spin is carried from the zero-momentum magnon, which is created by the FMR, to the TI surface state through the exchange interaction, 
due to the spin-momentum locking this carried spin is converted into the momentum. Then correspondingly, the electric current is induced in the direction perpendicular to the precession axis of FMR, namely,
spin-pumping-induced electric current. 
It is the quadratic response to the ac magnetic field whereas it scales linearly to the strength of exchange coupling.   
On the other side, the associated electric voltage has a structure represented by the spectrum of zero-momentum magnon
which clearly reflects that the driving force of this electric voltage is the spin pumping.
Our result enables us to understand clearly not only the mechanism of the spin-pumping-induced electric current and its characteristic,
but it also gives us  a qualitative explanation for the experimental results reported previously \cite{ShiomiTISP,expFermiEdependencemagneticTI2}. 

This paper is organized as follows. 
In Sec. \ref{microscopictheory}, we present our microscopic model of the composite system of ferromagnet and TI surface state.
Then, we construct the formalism for describing the time evolution of this system using the Keldysh Green's function approach. 
Based on it, we present a mathematical representation for the electric current of TI surface state at the non-equilibrium steady state. 
Next, to calculate this electric current we establish the perturbation theory for the Keldysh Green's function where the external ac magnetic field and the exchange interaction are regarded as perturbative terms.
In Sec. \ref{SPINLEcurrent}, which presents the main result of this paper, 
we discuss in detail the generation of electric current of TI surface state induced by the spin pumping as well as the associated electric voltage. 
By analyzing the structure of Feynman diagram for the perturbative Green's function, we discuss the mechanism of the spin-pumping-induced electric current as well as its characteristics. 
Then, we make a comparison between our result and the experimental results  \cite{ShiomiTISP,expFermiEdependencemagneticTI2} through the characteristic of electric voltage.   
Sec. \ref{CON} is devoted to the conclusion and outlook of this paper.

\section{Microscopic Theory}\label{microscopictheory}  

In this section, we first present our microscopic model for the ferromagnetic TI. 
Based on it, we establish the formalism to describe the time evolution of this system generated by the spin pumping.
Then we evaluate the electric current of the TI surface state at the non-equilibrium steady state using the Keldysh Green's function approach.
We do this by constructing the perturbation theory for the Keldysh Green's function so that the ac magnetic field and the exchange interaction are treated as perturbative terms. 

\subsection{Modeling and Formalism}\label{modelformalism}
\begin{figure}[t] 
\includegraphics[width=0.4 \textwidth]{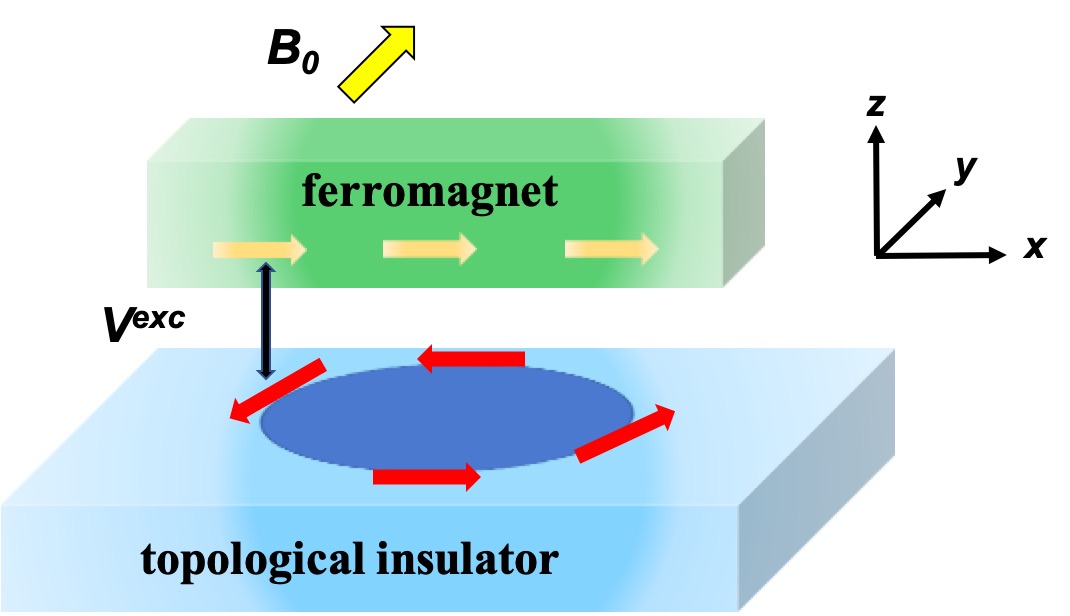}
\caption{ Schematic illustration of the ferromagnetic TI. FM is represented by the Hamiltonian $H^{\rm{FM}}$ while the surface state of TI is expressed by $H^{\rm{TI}}$.  
The localized spin in FM couples with the spin of TI surface state through the exchange interaction $ V^{\rm{exc}} $.  
The total Hamiltonian of this system is given by $H= H^{\rm{FM}} + H^{\rm{TI}} + V^{\rm{exc}}.$ } 
\label{ferromagneticTI} 
\end{figure}  
The ferromagnetic TI is the composite system of a ferromagnet (FM) and the three-dimensional TI. 
We take a spatial cartesian coordinate so that the $xy$ plane is parallel to the interface between the FM and TI whereas the $z$ axis perpendicular to it. 
 The surface state of TI appears in the $xy$ plane.  
The TI surface state and a localized spin in the FM are coupled through the exchange interaction.  
The illustration of ferromagnetic TI is presented in Fig. \ref{ferromagneticTI}.
Experimentally, this system is created by doping the magnetic atoms (for instance, Cr, V, and Mn) to the TI or 
implementing the heterostructure of ferromagnetic materials (e.g., a ferromagnetic insulator as well as metal including EuS, EuO, YIG, and permalloy such as $\rm{Ni}_{81}\rm{Fe}_{19}$ and CoFeB) and the TI \cite{Checkelskyetal1,Henketal,Changetal,Mooderagroup1,MellniketalTISTT,Fanetal,ISHETIPRB2014,ShiomiTISP,SPTISR2015,Leeetal,Kouetal,Mogietal,SPISHETINanoLett2015,ChangandLi,magneticTISPprl2016,expFermiEdependencemagneticTI1,Vdoping1,Vdoping2,Mogietal2,SHETINatMat2018,ChiTangetal2,expFermiEdependencemagneticTI2}. 
The examples of three-dimensional TI include tetradymites  $\rm{Bi}_2\rm{Se}_3$ and $\rm{Bi}_2\rm{Te}_3$ \cite{HansanKaneRMP2010,QiZhangTITSCRMP2011,AndoTIreview,TIbook,TIcrystaldata}.  
Hereinafter, let us focus on the interface between FM and TI and model the composite system of localized spin at this interface and the TI surface state (let us call it the surface of ferromagnetic TI).
 The spin pumping and the associated inverse spin Hall effect in the heterostructure systems composed of ferromagnetic metal (or ferromagnetic insulator such as YIG) and Pt or NiPd alloy 
have been modeled in \cite{yohnumaetalsspinpumpingPRB2014}. By referring to it, we model the surface of ferromagnetic TI for describing the spin pumping process and the associated electric-current generation. It is described by the Hamiltonian 
$H= H_0+V^{\rm{exc}},$ where $H_0=H^{\rm{FM}}+H^{\rm{TI}}$ with  $ H^{\rm{TI}} = \bar{H}^{\rm{TI}}_0 + H^{\rm{imp}}$. 
The Hamiltonian $\bar{H}^{\rm{TI}}_0$ is the unperturbed Hamiltonian of the TI surface state consisting of the spin-momentum-locking term with the dispersion relation being measured from the chemical potential $ \mu_{\rm{TI}} $:
 $ \bar{H}^{\rm{TI}}_0 = ( H^{ \rm{TI} }_0 -  \mu_{\rm{TI}}  N^{\rm{TI}})$. The operator $N^{\rm{TI}}$ is the number operator of TI surface state. Hereinafter, let us take the chemical potential $ \mu_{\rm{TI}} $ to be equal to the Fermi energy of TI
 and denote it  as $\epsilon_{ \rm{F}}$.  $H^{\rm{imp}}$ is an impurity potential term and assume to be spin independent (non magnetic). 
 $H^{\rm{FM}}$ is the unperturbed Hamiltonian of FM and take its chemical potential to be zero $( \mu_{\rm{FM}} = 0).$
 $V^{\rm{exc}}$ is the exchange interaction between the localized spin in FM and TI surface state. 
The Hamiltonians $\bar{H}^{\rm{TI}}_0$ and $H^{\rm{imp}}$ are given by 
\begin{align}
\bar{H}^{\rm{TI}}_0 &= \int d^2x \psi^\dagger_{\alpha^\prime}(\boldsymbol{x}) \left(
\mathcal{H}^{(0)}_{\rm{TI}}  (\boldsymbol{x}) - \epsilon_{ \rm{F}}  \boldsymbol{1} \right)_{\alpha^\prime \alpha}
\psi_{\alpha}(\boldsymbol{x}), \label{TIhamiltonian1} \\
H^{\rm{imp}} & =    \int d^2x   \psi^\dagger_{\alpha^\prime}(\boldsymbol{x})    \mathcal{H}^{\rm{imp}} _{\alpha^\prime \alpha} (\boldsymbol{x})    \psi_{\alpha}(\boldsymbol{x}),  \label{impurityhamiltonian1} 
\end{align}
where
\begin{align}
\left(
\mathcal{H}^{(0)}_{\rm{TI}}  (\boldsymbol{x}) \right)_{\alpha^\prime \alpha} & =  -i \hbar v_{\rm{F}}(\sigma^y \partial_x  -\sigma^x \partial_y)_{\alpha^\prime \alpha}, \label{TIhamiltoniandensity1} \\
 \mathcal{H}^{\rm{imp}} _{\alpha^\prime \alpha} (\boldsymbol{x})  &= \sum_{i_{  {\rm{imp}}  } = 1}^{N_{\rm{imp}}} V_{\rm{imp}}(\boldsymbol{x}-\boldsymbol{X}^{\rm{imp}}_{ i_{\rm{imp}}    })
 \cdot \boldsymbol{1}_{\alpha^\prime \alpha}.   \label{impurityHamiltoniandensity}
\end{align}
The operators $\psi_{\alpha}(\boldsymbol{x})$ and $\psi^\dagger_{\alpha}(\boldsymbol{x})$ are the annihilation and creation operators of the TI surface state at the two-dimensional spatial coordinate $\boldsymbol{x}=(x,y)$, respectively.
The index $\alpha=\uparrow,\downarrow$ describes the spin degrees of freedom of TI surface state.    
The summation is taken for two repeated indices $\alpha$ and $\alpha^\prime$ in Eqs.  \eqref{TIhamiltonian1} and \eqref{impurityhamiltonian1}.
$v_{\rm{F}}\sim5.0\times10^5$ m/s is the Fermi velocity while
 $\sigma^{x}$ and $\sigma^{y}$ are the Pauli matrices. $V_{\rm{imp}}(\boldsymbol{x}-\boldsymbol{X}^{\rm{imp}}_ {i_{\rm{imp}}}  )$ in Eq. \eqref{impurityHamiltoniandensity} is the impurity potential 
and the vector $\boldsymbol{X}^{\rm{imp}}_{i_{\rm{imp}}} = (X^{\rm{imp}}_{i_{\rm{imp}}},Y^{\rm{imp}}_{i_{\rm{imp}}})$ is the coordinate of ${i_{\rm{imp}}}$-th impurity. 
$N_{\rm{imp}}$ is the total number of impurities. $ \boldsymbol{1}_{\alpha^\prime \alpha} $ is the two by two unit matrix.
For the details of TI-surface-state field operators $\psi_{\alpha}$ and $\psi^\dagger_{\alpha^\prime}$ , see subSec. \ref{AppA-1} in Appendix \ref{appendixA}. 
The Hamiltonian  $H^{\rm{FM}}$ is given by
\begin{align}
H^{\rm{FM}} &=-J_{\rm{nx}}\sum_{\langle i j \rangle}\boldsymbol{S}_i\cdot\boldsymbol{S}_j + \hbar\gamma  \sum_i B_0 S^y_i.  \label{FMhamiltonian1}
\end{align}
The three-component vector $ \boldsymbol{S}_i = (S^x_i,S^y_j,S^z_i)$ represents the localized spin of the FM at the spatial coordinate $\boldsymbol{r}_i=(r^x_i,r^y_i).$ 
The indices $i$ and $j$ runs from 1 to $N_{\rm{loc}}$ with $N_{\rm{loc}}$ denoting the total number of localized spin at the interface between FM and TI. 
$J_{\rm{nx}}$ is the strength of the nearest-neighboring exchange interaction. The summation $\sum_{\langle i j \rangle}$ is taken for nearest-neighboring pairs.
For any $i$, the localized spin $\boldsymbol{S}_i$ satisfies  $\boldsymbol{S}_i^2=(S^x_i)^2+(S^y_i)^2+(S^z_i)^2=S_0(S_0+1)$ with $S_0$ its spin magnitude. 
$\gamma$ is the gyromagnetic ratio of localized spin. 
The static magnetic field $B_0$ is applied to the $y$ direction and the saturation magnetization is created along this direction.  
Hereinafter we will not include the demagnetizing coefficient for simplicity.
The exchange interaction $V^{\rm{exc}}$ has the form 
\begin{align}
    V^{\rm{exc}} = -J^{\rm{exc}} \sum_{ i } \sum_{a=x,y,z}  s^{ a }(\boldsymbol{r}_{ i }) S^{ a }_{ i }, \label{exchangeint}
\end{align}
where $J^{\rm{exc}}$ is the strength of the exchange interaction. 
$s^{ a }(\boldsymbol{r}_{ i }) =\psi^\dagger_{\alpha^\prime}(\boldsymbol{r}_{ i })(\sigma_{\alpha^\prime \alpha}^{ a }/2)\psi_\alpha(\boldsymbol{r}_{ i })$ is the spin density of TI surface state 
at the coordinate $\boldsymbol{r}_{ i }$.

\begin{figure}[t] 
\includegraphics[width=0.33\textwidth]{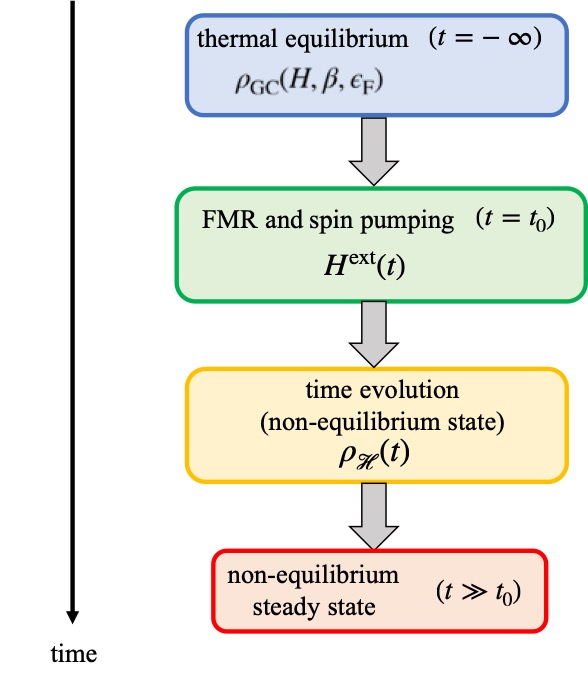}
\caption{ Diagrammatic representation of the time evolution of ferromagnetic TI surface. At far past ($t = - \infty$), 
the ferromagnetic TI surface is in the thermal equilibrium state described by  the grand-canonical ensemble $ \rho_{\rm{GC}}(H,\beta,\epsilon_{ \rm{F} } ). $
After then, at $t = t_0$ the external magnetic field $H^{\rm{ext}}(t)$ is applied and the FMR as well as the spin pumping are driven. 
The time evolution of this system is represented by the density matrix $\rho_{\mathcal{H}}(t)$. At sufficiently a long time ($t \gg t_0$), the surface of ferromagnetic TI is in the non-equilibrium steady state 
and the associated quantum transport of TI surface state is generated.  }  
\label{timeevolution} 
\end{figure} 

Next, let us discuss the time evolution of this system.
 At initial time ($t=-\infty$), the ferromagnetic TI is in the thermal equilibrium state with the temperature $T$.
It is represented by the grand-canonical ensemble with its density matrix 
\begin{align}
\rho_{\rm{GC}}(H,\beta, \epsilon_{ \rm{F}} )=\frac{\exp\left(-\beta H \right)}
{ {\rm{Tr}} \left(  \exp\left(-\beta H  \right)   \right)}, \label{Cstatisticaloperator}
\end{align}
where  $\beta^{-1}=k_{\rm{B}}T$ with $k_{\rm{B}}$ the Boltzmann constant. 
Note that the TI-Fermi-energy dependence is included in the Hamiltonian $H.$
At $t=t_0$, we apply an ac external  magnetic field  $\boldsymbol{B}^{\rm{ext}}(t)= B^{\rm{ext}}\left(\sin \big{(}  { \rm{sgn} }(B_0) \cdot  \omega^{\rm{ext}}t  \big{)}  ,0, \cos \big{(}  { \rm{sgn} }(B_0) \cdot  \omega^{\rm{ext}}t  \big{)}   \right)$, 
where  ${ \rm{sgn} }(B_0) = +1$ $(-1)$ when $B_0>0$ $(<0)$. Here we have taken a circular polarized light.
$B^{\rm{ext}}$  and $ \omega^{\rm{ext}} $ are its amplitude and frequency, respectively. 
This triggers the ferromagnetic resonance (FMR). 
The system at $t>t_0$ is going to be described by the total Hamiltonian $\mathcal{H}(t)= H+H^{\rm{ext}}(t)$ where $H^{\rm{ext}}(t)$  is given by
 \begin{align}
H^{\rm{ext}}(t) &= \hbar\gamma  \sum_{ a =x,z} \sum_{ i } B_{ a }^{\rm{ext}}(t)   S^{ a }_{ i }.  \label{externalhamiltonian}
\end{align} 
For later convenience, we decompose the total Hamiltonian $\mathcal{H}(t)$ into the form  $\mathcal{H}(t)= H_0+H^\prime(t)$ with $H^\prime(t) = V^{\rm{exc}} + H^{\rm{ext}}(t).$
The precession axis of the FMR is along the $y$ direction owing to the static magnetic field $B_0$.  
Once the FMR is triggered, a spin transfer occurs from FM to the TI surface state mediated by the exchange interaction $V^{\rm{exc}}$, i.e.,  spin pumping. 
As a result,  a spin polarization as well as an associated non-equilibrium state is generated on the surface of TI. 
Such a physical process (the time evolution of the system at  $ t > t_0 $) is represented by the density matrix  \cite{NEQGreensfunctionRMPandtxtbook1,NEQGreensfunctiontxtbook2}
\begin{align}
\rho_{\mathcal{H}}(t) &=U(t,t_0)   \rho_{ \rm{C} }(H,\beta,\epsilon_{ \rm{F}   }   )  U^\dagger(t,t_0), \label{DMtimeevolt1}
\end{align}
where the time-evolution operator $U(t,t_0)$ is given by
\begin{align}
&U(t,t_0)=T \exp \left( -\frac{i}{\hbar}  \int^t_{t_0}  \mathcal{H}(t^\prime) dt^\prime   \right),
\label{TEoperator1}
\end{align}
with the symbol $T$ denoting the time-ordering product of real time.  By using the density matrix in Eq. \eqref{DMtimeevolt1}, the expectation of a physical operator $A$ at $t>t_0$ is expressed by
\begin{align}
\langle A (t) \rangle ={\rm{Tr}} \left[
A_{\mathcal{H}}(t)  \rho_{\rm{C}}(H,\beta, \epsilon_{ \rm{F} } )  \right] ,  \label{HPoperators1}
\end{align}
where $A_{\mathcal{H}}(t)=U^\dagger(t,t_0)AU(t,t_0)$. 
 $\big{\langle}  X     \big{\rangle}={\rm{Tr}}(X \rho_{ \rm{C} } (H,\beta,  \epsilon_{ \rm{F} }  ))$ represents the thermal average taken with respect to the Hamiltonian $H$.
At sufficiently a long time ($t \gg t_0$), the non-equilibrium steady state is realized and the quantum transport phenomena of TI surface state is generated.
To summarize the above description, in Fig. \ref{timeevolution}  we present the diagrammatic structure of time evolution of the ferromagnetic TI surface.

Since the microscopic formalism for the time evolution of the ferromagnetic TI surface as well as that for the expectation of the physical operators have been established, 
let us discuss the quantum transport phenomena on the surface of ferromagnetic TI at the non-equilibrium steady state. 
When the spin pumping is driven by the FMR, the $y$-polarized spin is injected from FM to TI surface.
We write the spin current associated with this spin pumping process as $J^{\rm{spin}}_{y,z}$. The first subscript $y$ denotes the direction of the spin polarization whereas the second subscript $z$ describes the flowing direction of spin current.  
Through the exchange interaction $V^{\rm{exc}}$, the spin current flows toward the TI surface.
Some portion of $J^{\rm{spin}}_{y,z}$ is going to be converted into the momentum (the electric current flowing on the surface of TI) due to the spin-momentum locking.
Besides that, it might be converted into other types of phenomena, for instance,  a dissipation process like a spin relaxation process 
or the spin current which bounces back to FM. 
From such a consideration, the exact evaluation of the spin current $J^{\rm{spin}}_{y,z}$ and how efficiently it is converted into the electric current of TI surface state are very difficult tasks.  
This is because it is hard to mathematically define the spin current since the spin is not the conserved quantity or the spin current is not the conserved current in the macroscopic system like a  mili-meter-scale system. 
 On the other hand, what has been observed in the experiment is the electric voltage induced by the spin pumping \cite{ShiomiTISP,expFermiEdependencemagneticTI2}.     
By taking into account of this fact, although there are some theoretical approaches which treat mathematically the spin current and calculate the  spin-to-charge conversion efficiency using a concept such as spin-mixing conductance \cite{spintronicsRMP2005,yohnumaetalsspinpumpingPRB2014,interfacemagnetismRMP2017},  
we do not take such approaches. Instead,
we consider that the $y$-polarized spin carried from FM to TI surface via spin-pumping process is going to be mainly converted into the electric charge current of TI surface state.
Therefore, instead of calculating the spin current $J^{\rm{spin}}_{y,z}$ directly and analyze how efficiently it is converted into the electric charge current,    
we calculate directly the electric charge current of the TI surface state and analyze how it is created by the ac magnetic field and the exchange interaction. 
Here we calculate the $x$-component of electric charge current density $j_x(\boldsymbol{x})$. It is given by
 $j_x(\boldsymbol{x})=-ev_{\rm{F}} \big{(}\psi^\dagger_{\alpha^\prime} (\boldsymbol{x}) \sigma^y_{\alpha^\prime\alpha} \psi_{\alpha} (\boldsymbol{x}) \big{)}
=-2ev_{\rm{F}} s^y(\boldsymbol{x}),$  with   $-e$ $(<0)$ the electric charge and $s^y(\boldsymbol{x})$ is the $y$-component spin density of TI surface state at the coordinate $\boldsymbol{x}$. 
Such an equivalence of the $x$-component of electric charge current and the $y$-component spin originates in the spin-momentum locking.
By denoting the annihilation and creation operators of TI surface state field in the Heisenberg picture with respect to $\mathcal{H}(t)$ as $\psi_{\mathcal{H}\alpha}(\boldsymbol{x},t)$ and $\psi^\dagger_{\mathcal{H}\alpha}(\boldsymbol{x},t)$, 
respectively,  from Eq. \eqref{HPoperators1} the expectation of the $x$-component electric current density  at time $t$ is given by
\begin{align}
\langle j_x(\boldsymbol{x},t) \rangle =
-ev_{\rm{F}}\sigma^y_{\alpha^\prime\alpha} 
\big{\langle} \psi^\dagger_{\mathcal{H}\alpha^\prime} (\boldsymbol{x},t) \psi_{\mathcal{H}\alpha} (\boldsymbol{x},t)  
 \big{\rangle}. \label{TIECCexpv1}
\end{align}

\subsection{ Keldysh Green's Function and Perturbation Theory } \label{PNeqGreensfunc}

Our next task is to rewrite the expectation value of electric current density in Eq. \eqref{TIECCexpv1} with the Keldysh Green's function and evaluate it by constructing the perturbation theory
where the perturbative term is $H^\prime(t)=H^{\rm{ext}}(t)+V^{\rm{exc}}$.
Then, what we evaluate  at the end  is the spatial and temporal averaged electric current density at the non-equilibrium steady state. 
It is defined by 
\begin{align}
\bar{ j}_x =  \int \frac{d^2x}{V}   \int_{t_0}^{t_0+T} \frac{dt}{T}  
 \langle j_x(\boldsymbol{x},t) \rangle,
  \label{STaveragecurrent}
\end{align}
where $V$ the area of TI surface. The time $T$ is given by $T = 2\pi N_{\rm{time}} /  \omega^{\rm{ext}}$ with $N_{\rm{time}}$ a positive integer. 
We assume it to be very large to describe that we are taking the long-time average $( N_{\rm{time}} \gg 1 )$.
By analyzing the structure of perturbative Keldysh Green's function,  we investigate how the TI-surface-state electric current $\bar{ j}_x $ is generated by the spin pumping in terms of the ac external magnetic field  
and the exchange interaction.  

First, we rewrite the expectation value of electric current density in Eq. \eqref{TIECCexpv1} by the field operators in the interaction picture. 
We denote the creation and annihilation operators of TI surface state in the interaction picture as $\psi^\dagger_{H_0\alpha}$ and $\psi_{H_0\alpha}$, respectively. 
The expectation value of $x$-component electric current density at the non-equilibrium steady state becomes  \cite{FetterWaleckaQMPtxtbook}
 \begin{figure}[t] 
\includegraphics[width=0.4\textwidth]{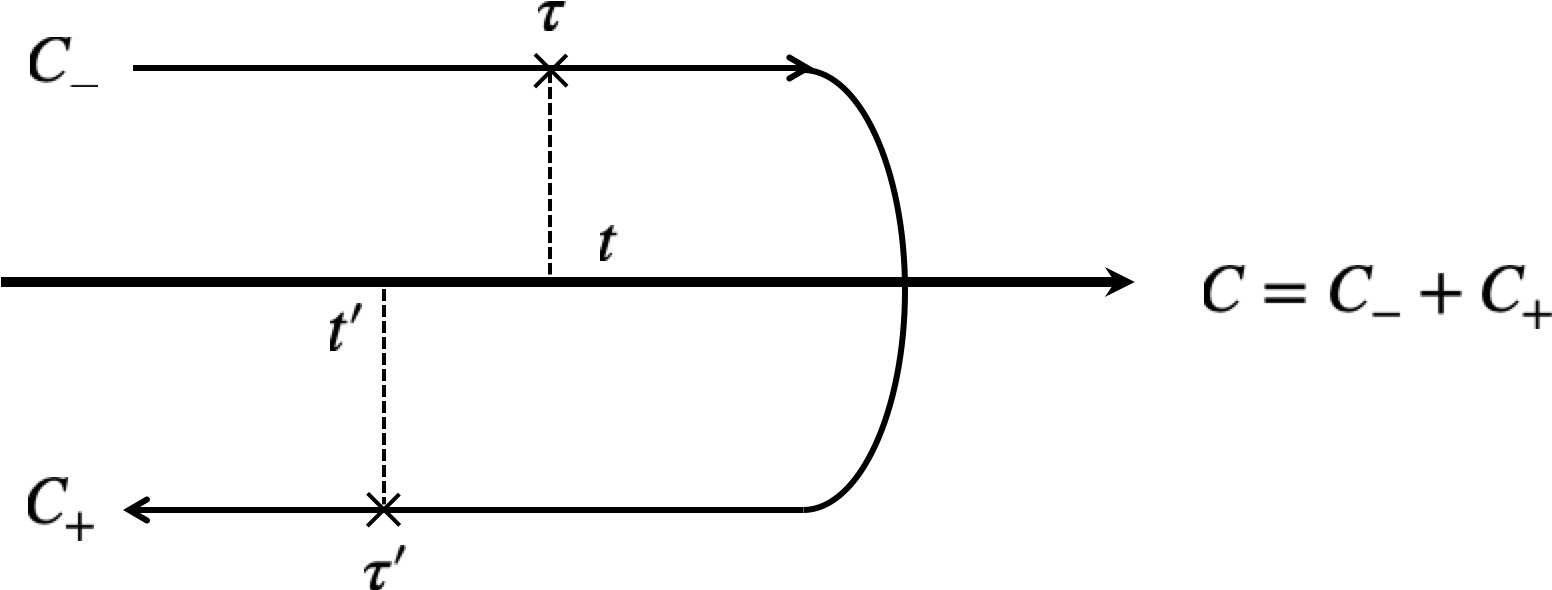}
\caption{ Schematic for the closed contour $C$. It consists of two sub contours $C_-$ and $C_+.$ 
The sub contour $C_-$ starts from $\tau = -\infty$ and ends at  $\tau = +\infty$  whereas  $C_+$ begins from  $\tau = +\infty$ and reaches $\tau = -\infty$.
The variables $t$ and $t^\prime$ are the real times which are obtained by performing the real-time projection on the contour times $\tau$ and $\tau^\prime$, respectively. 
As shown in the diagram in Fig. \ref{timeevolution}, the contour $C$ describes the time evolution of the ferromagnetic TI surface such that 
at the far past ($t \to -\infty$) the thermal-equilibrium state represented by $ \rho_{ \rm{GC} }(H,\beta,\epsilon_{ \rm{F}   }   )$ was realized, and due to the external field $ H^{\rm{ext}}(t)$, at sufficiently a long time ($t \to +\infty$) 
the non-equilibrium state is generated. } 
\label{SchwingerKeldyshcontour} 
\end{figure}  
\begin{align}
\big{\langle} j_x (\boldsymbol{x},t)  \big{\rangle} 
=  iev_{\rm{F}}\sigma^y_{ \alpha^\prime\alpha }   \lim_{\substack{t^\prime \to t^+  \\  \boldsymbol{x}^\prime \to \boldsymbol{x} }}
G^<_{  \alpha \alpha^\prime  } (  \boldsymbol{x}t ;  \boldsymbol{x}^\prime t^\prime  ) ,
 \label{TIECCexpv2}
\end{align}
where $G^<_{  \alpha \alpha^\prime  } (  \boldsymbol{x}t ;  \boldsymbol{x}^\prime t^\prime  )$ is the lesser component of full real-time Green's function. 
$t^+$ is the time which is infinitesimally later than $t$: $t^+=t+\epsilon_t^+$ with $\epsilon_t^+$ a positive infinitesimal.
The lesser Green's function $G^<_{  \alpha \alpha^\prime  } (  \boldsymbol{x}t ;  \boldsymbol{x}^\prime t^\prime  )$ is redescribed by the Keldysh (contour-time)  Green's function defined by  \cite{NEQGreensfunctionRMPandtxtbook1,NEQGreensfunctiontxtbook2}
\begin{align}
 i G_{C,\alpha\alpha^\prime}(\boldsymbol{x} \tau;\boldsymbol{x}^\prime \tau^\prime)
& =   \Big{\langle} T_C \big{[} \mathcal{U}^{ \rm{exc} }_{C} \mathcal{U}^{\rm{ext}}_{C}     
\psi_{ H_0 \alpha} (\boldsymbol{x} \tau)  \psi^\dagger_{ H_0 \alpha^\prime} (\boldsymbol{x}^\prime \tau^\prime)   \big{]} \Big{\rangle}_{0} , \label{KGreen'sfunctiondef}
\end{align}
where $\big{\langle}  X     \big{\rangle}_0= {\rm{Tr}} (X \rho_{\rm{GC}}(H_0,\beta,\epsilon_{ \rm{F} }))$ is the thermal average taken with respect to the unperturbed Hamiltonian $H_0$.
The contour $C$ is the closed path  as shown in Fig.  \ref{SchwingerKeldyshcontour} and is represented by the time variable called the contour time. 
Let us denote it as $\tau$. The symbol $T_{C}$ represents the time-ordering operator for contour times belonging to $C$.
 For instance, if $\tau_1<\tau_2$ we have $T_{C} [A_1(\tau_1)A_2(\tau_2)]=\pm A_2(\tau_2)A_1(\tau_1)$.
 We obtain the positive sign after we exchanged the order between  $A_1(\tau_1)$ and $A_2(\tau_2)$
if this exchange was bosonic (exchanging even numbers of fermionic operators) while we get the negative sign if the exchange was fermionic (exchanging odd numbers of fermionic operators).
The contour $C$ consists of two sub contours $C_-$ and $C_+.$  
The sub contour $C_-$ starts from $\tau=-\infty$ and reaches $\tau=+\infty$ while the sub contour $C_+$ begins from $\tau=+\infty$ and ends at $\tau=-\infty$. 
Such a structure represents that the timescale of dynamics we are focusing on is when the non-equilibrium steady state is realized. 
It is when sufficiently a long time has passed since we applied the external field $H^{\rm{ext}}(t)$ (at time $t_0$).
In order to describe such a situation, the limit $t_0\rightarrow-\infty$ is going to be taken while for time $t$, which is the time when the non-equilibrium state we are focusing on is realized,
we take $t \rightarrow \infty$. The reason we have the two sub contours $C_-$ and $C_+$ is because, as described in Eq. \eqref{HPoperators1}, 
 the physical operators are sandwiched between the two time evolution operators $U^\dagger(t,t_0)$ and $U(t,t_0)$. 
 Note that the temporal structure of contour $C$ is equivalent to the structure of time evolution presented in Fig. \ref{timeevolution}.  
 The contour times $\tau$ and $\tau^\prime$ in Eq.   \eqref{KGreen'sfunctiondef} belong to $C_-$ and $C_+$,  respectively. 
 In such a temporal configuration, the Keldysh function $G_{C,\alpha\alpha^\prime}(\boldsymbol{x} \tau;\boldsymbol{x}^\prime \tau^\prime)$ becomes the lesser Green's function via real-time projection.
 For more details on the real-time projection of the Keldysh Greens' function formalism see subSec. \ref{real-timeP} in appendix \ref{appendixB}. 
 
 The operators $\mathcal{U}^{ \rm{exc} }_{C}$ and $\mathcal{U}^{\rm{ext}}_{C}$ are the time-evolution operators along the contour $C$ generated by $V^{\rm{exc}}$ and $H^{\rm{ext}}$, respectively.
 They are defined by
 \begin{align}
& \mathcal{U}^{\rm{ext}}_C=\exp\left(-\frac{i}{\hbar}\int_C d \tilde{\tau}  H^{\rm{ext}}_{H_0}( \tilde{\tau}) \right),\notag\\ 
& \mathcal{U}^{ \rm{exc }}_C=\exp\left(-\frac{i}{\hbar}\int_C d \check{\tau}  V^{\rm{exc}}_{H_0}( \check{\tau} ) \right).
\label{Uoperators} 
\end{align}
$H^{\rm{ext}}_{H_0}(\tilde{\tau})$ and $V^{\rm{exc}}_{H_0}( \check{\tau} )$ in the above equation are written by the field operators in the interaction picture at the contour time $\tilde{\tau}$ or $\check{\tau}$.
In order to perform the perturbative calculation, we rewrite the Hamiltonians $ H^{\rm{ext}}_{H_0}( \tilde{\tau})$ and  $V^{\rm{exc}}_{H_0}( \check{\tau} )$ in the momentum representation 
and reorganize the unperturbed and perturbed terms. For doing this, let us introduce the Fourier transform of the spin density for TI surface state.  
It is given by $s^a (\boldsymbol{x}) = V^{-1} \sum_{\boldsymbol{k}}  e^{ i \boldsymbol{k} \cdot \boldsymbol{x} }  s^a (\boldsymbol{k})  $ where $\boldsymbol{k} = (k^x , k^y  )$ is the two-dimensional wavevector of TI surface state
and  $s^a (\boldsymbol{k}) = \sum_{\boldsymbol{k}^\prime}  \psi^\dagger_{  \alpha^\prime} (\boldsymbol{k}^\prime)  (\sigma^a_{ \alpha^\prime   \alpha } / 2 )   \psi_{  \alpha } (\boldsymbol{k}^\prime + \boldsymbol{k}) $
with $a=x,y,z$. 
The (inverse) Fourier transform of the field operator of TI surface state is given by  
$   \psi_{ H_0 \alpha } (\boldsymbol{x}t) = \big{(} 1 / \sqrt{V} \big{)}  \sum_{\boldsymbol{k}} e^{ i \boldsymbol{k} \cdot  \boldsymbol{x} }     \psi_{ H_0 \alpha } (\boldsymbol{k}t),
 \psi_{ H_0 \alpha } (\boldsymbol{k}t) = \big{(} 1 / \sqrt{V} \big{)} \int d^2x  e^{ - i \boldsymbol{k} \cdot  \boldsymbol{x} }     \psi_{ H_0 \alpha } (\boldsymbol{x}t).$
Besides the TI-surface-state field operator in the momentum representation, we introduce the magnon field operators represented in the momentum space by re-expressing the localized spin with them (Holstein-Primakoff tranformation).
They are given by 
\begin{align}
S^y_i & = - { \rm{sgn} }(B_0) \left( S_0 - \frac{1}{ N_{\rm{loc}} } \sum_{ \boldsymbol{p}  \boldsymbol{p}^\prime   } a^\dagger(\boldsymbol{p}^\prime ) a(\boldsymbol{p} )   e^{  i  ( \boldsymbol{p} -\boldsymbol{p}^\prime )   \cdot \boldsymbol{r}_i     } \right)   , \notag\\  
   S^-_i & = S_i^z-iS_i^x =
   \left\{
    \begin{array}{l}
     \sqrt{ \frac{2S_0}{ N_{\rm{loc}} }   }\sum_{\boldsymbol{p}} e^{-i\boldsymbol{p}\cdot \boldsymbol{r}_i }a^\dagger(\boldsymbol{p}), \quad  (B_0 < 0)  \\
     \sqrt{ \frac{2S_0}{ N_{\rm{loc}} }   }\sum_{\boldsymbol{p}} e^{i\boldsymbol{p}\cdot \boldsymbol{r}_i }a(\boldsymbol{p}), \quad  (B_0 > 0) 
    \end{array}
  \right. , \notag\\ 
  S^+_i & = S_i^z + iS_i^x =
   \left\{
    \begin{array}{l}
     \sqrt{ \frac{2S_0}{ N_{\rm{loc}} }   }\sum_{\boldsymbol{p}} e^{i\boldsymbol{p}\cdot \boldsymbol{r}_i }a(\boldsymbol{p}), \quad  (B_0 < 0)  \\
     \sqrt{ \frac{2S_0}{ N_{\rm{loc}} }   }\sum_{\boldsymbol{p}} e^{-i\boldsymbol{p}\cdot \boldsymbol{r}_i }a^\dagger(\boldsymbol{p}), \quad  (B_0 > 0)  
    \end{array}
  \right. ,
\end{align} 
where  $a(\boldsymbol{p})$ ($a^\dagger(\boldsymbol{p})$) denotes the annihilation (creation) operator of magnon with momentum $\boldsymbol{p}=(p^x,p^y)$.
The annihilation and creation operators of magnon satisfy the commutation relation $[a(\boldsymbol{p}), a^\dagger(\boldsymbol{q})] = \delta( \boldsymbol{p}- \boldsymbol{q} )$ with all others being zero.
By using the spin density $s^a (\boldsymbol{k})$ and the magnon field operators  $a(\boldsymbol{p} )$ and $a^\dagger(\boldsymbol{p}^\prime )$, 
the Hamiltonian of the surface of ferromagnetic TI is re-expressed in the momentum space as
\begin{widetext}
\begin{align}
\bar{H}^{\rm{TI}}_0 & =  \hbar v_{\rm{F}} \sum_{\boldsymbol{k}}    \psi_{  \alpha^\prime } (\boldsymbol{k}) 
\big{(}     \sigma^y (k^x + k^x_0)  -\sigma^x k^y - \epsilon_{ \rm{F}} \boldsymbol{1} \big{)}_{\alpha^\prime \alpha}   \psi_{  \alpha } (\boldsymbol{k}),     \label{momentumTIhamiltonian1} \\
H^{\rm{imp}} &   =    \frac{1}{V}
\sum_{ \boldsymbol{k} \boldsymbol{q} \alpha  }  v_{\rm{imp}}(\boldsymbol{q} ) \rho_{\rm{imp}}(\boldsymbol{q} )
\psi^\dagger_{\alpha}( \boldsymbol{k}  + \boldsymbol{q}  ) \psi_{\alpha}(\boldsymbol{k}),  \label{momentumimpurityhamiltonian1} \\
H^{\rm{FM}}_0 & = \sum_{\boldsymbol{p}}  \epsilon^{\rm{FM}} _{\boldsymbol{p}}   a^\dagger( \boldsymbol{p} ) a( \boldsymbol{p} ), \notag\\
V^{\rm{exc} } & = 
  \left\{
    \begin{array}{l}
- \sqrt{ \frac{S_0}{ 2N_{\rm{loc}} V^2 }   } \sum_{  \boldsymbol{q} \boldsymbol{p} }  \Big{(} J^{\rm{exc}}_{ (\boldsymbol{q} \boldsymbol{p})   }   s^-(\boldsymbol{q}) a( \boldsymbol{p})       
+ J^{\rm{sd} \ast} _{(\boldsymbol{q} \boldsymbol{p} ) }  s^+(-\boldsymbol{q})a^\dagger( \boldsymbol{p} ) \Big{)}  
+  \frac{J^{\rm{exc}} }{V}  \sum_{  \boldsymbol{p} \boldsymbol{p}^\prime}       a^\dagger(\boldsymbol{p}^\prime)   a( \boldsymbol{p})     s^y(  \boldsymbol{p}^\prime - \boldsymbol{p} ), \ (B_0 < 0)  \\
- \sqrt{ \frac{S_0}{ 2N_{\rm{loc}} V^2 }   } \sum_{  \boldsymbol{q} \boldsymbol{p} }  \Big{(} J^{\rm{exc}}_{( \boldsymbol{q} \boldsymbol{p} )  }   s^+(\boldsymbol{q}) a( \boldsymbol{p})      
+ J^{\rm{sd} \ast} _{(\boldsymbol{q} \boldsymbol{p})  }  s^-(-\boldsymbol{q})a^\dagger( \boldsymbol{p} ) \Big{)}  
-  \frac{J^{\rm{exc}} }{V}  \sum_{  \boldsymbol{p} \boldsymbol{p}^\prime }       a^\dagger(\boldsymbol{p}^\prime)   a( \boldsymbol{p})     s^y(  \boldsymbol{p}^\prime - \boldsymbol{p} ), \ (B_0 > 0)
 \end{array}
  \right. , \label{momentumexchangehamiltonian1} \\              
  H^{\rm{ext}}(t) & = \hbar\gamma B^{\rm{ext}}\sqrt{\frac{N_{\rm{loc}}S_0}{2}} 
\left( a^\dagger (\boldsymbol{0}) e^{ - i\omega^{\rm{ext}}t}+a(\boldsymbol{0})  e^{ i\omega^{\rm{ext}}t} \right),  \label{momentumacmagnetichamiltonian1}
\end{align}
\end{widetext}
where $  v_{\rm{imp}}( \boldsymbol{q} ) = \int d^2 x  e^{- i  \boldsymbol{q}  \cdot  \boldsymbol{x} } V_{\rm{imp}}( \boldsymbol{x} ) $ 
and $ \rho_{\rm{imp}} ( \boldsymbol{q} ) =  \sum_{i=1}^{N_{\rm{imp}}}  e^{- i  \boldsymbol{q}  \cdot  \boldsymbol{X}_i }$. We will take $  v_{\rm{imp}}( \boldsymbol{0} ) = 0$.
$ \epsilon^{\rm{FM}} _{\boldsymbol{p}} =  zJ_{\rm{nx}} S_0 (1-\gamma_{\boldsymbol{p}}) + \hbar\gamma_{\rm{e}} |B_0| $ is the dispersion relation of magnon with $\gamma_{\boldsymbol{p}}=z^{-1}\sum_{\boldsymbol{\rho}}e^{ - i\boldsymbol{p}\cdot \boldsymbol{\rho}}.$
$z$ is the number of nearest-neighboring sites for localized spins and $\boldsymbol{\rho}$ represents the nearest-neighboring-site vector.
The quantity  $J^{\rm{exc}} _{ (\boldsymbol{q} \boldsymbol{p})}$  is defined by $ J^{\rm{exc}} _{(\boldsymbol{q}\boldsymbol{p})}  = \sum_i J^{\rm{exc}}   e^{i( \boldsymbol{q} + \boldsymbol{p} ) \cdot \boldsymbol{r}_i}  .  $
By comparing the Hamiltonian $\bar{H}^{\rm{TI}}_0$ in Eq. \eqref{TIhamiltonian1} and that in Eq.  \eqref{momentumTIhamiltonian1}, 
we see that because of the exchange interaction the Dirac point of TI surface state (the point where the dispersion of TI surface state becomes zero) is shifted to the momentum $\boldsymbol{k}_0=(k^x_0,0)$ with 
 $k^x_0 = { \rm{sgn} } (B_0) (J^{\rm{exc}} S_0 n^{\rm{2D}}_{\rm{loc}} ) / (2 \hbar  v_{\rm{F}} ) $. Here $ n^{\rm{2D}}_{\rm{loc}} =  N_{\rm{loc}}/V $ is the two-dimensional number density of localized spins.
For the convenience, we perform the Fourier transformation on the field operators of TI surface state $\psi_{  \alpha } (\boldsymbol{x}) $ and 
  $\psi^\dagger_{  \alpha^\prime } (\boldsymbol{x}) $ by  the shifted momentum $ \tilde{\boldsymbol{k}} = \boldsymbol{k} + \boldsymbol{k}_0.$
 As a result, the formula of Hamiltonian $H^{\rm{TI}}_0$ in Eq.  \eqref{momentumTIhamiltonian1} described by the shifted momentum $ \tilde{\boldsymbol{k}} $
 is going to be equivalent to that in Eq. \eqref{TIhamiltonian1}  represented by the original momentum $ \boldsymbol{k} $.
 Hereinafter we just simply write the shifted momentum $ \tilde{\boldsymbol{k}}$ as $ \boldsymbol{k}$. 
 Note that without the impurity effect, the TI surface state exhibits the linear dispersion relation $ \epsilon^{\rm{TI}} _{\boldsymbol{k}} =   \hbar v_{\rm{F}} k $ 
 with $k=\sqrt{(k^x)^2 + (k^y)^2  } $. Consequently, the surface of ferromagnetic TI is remodeled as the hybrid quantum system of magnon and TI surface state 
 with the Hamiltonians in Eqs. \eqref{momentumTIhamiltonian1} - \eqref{momentumacmagnetichamiltonian1}.   

Next, in order to construct the perturbation theory for the Green's function $G_{C,\alpha\alpha^\prime}(\boldsymbol{x} \tau;\boldsymbol{x}^\prime \tau^\prime)$ in Eq. \eqref{KGreen'sfunctiondef} 
let us perform the Fourier transformation with taking the limit  $ \boldsymbol{x}^\prime \to \boldsymbol{x} $.
We have $ G_{C,\alpha\alpha^\prime}(\boldsymbol{x} \tau;\boldsymbol{x}^\prime \tau^\prime)   = V^{-1}\sum_ { \boldsymbol{k} \boldsymbol{k}^\prime }  e^{  i  ( \boldsymbol{k}  - \boldsymbol{k}^\prime ) \boldsymbol{x}   }
 G_{C,\alpha\alpha^\prime}(\boldsymbol{k} \tau;\boldsymbol{k}^\prime \tau^\prime) $.  
 Here $   G_{C,\alpha\alpha^\prime}(\boldsymbol{k} \tau;\boldsymbol{k}^\prime \tau^\prime) $ is given by
 $  G_{C,\alpha\alpha^\prime}(\boldsymbol{k} \tau;\boldsymbol{k}^\prime \tau^\prime)  = 
 - i \Big{\langle}  T_{C}\big{[} \mathcal{U}^{ \rm{exc} }_{C} \mathcal{U}^{\rm{ext}}_{C}  \psi_{ H_0 \alpha} (\boldsymbol{k} \tau)  \psi^\dagger_{ H_0 \alpha^\prime} (\boldsymbol{k}^\prime \tau^\prime)  \big{]}  \Big{\rangle}_{0}  $.
 Then, we perform the perturbative expansion on $ G_{C,\alpha\alpha^\prime}(\boldsymbol{k} \tau;\boldsymbol{k}^\prime \tau^\prime) $
 by expanding the two operators $ \mathcal{U}^{\rm{ext}}_C$ and $ \mathcal{U}^{\rm{exc}}_C$ in Eq. \eqref{Uoperators} 
with respect to $H^{\rm{ext}}_{H_0}(\check{\tau})$ and $V^{\rm{exc}}_{H_0}(\tilde{\tau})$, respectively. 
It is going to be represented in the form 
\begin{align}
G_{C,\alpha\alpha^\prime}(\boldsymbol{k}\tau; \boldsymbol{k}^\prime \tau^\prime) = \sum_{n=0}^\infty \sum_{ n^\prime=0}^\infty G^{(n, n^\prime )}_{C,\alpha\alpha^\prime}(\boldsymbol{k}\tau;\boldsymbol{k}^\prime \tau^\prime).
\label{perturbativeexpansionlesserGreensfunction}
\end{align}
We have used the superscript ($n, n^\prime )$ in the right-hand side of  Eq. \eqref{perturbativeexpansionlesserGreensfunction} 
 to describe that the perturbative Green's function $G^{(n, n^\prime )}_{C,\alpha\alpha^\prime}(\boldsymbol{k} \tau;\boldsymbol{k}^\prime \tau^\prime)$ is in the $n$-th order of  $H^{\rm{ext}}$
while it is in the $n^\prime$-th order of  $V^{\rm{exc}}$. 
Note that the Green's function $ \sum_{ n^\prime =0}^\infty  G^{(0, n^\prime )}_{C,\alpha\alpha^\prime}(\boldsymbol{k}\tau; \boldsymbol{k}^\prime \tau^\prime) $ is the full thermal-equilibrium Green's function
since it does not contain the external-field Hamiltonian $H^{\rm{ext}}$. 
At the non-equilibrium steady state, what we observed in the experiment is the deviation (fluctuation) from the thermal-averaged value at thermal equilibrium. 
Thus, we calculate and show the expectation value of $\big{\langle} j_x (\boldsymbol{x},t)  \big{\rangle} $   in Eq. \eqref{TIECCexpv2}  
as well as the spatial and temporal averaged electric current $\bar {j}_x $  in Eq. \eqref{STaveragecurrent} for $ n \geq 1.$ 
\\ \noindent
As a result, the perturbative Green's function $G^{(n, n^\prime )}_{C,\alpha\alpha^\prime}(\boldsymbol{k} \tau;\boldsymbol{k}^\prime \tau^\prime)$ is  expressed by the unperturbed Keldysh Green's functions 
of TI surface state and magnon given by 
\begin{align}
&i  \mathcal{G}^0_{C,\alpha \alpha^\prime}(\boldsymbol{x} \tau ; \boldsymbol{x}^\prime \tau^\prime) = \Big{\langle}  T_{C}\big{[}   
\psi_{H_0\alpha} (\boldsymbol{x}\tau) \psi^\dagger_{H_0  \alpha^\prime} (\boldsymbol{x}^\prime\tau^\prime)  \big{]}
\Big{\rangle}_0,  \label{TICGreensfunction1}   \\
&i \mathcal{D}^0_{C}(\boldsymbol{q} \tau ; \boldsymbol{q}^\prime \tau^\prime)= \Big{\langle}  T_{C}\big{[}   
a_{H_0} (\boldsymbol{q}\tau) a^\dagger_{H_0 } (\boldsymbol{q}^\prime\tau^\prime)  \big{]}
\Big{\rangle}_0.  \label{magnonCGreensfunction1} 
\end{align}
$ \mathcal{G}^0_{C,\alpha \alpha^\prime}(\boldsymbol{x} \tau ; \boldsymbol{x}^\prime \tau^\prime)$ is the unperturbed Keldysh Green's function of the TI surface state while
$ \mathcal{D}^0_{C}(\boldsymbol{q} \tau ; \boldsymbol{q}^\prime \tau^\prime)$ is that of magnon. 
The Fourier transform of $ \mathcal{G}^0_{C,\alpha \alpha^\prime}(\boldsymbol{x} \tau ; \boldsymbol{x}^\prime \tau^\prime)$ is given as 
$ \mathcal{G}^0_{C,\alpha \alpha^\prime}(\boldsymbol{x} \tau ; \boldsymbol{x}^\prime \tau^\prime) = V^{-1}\sum_ { \boldsymbol{k} \boldsymbol{k}^\prime }  e^{  i  ( \boldsymbol{k}  - \boldsymbol{k}^\prime ) \boldsymbol{x}   } 
   \mathcal{G}^0_{C,\alpha \alpha^\prime}(\boldsymbol{k} \tau ; \boldsymbol{k}^\prime \tau^\prime)$. 
   Note that both  $\mathcal{G}^0_{C,\alpha \alpha^\prime}(\boldsymbol{k} \tau ; \boldsymbol{k}^\prime \tau^\prime)$ and $\mathcal{D}^0_{C}(\boldsymbol{q} \tau ; \boldsymbol{q}^\prime \tau^\prime)$ are diagonal in momentum:
 $\mathcal{G}^0_{C,\alpha \alpha^\prime}(\boldsymbol{k} \tau ; \boldsymbol{k}^\prime \tau^\prime) = \mathcal{G}^0_{C,\alpha \alpha^\prime}(\boldsymbol{k}; \tau,  \tau^\prime) \delta_{ \boldsymbol{k} \boldsymbol{k}^\prime }$
 and  $\mathcal{D}^0_{C}(\boldsymbol{q} \tau ; \boldsymbol{q}^\prime \tau^\prime) = \mathcal{D}^0_{C}(\boldsymbol{q}; \tau, \tau^\prime) \delta_{ \boldsymbol{q} \boldsymbol{q}^\prime }$. 
 To obtain the physical observables like the electric current of TI surface state,  we project the contour times onto the real-time axis. 
Then, the Keldysh Green's functions $ \mathcal{G}^0_{C,\alpha \alpha^\prime}(\boldsymbol{x} \tau ; \boldsymbol{x}^\prime \tau^\prime)$ and $ \mathcal{D}^0_{C}(\boldsymbol{q} \tau ; \boldsymbol{q}^\prime \tau^\prime)$
are rewritten by the unperturbed real-time Green's functions: 
$ \mathcal{G}^0_{C,\alpha \alpha^\prime}(\boldsymbol{k}; \tau,  \tau^\prime) \to \bar{g}^{\nu} _{\alpha \alpha^\prime} (  \boldsymbol{k}, t - t^\prime) $ and 
$ \mathcal{D}^0_{C}(\boldsymbol{q}; \tilde{\tau}, \tilde{\tau}^\prime) \to  \bar{D}^{ \tilde{\nu}}  (  \boldsymbol{q}, \tilde{t} - \tilde{t}^\prime   ) $. 
Here   $\nu,\tilde{\nu} = \rm{t}, <, >,\tilde{\rm{t}}$ denoting the time-ordered, lesser, greater, and anti-time-ordered components, respectively. 
$t, t^\prime, \tilde{t}$, and $\tilde{t}^\prime$ are real-time variables introduced by the real-time projection and 
correspond to $\tau,  \tau^\prime \tilde{\tau}$, and $\tilde{\tau}^\prime$, respectively. After the Keldysh Green's functions are transformed into the real-time Green's functions they are represented by the differences of two real-time variables.  
As a result, the perturbative Keldysh Green's function $G^{(n, n^\prime )}_{C,\alpha\alpha^\prime}(\boldsymbol{k}\tau;\boldsymbol{k}^\prime \tau^\prime)$ is redescribed as products of unperturbed real-time Green's functions.
Its formula can be organized with the retarded and advanced components in the momentum-frequency representation given by
\begin{widetext}
\begin{align}
& \bar{g}^{\rm{r}} _{\alpha \alpha^\prime} (  \boldsymbol{k}, \omega   )  = \frac{( \boldsymbol{1} + \tilde{\mathcal{H}}_0 )_{\alpha \alpha^\prime} }{ 2\left( \omega + \omega_{\rm{F}} -  \omega^{\rm{TI}}_{\boldsymbol{k}} 
+ \frac{i}{2\tau^{\rm{rel}}_{\rm{TI}} } \right)  }  
 +
\frac{(\boldsymbol{1}-\tilde{\mathcal{H}}_0)_{\alpha \alpha^\prime} }{ 2\left( \omega + \omega_{\rm{F}} +  \omega^{\rm{TI}}_{\boldsymbol{k}} + \frac{i}{2\tau^{\rm{rel}}_{\rm{TI}}} \right)  }  \notag\\
&  \bar{g}^{\rm{a}} _{\alpha \alpha^\prime} (  \boldsymbol{k}, \omega   ) = \frac{(\boldsymbol{1}+\tilde{\mathcal{H}}_0)_{\alpha \alpha^\prime} }{ 2\left( \omega + \omega_{\rm{F}} -  \omega^{\rm{TI}}_{\boldsymbol{k}} -\frac{i}{2\tau^{\rm{rel}}_{\rm{TI}}} \right)  }
  +
\frac{(\boldsymbol{1}-\tilde{\mathcal{H}}_0)_{\alpha \alpha^\prime} }{ 2\left( \omega + \omega_{\rm{F}} +  \omega^{\rm{TI}}_{\boldsymbol{k}}  -\frac{i}{2\tau^{\rm{rel}}_{\rm{TI}}} \right)  }  ,  \label{unperturbedTIGreen'sfunctions}\\
& \bar{D}^{\rm{r}}  (  \boldsymbol{0}, \omega   ) = \frac{1}{ \omega  -  \omega^{\rm{FM}}_{\boldsymbol{0}}   +  i  \alpha  \omega }, \quad
\bar{D}^{\rm{a}}  (  \boldsymbol{0}, \omega   ) = \frac{1}{   \omega  -  \omega^{\rm{FM}}_{\boldsymbol{0}}  -    i \alpha  \omega},  \label{unperturbedmagnonGreen'sfunctions}
\end{align}
\end{widetext}
   where $ \boldsymbol{1} $ is the two by two unit matrix and $ \tilde{\mathcal{H}}_0$ is given by  
 \begin{align}
 \tilde{\mathcal{H}}_0 = 
 \left(
\begin{array}{cc}
0 &  - \frac{ i( k^x -i k^y) }{k} \\
 \frac{ i(k^x + i k^y) }{k}  & 0 \\
\end{array}
\right).    \label{dimensionlessDiracHamiltonian} 
\end{align}
The Green's functions  $\bar{g}^{\rm{r(a)}} _{\alpha \alpha^\prime} (  \boldsymbol{k}, \omega) $ and  $\bar{D}^{\rm{r(a)}}  (  \boldsymbol{0}, \omega   )$ in Eqs. \eqref{unperturbedTIGreen'sfunctions} and \eqref{unperturbedmagnonGreen'sfunctions}
 are the retarded (advanced) components of TI-surface-state and zero-momentum magnon Green's functions, respectively.
 The frequencies $\omega^{\rm{TI}}_{\boldsymbol{k}},  \omega_{\rm{F}} $, and $ \omega^{\rm{FM}}_{\boldsymbol{0}} $ 
 are defined by $\omega^{\rm{TI}}_{\boldsymbol{k}} = \hbar^{-1} \epsilon^{\rm{TI}}_{\boldsymbol{k}}$, $  \omega_{\rm{F}} = \hbar^{-1}   \epsilon_{\rm{F}} $, 
 and $ \omega^{\rm{FM}}_{\boldsymbol{0}} = \hbar^{-1} \epsilon^{\rm{FM}}_{\boldsymbol{0}} = \gamma |B_0| $, respectively.
 $\tau^{\rm{rel}}_{\rm{TI}}$ is the relaxation time of the TI surface state due to the impurity effect $H^{\rm{imp}}$ while the constant $\alpha$ appearing in the magnon Green's function
 is the Gilbert damping constant. 
 We put bars on top of these Green's functions to express that we have taken into account the impurity and damping effects.  
 \\ \noindent
 The perturbation theory for the Keldysh Green's function in the above way enables us to clearly explore how the electric current on the surface of TI is induced by the spin pumping 
 in terms of the external ac magnetic field and the exchange interaction. 
 In Appendix  \ref{appendixA}, we present the details for the real-time Green's functions of TI surface state
  as well as the derivation of retarded and advanced components of impurity-averaged Green's functions given in Eq.  \eqref{unperturbedTIGreen'sfunctions} using the imaginary-time Green's function formalism.
 Moreover, we describe the real-time Green's functions of magnon and then  discuss the derivation of retarded and advanced Green's functions in Eq. \eqref{unperturbedmagnonGreen'sfunctions} using the Landau-Lifshitz-Gilbert equation.
In Appendix \ref{appendixB}, we present the detailed description for the Keldysh Green's function formalism as well as the relation between Keldysh Green's function and real-time Green's function.
Further, we show some formulas of Keldysh Green's function formalism and by applying them 
we demonstrate the derivation of impurity-averaged Green's functions of TI surface state for the retarded, advanced, lesser, and greater components. 

\section{Spin-Pumping-Induced Non-Linear Electric Current} \label{SPINLEcurrent}  

Since we have established the perturbation theory, we now evaluate the electric current of the TI surface state induced by the spin pumping.
Let us first  present the diagrammatic representation of our perturbative Green's function. 
Based on it, we microscopically analyze how the electric current is generated by the external magnetic field and the exchange interaction.
Then, we show the structure of spin-pumping-induced electric current as well as the associated electric voltage represented by the static external magnetic field, the amplitude and the frequency of ac magnetic field,
the exchange-interaction strength,
 the relaxation time originating in the non-magnetic impurity, and the Gilbert-damping constant. Finally, we compare our result of the electric voltage with the experimental results \cite{ShiomiTISP,expFermiEdependencemagneticTI2}.   

Let us evaluate the right-hand side of Eq. \eqref{perturbativeexpansionlesserGreensfunction}.
We denote the expectation of spatial and temporal averaged $x$-component electric current corresponding to the term $ G^{(n, n^\prime )}_{C,\alpha\alpha^\prime}(\boldsymbol{k}\tau;\boldsymbol{k}^\prime \tau^\prime) $ as
$\bar{ j}^{ (n, n^\prime) }_x.$ 
First, we can show that the spatial and temporal averaged electric current $\bar{ j}^{ (1, 1) }_x$ is zero. This implies that the surface of ferromagnetic TI does not show the linear response in the ac external magnetic field. 
Next, let us present the next-leading-order term $\bar{ j}^{ (2, 1) }_x$. 
 In order to obtain this, we calculate the perturbative Green's function 
$G^{(2, 1 )}_{C,\alpha\alpha^\prime}(\boldsymbol{k}\tau;\boldsymbol{k}^\prime \tau^\prime)$ given in the right-hand side of Eq. \eqref{perturbativeexpansionlesserGreensfunction}. 
First, we expand  the time-evolution operators  $\mathcal{U}^{\rm{ext}}_C$ and $\mathcal{U}^{\rm{exc}}_C$ with respect to $H^{\rm{ext}}_{H_0}(\tilde{\tau})$ and $V^{\rm{exc}}_{H_0}(\check{\tau})$, respectively. 
With using the Wick's theorem the perturbative Green's function $G^{(2, 1 )}_{C,\alpha\alpha^\prime}(\boldsymbol{k}\tau;\boldsymbol{k}^\prime \tau^\prime)$ is given in terms of the unperturbed Keldysh Green's functions of
TI surface state and magnon as
\begin{widetext} 
\begin{align}
G^{(2, 1 )}_{C,\alpha\alpha^\prime}( \boldsymbol{k} \tau;\boldsymbol{k}^\prime \tau^\prime) & =  
i  { \rm{sgn}}(B_0) \left( - \frac{i}{\hbar} \right)^3  \left( \hbar \gamma B^{\rm{ext}}       \right)^2   \frac{  J^{\rm{exc}}  n^{\rm{2D}}_{\rm{loc}}  S_0    }{4}
  \int_C d \tilde{\tau}_1 d \tilde{\tau}_2 d \check{\tau}_1 e^{- i \omega^{\rm{ext}} ( \tilde{\tau}_1 -  \tilde{\tau}_2 )   }      \sum_{ \boldsymbol{p} \boldsymbol{p}^\prime  \boldsymbol{k}_1 }     \notag\\
 & \times  
   \left[
 \mathcal{D}^0_{C}(\boldsymbol{0}; \tilde {\tau}_1 ,  \tilde {\tau}_2 )    \mathcal{D}^0_{C}(\boldsymbol{p}; 0^+ ) \delta_{  \boldsymbol{p}, \boldsymbol{p}^\prime } +
  \mathcal{D}^0_{C}(\boldsymbol{0}; \tilde {\tau}_2 , \check{\tau}_1 )  \mathcal{D}^0_{C}(\boldsymbol{0};  \check{\tau}_1 ,  \tilde {\tau}_1 )  \delta_{  \boldsymbol{p} , \boldsymbol{0}}  \delta_{ \boldsymbol{p}^\prime , \boldsymbol{0} }
 \right] \notag\\
 & \times   
   \Big{[}  
 \mathcal{G}^0_{C, \alpha\alpha^\prime_1 }(\boldsymbol{k}; \tau ,  \check{\tau}_1 )    \sigma^y_{ \alpha^\prime_1 \alpha_1 }     \mathcal{G}^0_{C,\alpha_1\alpha^\prime }(\boldsymbol{k}^\prime; \check{\tau}_1 , \tau^\prime )
 \delta_{  \boldsymbol{k} , \boldsymbol{k}_1  }  \delta_{ \boldsymbol{k}^\prime ,   \boldsymbol{k}_1+  \boldsymbol{p}^\prime -  \boldsymbol{p}     }  \notag\\
 &  -   \mathcal{G}^0_{C, \alpha\alpha^\prime }(\boldsymbol{k}; 0^+ )    \sigma^y_{ \alpha^\prime_1 \alpha_1 }   \mathcal{G}^0_{C,\alpha_1\alpha^\prime_1 }(\boldsymbol{k}_1; 0^+ ) 
 \delta_{  \boldsymbol{k} , \boldsymbol{k}^\prime  } \delta_{   \boldsymbol{k}_1,    \boldsymbol{k}_1 +  \boldsymbol{p}^\prime -  \boldsymbol{p}  } 
 \Big{]}  . 
\label{21greensfunction1}
\end{align}
\end{widetext}
We note that in the above equation the positive infinitesimal time difference $0^+$ for $\mathcal{G}^0_{C, \alpha\alpha^\prime }(\boldsymbol{k}; 0^+ )$ is equal to $t^\prime - t.$ 
Since  $\tau$ $(=t)$ $\in C_-$ while $\tau^\prime$ $(=t^\prime)$ $\in C_+$, the Green's function $\mathcal{G}^0_{C, \alpha\alpha^\prime }(\boldsymbol{k}; 0^+ )$ is the lesser Green's function. 
On the other hand, $0^+$ for $\mathcal{G}^0_{C,\alpha_1\alpha^\prime_1 }(\boldsymbol{k}_1; 0^+ ) $ is equal to $ \tau_1^+ - \tau_1.$ 
The contour times $ \tau_1^+ $ and  $\tau_1$ both belong to the same sub contour $C_\mu$ $(\mu=-,+).$  

Second, we perform the real-time projection on the contour times  $  \tilde{\tau}_1,   \tilde{\tau}_2$, and $ \check{\tau}_1$ and rewrite the right-hand side of Eq. \eqref{21greensfunction1} by the real-time Green's functions of TI surface state and magnon.
Then, the perturbative Keldysh Green's function $G^{(2, 1 )}_{C,\alpha\alpha^\prime}(\boldsymbol{k}\tau;\boldsymbol{k}^\prime \tau^\prime)$ in the right-hand side of Eq. \eqref{21greensfunction1}
becomes the lesser real-time Green's function which we write as $G^{< (2, 1 )}_{\alpha\alpha^\prime}(\boldsymbol{k}t;\boldsymbol{k}^\prime t^\prime)$ with $t^\prime = t^+$ (see also Eq. \eqref{TIECCexpv2}).
Let us denote the real-time variables corresponding to $  \tilde{\tau}_1,   \tilde{\tau}_2$, and $ \check{\tau}_1$ as $  \tilde{t}_1,   \tilde{t}_2$, and $ \check{t}_1$, respectively. 
Then, by using the first formula in Eq.  \eqref{CtoRtimeintegral6}, we obtain
 \begin{widetext} 
 \begin{align}
 G^{< (2, 1 )}_{\alpha\alpha^\prime}(\boldsymbol{k}t ;\boldsymbol{k}^\prime t^\prime)&= 
 i   { \rm{sgn}}(B_0) \left( - \frac{i}{\hbar} \right)^3  \left( \hbar \gamma B^{\rm{ext}}       \right)^2   \frac{  J^{\rm{exc}}  n^{\rm{2D}}_{\rm{loc}}  S_0    }{4}
  \int d\tilde{t}_1   d\tilde{t}_2  d\check{t}_1  e^{ - i \omega^{\rm{ext}} ( \tilde{t}_1 -  \tilde{t}_2 )   }  
     \bar{D}^{\rm{a}}  (  \boldsymbol{0}, \tilde{t}_2 -  \check{t}_1   )  \bar{D}^{\rm{r}}  (  \boldsymbol{0}, \check{t}_1 -  \tilde{t}_1   ) \notag\\
 & \times 
  \left( \bar{g}^{\rm{r}} _{\alpha \alpha^\prime_1} (  \boldsymbol{k},  t - \check{t}_1   )  \sigma^y_{ \alpha^\prime_1 \alpha_1 }    \bar{g}^{ < } _{\alpha_1 \alpha^\prime} (  \boldsymbol{k},  \check{t}_1 - t   )    
 +
 \bar{g}^{ < } _{\alpha \alpha^\prime_1} (  \boldsymbol{k}, t - \check{t}_1  )  \sigma^y_{ \alpha^\prime_1 \alpha_1 }    \bar{g}^{ \rm{a} } _{\alpha_1 \alpha^\prime} (  \boldsymbol{k}, \check{t}_1 - t   )  \right)  \delta_{ \boldsymbol{k}  \boldsymbol{k}^\prime } , 
\label{21greensfunction2}
\end{align}
\end{widetext}
where we have used  $ \int_C d \tilde{\tau}_1 d \tilde{\tau}_2    \mathcal{D}^0_{C}(\boldsymbol{0}; \tilde {\tau}_1 ,  \tilde {\tau}_2 )  
=  \int d \tilde{t}_1 d \tilde{t}_2  \Big{(}  \bar{D}^{\rm{t}} -  \bar{D}^{<}  -  \bar{D}^{>}  +  \bar{D}^{\rm{\tilde{t}}} 
 \Big{)} (\boldsymbol{0}, \tilde {t}_1 -  \tilde {t}_2 )  
=0.$ Further, the term $ \mathcal{G}^0_{C,\alpha \alpha^\prime }(\boldsymbol{k}; 0^+ )   \mathcal{G}^0_{C,\alpha_1\alpha^\prime_1 }(\boldsymbol{l}; 0^+ ) $ in Eq. \eqref{21greensfunction2} 
vanishes since 
 it describes the disconnected diagram: We denote the real time variables $t_1$ and $t_1^\prime$ which are the real-time projection of the contour times $\tau_1$ and $\tau_1^\prime$, respectively.
They satisfy $\tau_1 < \tau_1^\prime$ because in the exchange-interaction Hamiltonian $V^{\rm{sd} }$ the operator $\psi^\dagger_{\alpha^\prime_1 }(\boldsymbol{l}, \tau_1^\prime)$ comes to the left side of $\psi_{\alpha_1 }(\boldsymbol{l}, \tau_1)$.
When $\tau_1, \tau^\prime_1 \in C_-$ we have $t_1 < t_1^\prime$ and 
$ \mathcal{G}^0_{C,\alpha_1\alpha^\prime_1 }(\boldsymbol{l}; \tau_1 , \tau_1^\prime ) =  \bar{g}^{ < } _{\alpha_1 \alpha^\prime_1} (  \boldsymbol{l}, t_1 - t_1^\prime  ) $
whereas for $\tau_1, \tau^\prime_1 \in C_+$ we obtain $t_1 > t_1^\prime$ and 
$ \mathcal{G}^0_{C,\alpha_1\alpha^\prime_1 }(\boldsymbol{l}; \tau_1 , \tau_1^\prime ) =  \bar{g}^{ < } _{\alpha_1 \alpha^\prime_1} (  \boldsymbol{l}, t_1 - t_1^\prime  )$.
Hence, we  have $ \int_C d\tau_1   \lim_{\substack{ \tau_1^\prime \to \tau_1^+ }}   \mathcal{G}^0_{C,\alpha_1\alpha^\prime_1 }(\boldsymbol{l}; \tau_1 , \tau_1^\prime ) = 
\int dt_1 \big{[} \bar{g}^{ < } _{\alpha_1 \alpha^\prime_1} (  \boldsymbol{l}, 0^-  ) - \bar{g}^{ < } _{\alpha_1 \alpha^\prime_1} (  \boldsymbol{l}, 0^+  ) \big{]} = 0$. 
Here $0^-$ is the negative infinitesimal. 
For the detail treatments on real-time projection as well as the real-time integration see subSec. \ref{real-timeP} in Appendix \ref{appendixB}. 
Third, what we do is we perform the Fourier transforms on the above Green's functions as, for instance,  
 $ \bar{D}^{ \rm{a} }  (  \boldsymbol{0}, \tilde{t}_1 -  \check{t}_1   )  = \int \frac{d \tilde{\omega}}{2\pi} e^{ -i\tilde{\omega} ( \tilde{t}_1 -  \check{t}_1 ) }  \bar{D}^{ \rm{a} }  (  \boldsymbol{0}, \tilde{\omega}   )$ and 
$\bar{g}^{\rm{t}} _{\alpha \alpha^\prime_1} (  \boldsymbol{k},  t - \check{t}_1   ) =  \int \frac{d \omega}{2\pi} e^{- i \omega ( t -  \check{t}_1 ) } \bar{g}^{\rm{t}} _{\alpha \alpha^\prime_1} (  \boldsymbol{k},  \omega). $
 
 \begin{figure}[b]
\includegraphics[width=0.34\textwidth]{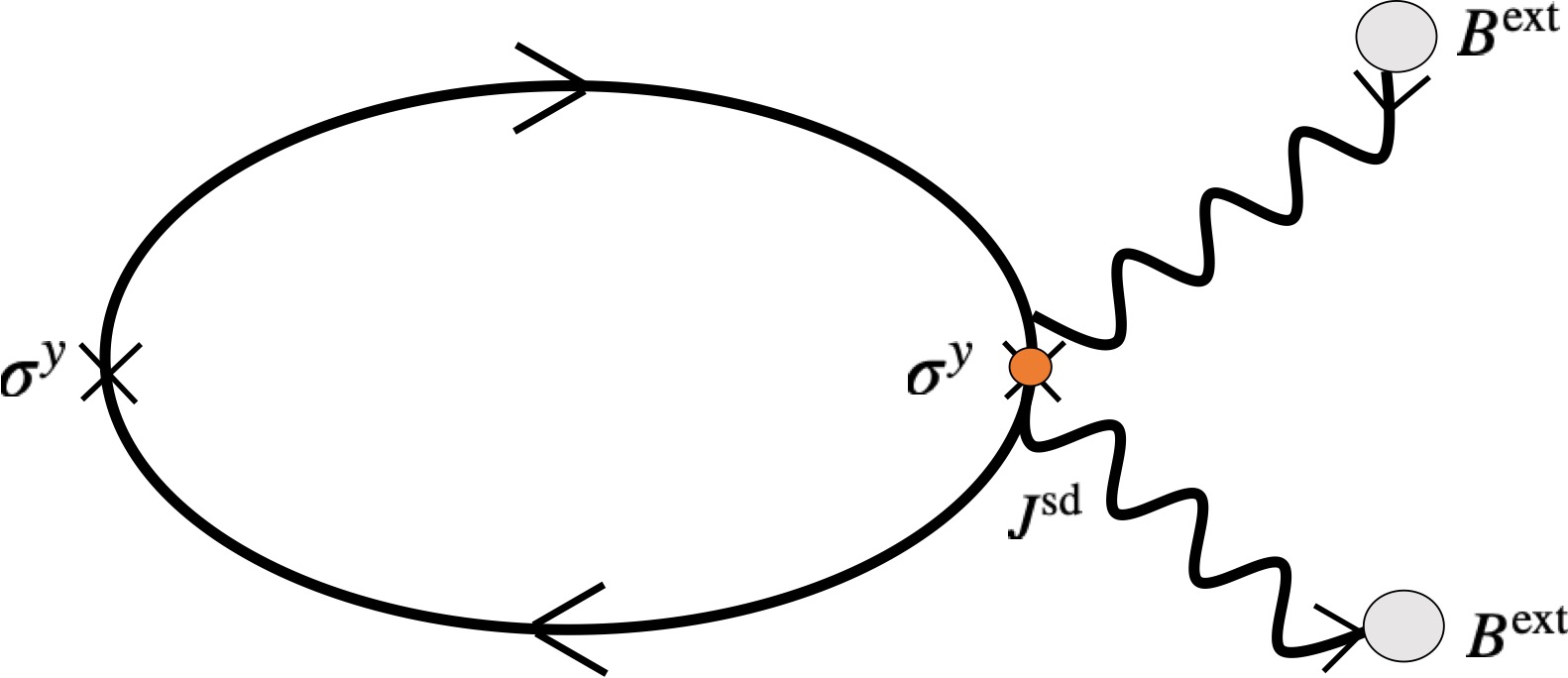}
\caption{Feynman diagram for  the spin-pumping-induced electric current described by Eq. \eqref{21electriccurrent1}. 
It consists of two TI-surface-state Green's functions (solid lines), two magnon Green's functions (wavy lines), and two vertices denoted by crosses. 
As described by the orange circle, the exchange interaction between the zero-momentum magnon and the TI surface state occurs at the right vertex leading to the generation of electric current on the surface of TI.}
\label{Feynmandiagram}  
\end{figure}

\begin{figure}[t]
\includegraphics[width=0.45 \textwidth]{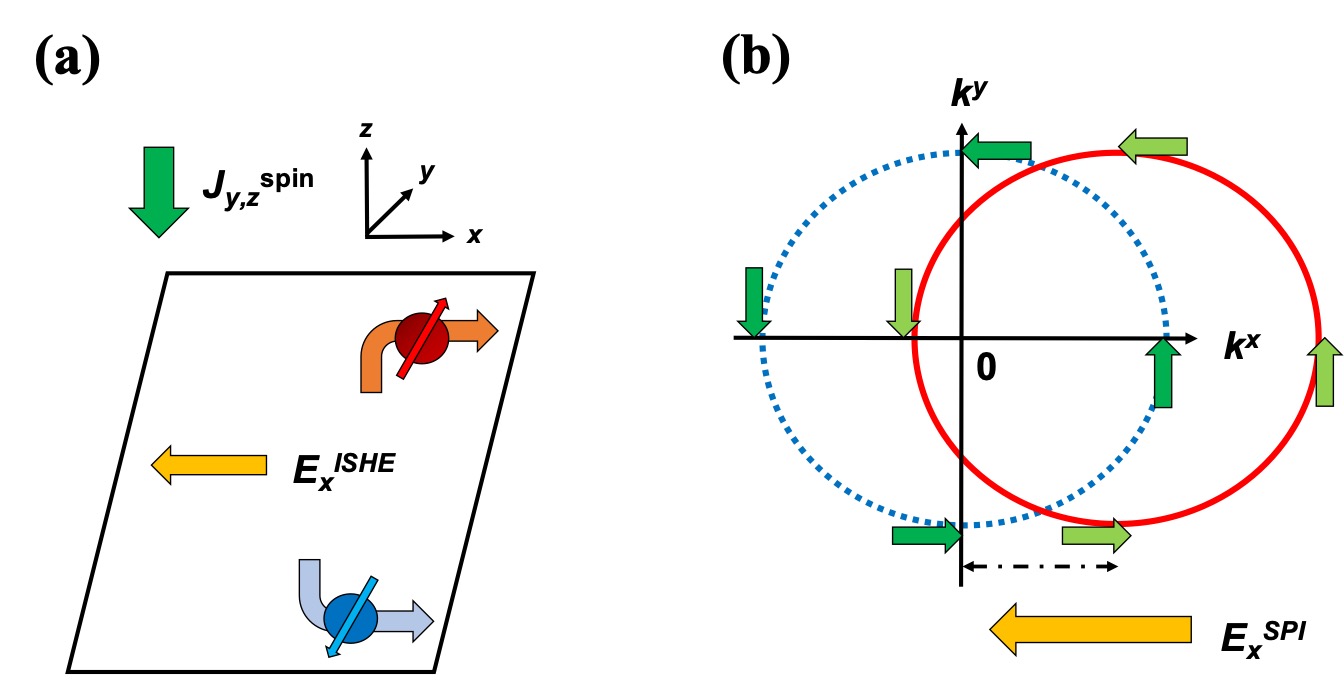}
\caption{ (a) Schematic of the generation of the electric field  $E^{\rm{ISHE}}_{x}$. When the spin current $J^{\rm{spin}}_{y,z}$ is injected to the non-magnetic metal, 
owing to the spin-orbit interaction the inverse spin Hall effect is generated so that the both electrons of spins polarized in the positive and negative $y$ directions accumulate on the edge of the sample.
 As a result, the electric field $E^{\rm{ISHE}}_{x}$ and the associated electric current  $j^{\rm{ISHE}}_{\rm{c}}$ emerge. 
(b)  Schematic of the generation of the spin-pumping-induced electric current $\bar{ j}^{(2,1)}_{x} $.
When we drive the FMR and the associated spin pumping, the zero-momentum magnon couples with the TI surface state through the exchange interaction. 
Due to the spin-momentum locking, the TI surface state effectively experiences the zero-momentum magnon as the electric field.     
As a result,  the spin-pumping-induced electric field $E^{\rm{SPI}}_{x}$ as well as the spin-pumping-induced electric current $\bar{ j}^{(2,1)}_{ x} $ are generated (inverse Edelstein effect).
 It is described as the flow of Fermi circle of TI surface state.} 
\label{ISHETI} 
\end{figure}

 Then, using  Eq. \eqref{impurityaveragedGreen'sfunctionrelations1}  the right-hand side of Eq. \eqref{21greensfunction2} is rewritten by the retarded and advanced Green's functions in Eqs. \eqref{unperturbedTIGreen'sfunctions} and  \eqref{unperturbedmagnonGreen'sfunctions}. By performing the temporal integrals $ \int d\tilde{t}_1   d\tilde{t}_2  d\check{t}_1$ and from Eq. \eqref{STaveragecurrent}, the $x$-component averaged electric current density of TI surface state becomes  
\begin{widetext}
\begin{align}
\bar{ j}^{ (2, 1) }_{  x} &=  \left( - ev_{\rm{F}}   { \rm{sgn}}(B_0)  \right)  \left( - \frac{i}{\hbar} \right)^3  \left( \hbar \gamma B^{\rm{ext}}       \right)^2 
\left( \frac{ J^{\rm{exc}} S_0 n^{\rm{2D}}_{\rm{loc}} }{4}  \right)
\bar{D}^{\rm{r}}  (  \boldsymbol{0}, \omega^{\rm{ext}}   )   \bar{D}^{\rm{a}}  (  \boldsymbol{0}, \omega^{\rm{ext}}   ) \notag\\
& \times \int \frac{d^2kd\omega}{ (2\pi)^3 } f( \hbar \omega )
 \left[  \left( \bar{g}^{\rm{a}} _{\alpha \alpha^\prime_1} (  \boldsymbol{k}, \omega   )  \sigma^y_{ \alpha^\prime_1 \alpha_1 }    \bar{g}^{\rm{a}} _{\alpha_1 \alpha^\prime} (  \boldsymbol{k}, \omega   )    
 -
 \bar{g}^{\rm{r}} _{\alpha \alpha^\prime_1} (  \boldsymbol{k}, \omega   )  \sigma^y_{ \alpha^\prime_1 \alpha_1 }    \bar{g}^{\rm{r}} _{\alpha_1 \alpha^\prime} (  \boldsymbol{k}, \omega   )  \right)
 \sigma^y_{ \alpha^\prime\alpha }    
 \right],
\label{21electriccurrent1}
\end{align}
 \end{widetext}
where  $f( \hbar \omega)=   \left(  1 + e^{ \beta   \hbar \omega  } \right)^{-1} $  and  we have taken a continuum limit $V^{-1} \sum_{ \boldsymbol{k} } \to \int d^2k / (2\pi)^2$.
The right-hand side of Eq. \eqref{21electriccurrent1} represents the way the electric current of TI surface state is induced by the spin pumping due to the external magnetic field $H^{\rm{ext}}$ and the exchange interaction $V^{\rm{exc}}$.
To see this clearly, let us describe $\bar{ j}^{ (2, 1) }_{x} $ diagrammatically and present this in Fig. \ref{Feynmandiagram}. 
The solid and the wavy lines represent the Green's function of TI surface state (fermion line) and that of magnon (boson line),  respectively.
The two vertices are described by crosses where the energy and momentum conserve. 
The gray and orange circles denote the amplitude of ac magnetic field  $ B^{\rm{ext}}$ and the exchange-interaction strength $J^{\rm{exc}}$, respectively. 
The Pauli matrix in the left side originates in the generator of $x$-component electric current while the right one is coming from the $y$-component exchange interaction.
The retarded Green's function $\bar{D}^{\rm{r}}  (  \boldsymbol{0}, \omega^{\rm{ext}}   ) $ appearing in this diagram describes the
emission process of magnon with the zero momentum and the energy $\hbar \omega^{\rm{ext}} $ going from orange to gray circles 
whereas the advanced Green's function $\bar{D}^{\rm{a}}  (  \boldsymbol{0}, \omega^{\rm{ext}}   ) $ represents the absorption process going from gray to orange.
In the diagram in Fig. \ref{Feynmandiagram}, the energy and momentum of TI surface state remains unchanged. 
This is because, first,  the emission and absorption processes of magnon of the energy $\hbar \omega^{\rm{ext}} $ occur with each process occurring once.    
Second, the magnon does not carry momentum since the ac magnetic field $H^{\rm{ext}}$ is spatially homogeneous and so does the Fourier transform of exchange interaction $J^{\rm{exc}}_{(\boldsymbol{q} \boldsymbol{p})} $:
for the $y$-component it is  described by the constant $J^{\rm{exc}}$ (see Eq.  \eqref{momentumexchangehamiltonian1}). 
Since we have overlooked at the structure of our diagram,  let us now analyze the mechanism of the spin-pumping-induced electric current.
Initially, the TI surface state is in the thermal equilibrium $\rho_{\rm{GC}}(H,\beta,\epsilon_{ \rm{F}  }) $ and the origin of Fermi sphere of TI surface state is at $ \boldsymbol{k}_0 = ( 0, 0 ) $.
When the FMR is triggered at $t=t_0$, the localized spin $ \boldsymbol{S}_i $ starts to show its dynamics described by the Landau-Lifshitz-Gilbert equation (see Eq. \eqref{LLGEq}) and
the magnon of zero momentum and frequency  $ \omega^{\rm{ext}} $ emerges.
It is the fluctuation of the saturation magnetization in the $y$ direction created by the external magnetic field $B_0$. 
After then, the spin pumping occurs associating with the spin current  $ J_{y,z}^{\rm{pump}}$ flowing from FM to the surface of TI.   
The zero-momentum magnon is going to be the carrier of it. In other words, the spin current  $ J_{y,z}^{\rm{pump}}$ is the flow of zero-momentum magnon.
The magnon couples with the TI surface state through the exchange interaction $V^{\rm{exc}}$. 
Then owing to the spin-momentum locking,  the magnon acts like an additional momentum of TI surface state. 
This means that effectively TI surface state experiences the coupling between the magnon as an electric field being applied and a non-equilibrium state of TI surface state is driven. 
Such a situation can be described as the deviation of the TI-surface-state Fermi circle from the origin (see also Fig. \ref{ISHETI} (b)). 
On the other side, the TI surface state is affected by the impurity potential $H^{\rm{imp}}$ given by Eq. \eqref{impurityhamiltonian1}.
Then as time goes by, the effect of effective electric field of magnon and the impurity effect  $H^{\rm{imp}}$ are going to get balanced. 
As a result,  the TI surface state and the magnon both relax to the non-equilibrium steady state and the static electric field is created on the surface of TI. 
Let us call it the spin-pumping-induced electric field  $ E_x^{\rm{SPI}} $. 
At the non-equilibrium steady state, the TI surface state experiences the  $ E_x^{\rm{SPI}} $ and the spin-pumping-induced electric current $\bar{ j}^{ (2, 1) }_{x} $  flows on the TI surface as the response to it.
To make the relation between $\bar{ j}^{ (2, 1) }_{x} $ and  $ E_x^{\rm{SPI}} $ clear, let us rewrite  $\bar{ j}^{ (2, 1) }_{x} $ with using the electrical conductivity $\sigma^{\rm{TI}}_{xx}$  as $ \bar{ j}^{(2,1)}_{ x}  =    \sigma^{\rm{TI}}_{xx}   E_x^{\rm{SPI}}$. 
As in the case of Dirac electrons in graphene, the electrical conductivity of TI surface state $ \sigma^{\rm{TI}}_{xx} $ can be calculated by using the Boltzmann equation \cite{Nomuragraphene}. 
We obtain $ \sigma^{\rm{TI}}_{xx} =  \frac{ \epsilon_{\rm{F}} \tau^{\rm{rel}}_{\rm{TI}} }{2 \hbar } \cdot  \frac{e^2}{2\pi \hbar} $.
On the other side, the formula of $ E_x^{\rm{SPI}}$ is obtained by evaluating the right-hand side of Eq. \eqref{21electriccurrent1}.
For doing this,  first we remark that the denominator in the right-hand side of Eq. \eqref{21electriccurrent1} is a function of the dispersion of TI surface state ($ \omega^{\rm{TI}}_{\boldsymbol{k}}= \hbar^{-1}\epsilon^{\rm{TI}}_{\boldsymbol{k}}$), 
and thus,  it is the function of the absolute $k = \sqrt{ ( k^x )^2 + ( k^y )^2  }$. Hence, it means that the denominator in the right-hand side of Eq. \eqref{21electriccurrent1} is an even and symmetric function of $k^x$ and $ k^y $. 
Due to this fact, for the numerator in the right-hand side of Eq. \eqref{21electriccurrent1}  the terms proportional to $k^x k^y $ as well as $ ( k^x )^2 - ( k^y )^2 $ vanish. 
As a result, the only terms which remain are the products of two diagonal elements of TI-surface state Green's function, i.e., $ \big{(} \bar{g}^{\rm{a(r)}} _{\uparrow\uparrow} (  \boldsymbol{k}, \omega ) \big{)}^2 $
and/or $ \big{(} \bar{g}^{\rm{a(r)}} _{\downarrow\downarrow} (  \boldsymbol{k}, \omega ) \big{)}^2 $. 
In the following evaluation, we only retain the first term of $\bar{g}^{\rm{r(a)}} _{\alpha \alpha^\prime} (  \boldsymbol{k}, \omega   ) $ in Eq. \eqref{unperturbedTIGreen'sfunctions} 
since only the electronic state in the vicinity of Fermi-energy level contributes to the electric transport. Next, we perform $ \boldsymbol{k}$ and $\omega$ integrals with using three types of approximations. 
We first do from the $\omega$ integral and rewrite the integrand with the derivative term  $\partial f ( \hbar \omega) /\partial (\hbar \omega)$.  
As the first approximation, we take the low-temperature limit ($\beta \to \infty$) and we obtain $ \partial f ( \hbar \omega) /\partial (\hbar\omega) = - \delta (\hbar\omega).$ 
By performing the $\omega$ integral, the integrand becomes the function of the relaxation time $\tau^{\rm{rel}}_{\rm{TI}}$, the TI-surface-state dispersion $\omega^{\rm{TI}}_{\boldsymbol{k}}$, 
and the Fermi energy $\omega_{\rm{F}}$ $(=\hbar^{-1} \epsilon_{\rm{F}})$ given as $ \frac{1/2\tau^{\rm{rel}}_{\rm{TI}}} {  \big{(}   ( \omega^{\rm{TI}}_{ \boldsymbol{k} }  - \omega_{\rm{F}} )^2 +  (1/2\tau^{\rm{rel}}_{\rm{TI}})^2   \big{)} } $.
Next,  we rewrite the $ \boldsymbol{k}$ integral in the following way:  
$ \frac{d^2k}{(2\pi)^2} = \tilde{N}(\xi_{ \boldsymbol{k} } ) d \xi_{ \boldsymbol{k} } $, 
where $ \tilde{N}(\xi_{ \boldsymbol{k} }) = \frac{ ( \xi_{ \boldsymbol{k} } + \epsilon_{\rm{F}} ) }{ 2\pi (\hbar v_{\rm{F}})^2 }$ is the density of states per volume
and $ \xi_{ \boldsymbol{k}} $ is the energy of the TI surface state measured with respect to the 
Fermi energy. It is defined by  $  \xi_{ \boldsymbol{k}} = \epsilon^{\rm{TI}}_{\boldsymbol{k}}  - \epsilon_{\rm{F}} $.  
As a result, the integrand becomes $ \tilde{N}(\xi_{ \boldsymbol{k} } ) \times  \frac{\hbar /2\tau^{\rm{rel}}_{\rm{TI}}} {   \xi_{ \boldsymbol{k} }^2 +  ( \hbar /2\tau^{\rm{rel}}_{\rm{TI}})^2 } $.   
Then as a second approximation, we regard only the electronic state in the vicinity of Fermi surface contributes to the electric current. 
In other words, the TI surface state depends weakly on the density of states $ \tilde{N}(\xi_{ \boldsymbol{k} } )$. Therefore, we take $ \tilde{N}(\xi_{ \boldsymbol{k} } ) \approx \tilde{N}(0).$  
On the other side, the lower limit of $ \xi_{ \boldsymbol{k}} $-integral is $-\epsilon_{\rm{F}}$.  We consider that on the surface of area $V$ a huge number of electrons are contained. 
Hence, as a third approximation we take the number density of TI surface state  $ n^{\rm{TI}}_{\rm{2D}} $ to be sufficiently large.  Since the number density $ n^{\rm{TI}}_{\rm{2D}} $ is related to the Fermi energy $\epsilon_{\rm{F}}$ as 
$\epsilon_{\rm{F}} = \hbar v_{\rm{F}} \sqrt{4\pi n^{\rm{TI}}_{\rm{2D}} }, $  we take $\epsilon_{\rm{F}}  \to \infty $ (see also the argument below Eqs.  \eqref{1BAselfenergy} and \eqref{retarded1BAselfenergy}).
By using these three types of approximations and performe the $ \xi_{ \boldsymbol{k}} $-integral, we have
\begin{widetext}
\begin{align}
E_x^{\rm{SPI}}  (\omega^{ \rm{ext} } ,  B_0) &= - { \rm{sgn}}(B_0)  \left(   \frac{ J^{\rm{exc}} S_0   n^{\rm{2D}}_{\rm{loc}}      }{ 4 e v_{\rm{F}} \tau^{\rm{rel}}_{\rm{TI}}  }  \right)
\frac{   ( \gamma B^{\rm{ext}}    )^2   }{  (  \omega^{\rm{ext}} -  \omega^{\rm{FM}}_{\boldsymbol{0}}   )^2 +  (  \alpha \omega^{\rm{ext}}  )^2 },
 \label{SPIelectricfield}
\end{align} 
\end{widetext}
where $ \omega^{\rm{FM}}_{\boldsymbol{0}}  = \gamma |B_0|$. 
Consequently, when the FMR occurs with the frequency  $\omega^{\rm{ext}}$,  
the electric current  $\bar{ j}^{(2,1)}_{x}$ as well as the electric field $E_x^{\rm{SPI}} $ are induced by the spin pumping  on the surface of TI. It flows perpendicular to the precession axis ($y$ axis) of FMR owing to the spin-momentum locking.
It is proportional to the square of the ac-magnetic-field amplitude $B^{\rm{ext}}$ describing that
it is the non-linear (quadratic) response to the external ac magnetic field. 
In other words, it is proportional to the power of the applied electromagnetic wave (microwave). 
Like the FMR (magnon) spectrum, the electric current $\bar{ j}^{(2,1)}_{x}$  (or the electric field $E_x^{\rm{SPI}} $) is described by the quantities $  \gamma B^{\rm{ext}} $, $\omega^{\rm{ext}}$, $\omega^{\rm{FM}}_{\boldsymbol{0}} $, 
and the Gilbert-damping constant $\alpha$. Indeed, the spectral function of magnon can be obtained by multiplying the factor $  \alpha \omega^{\rm{ext}}$ to 
the third factor of $E_x^{\rm{SPI}} $ in Eq.  \eqref{SPIelectricfield}: $ \alpha \omega^{\rm{ext}}\times \frac{   1   }{  (  \omega^{\rm{ext}} -  \omega^{\rm{FM}}_{\boldsymbol{0}}   )^2 +  (  \alpha \omega^{\rm{ext}}  )^2 }$. In other words, the retarded Green's function of magnon is equivalent to the magnetic susceptibility (see Eq. \eqref{magneticsusceptibility} and the argument below it). 
Physically, this represents the absorption energy of localized spin which we need to drive the FMR (see also the argument after Eq. (8) in \cite{spintronicsRMP2005}).  
The electric field $E_x^{\rm{SPI}} $ depends on both magnetic quantities including  $  \gamma B^{\rm{ext}}, \alpha$, the exchange interaction strength $J^{\rm{exc}}$, the density of localized spin $n^{\rm{2D}}_{\rm{loc}} $, 
and those of TI such as Fermi velocity $v_{\rm{F}}$,  and the relaxation time $\tau^{\rm{rel}}_{\rm{TI}}$.
This is natural and reasonable because the spin-pumping-induced electric field  $E_x^{\rm{SPI}} $ is realized at the non-equilibrium steady state
owing to the commensuration of the effective electric field  of magnon and the impurity effect $H^{\rm{imp}} $ mediated by the exchange interaction. 
Based on the Feynman diagram in Fig. \ref{Feynmandiagram}, we can understand why the spin-pumping-induced electric current $\bar{ j}^{(2,1)}_{ x}$ is the quadratic response to the ac magnetic field in the following way.  
First, the ac magnetic field is used to drive the FMR and the associated zero-momentum magnon which couples with the TI surface state through the exchange interaction.
 Second, to generate the transport phenomena of TI surface state we need to drive the magnon with the ac magnetic field once more. 
 As a result, the electric current of TI surface state becomes the quadratic response to the ac magnetic field such that both the emission and absorption processes of zero-momentum magnon occur. 
 Indeed, this naturally reflects that the spin-pumping-induced electric current is generated by the electromagnetic wave whose power is quadratic to the amplitude of ac magnetic field.
The spin-pumping-induced electric current $\bar{ j}^{(2,1)}_{x} $  (or the spin-pumping-induced electric field  $ E_x^{\rm{SPI}} $) gets larger by raising the ac-magnetic-field amplitude $B^{\rm{ext}}$ (or the power of electromagnetic wave) and by choosing the ferromagnetic material exhibiting a strong exchange coupling strength.   
Such a feature is reflecting that the surface of ferromagnetic TI has a high-performing functionality of generating the electric charge current by magnetic controlling. 

To make our understanding on the spin-pumping-induced electric current $\bar{ j}^{(2,1)}_{x} $ better, let us compare it with the electric field associated with the inverse spin Hall effect by using the illustrations presented in Fig. \ref{ISHETI}. 
The inverse spin Hall effect occurs in, for instance, the hybrid system comprise of FM and non-magnetic heavy metal exhibiting strong spin-orbit interaction, for example, the heterojunction of NiFe and Pt \cite{yohnumaetalsspinpumpingPRB2014,interfacemagnetismRMP2017,spintronicsreviewnpj2018,SHERMP2015}. 
When we inject the $y$-polarized spin current to the non-magnetic metal flowing in the $z$ direction, due to the spin-orbit coupling both the electrons whose spins are polarized in the positive and negative $y$ directions 
flow parallel into the $x$ direction and accumulate to the edge of sample. As a result, the electric field, namely, $E^{\rm{ISHE}}_{x}$  emerges in the $x$ direction. 
Simultaneously, the associated electric current $j^{\rm{ISHE}}_{\rm{c}}$ flows in the same direction (Fig. \ref{ISHETI}(a)).    
This is the phenomenon in a three-dimensional bulk system.  
In contrast, our spin-pumping-induced electric current $\bar{ j}^{(2,1)}_{x} $ is the phenomenon intrinsic in the two-dimensional surface system. 
The mechanism of its generation is not due to the accumulation of electrons on the edge of sample 
but due to the effective electric field of magnon via the spin-momentum locking.  
As illustrated in Fig. \ref{ISHETI}(b), the spin-pumping-induced electric current can be described as the flow of Fermi circle of TI surface state. 
It is nothing but the inverse Edelstein effect which is also realized in systems possessing Rashba interfaces \cite{interfacemagnetismRMP2017,spintronicsreviewnpj2018}. 

\begin{figure}[t] 
\includegraphics[width=0.45 \textwidth]{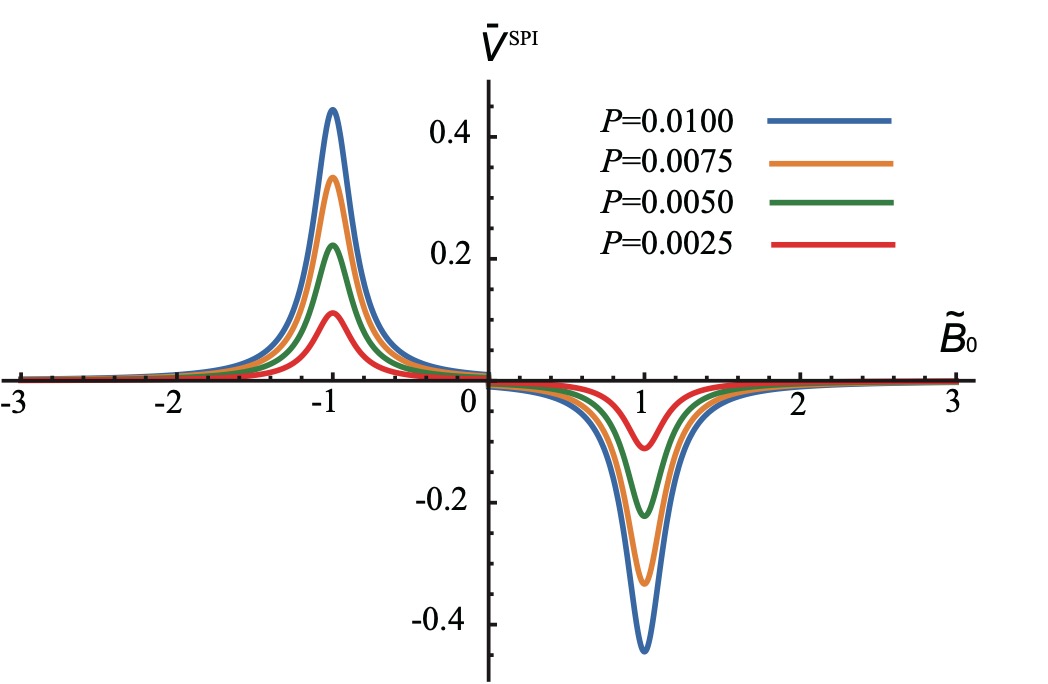}
\caption{ Plots of spin-pumping-inducedelectric voltage $\bar{V}^{\rm{SPI}}_{x} $ defined by Eq.  \eqref{effectivevoltage}. 
The vertical axis represents the dimensionless magnetic field  $     \tilde{B}_0  $ whereas the horizontal axis  represents the spin-pumping-induced  electric voltage $\bar{V}^{\rm{SPI}}_{x} $. 
The blue, orange, green, and red curves are for $ P = 0.0100, 0.0075, 0.0050$, and $ 0.0025$, respectively.  
For all four curves, we take the Gilbert damping constant $\alpha$ to be equal to 0.15. } 
\label{voltage} 
\end{figure}

Finally, let us make a qualitative comparison between our result and the experimental results \cite{ShiomiTISP,expFermiEdependencemagneticTI2}.
What has been measured in these experiments are the electric voltage emerged on the surface due to the spin pumping.
Therefore, we calculate the spin-pumping-induced electric voltage and compare its characteristic with the experimental results. 
Before we give a detailed argument, we note here that in \cite{ShiomiTISP} the direction of static magnetic field $\boldsymbol{B}_0$ (the precession axis of FMR)  
is taken to be parallel to the $y$ axis while in \cite{expFermiEdependencemagneticTI2} it is taken to be in the $x$ axis. 
Since the essence of physics does not change, as we did in Sec. \ref{modelformalism} we take the precession axis of FMR to be in the $y$ axis (thus, the electric current of TI surface state or electric voltage emerges in the $x$ direction).
To make our argument clear and simple, in the following we introduce the effective electric voltage by using $E^{\rm{SPI}}_{x}$.
First, as we see in Eq. \eqref{SPIelectricfield} the electric field $E^{\rm{SPI}}_{x}$ is spatially homogeneous along the $x$ direction.
Thus, by multiplying $E^{\rm{SPI}}_{x}$ with the length of TI surface in the $x$ direction  $l_x^{\rm{TI}}$, 
we obtain the electric voltage in the $x$ direction and call it as $V^{\rm{SPI}}_{x} $.
Next, we divide $V^{\rm{SPI}}_{x} $ by the factor $ \left(  - \frac{ J^{\rm{exc}} S_0   n^{\rm{2D}}_{\rm{loc}}  l_x^{\rm{TI}}    }{ 4  e v_{\rm{F}} \tau^{\rm{rel}}_{\rm{TI}} B_0^2 }  \right)$ 
because essentially its characteristic is represented by the Gilbert-damping constant $\alpha$, the external frequency $ \omega^{\rm{ext}} $, 
and the frequency $ \omega^{\rm{FM}}_{\boldsymbol{0}}  = \gamma |B_0| = \gamma\left( {\rm{sgn}}(B_0)B_0 \right) $. 
In addition, in the experiment the external frequency $ \omega^{\rm{ext}} $ is fixed whereas the magnetic field $B_0$ varies from positive to negative values.
By taking account of this, we take $ \omega^{\rm{ext}} $ to be the positive constant and introduce the ``spin-pumping-induced electric voltage $\bar{V}^{\rm{SPI}}_{x} $'' defined as the function of $B_0$ as
\begin{align}
\bar{V}^{\rm{SPI}}_{x}  ( \tilde{B}_0) & = - \Theta ( \tilde{B}_0  )
 \frac{   e P  }{  \left(  \tilde{B}_0  - 1 \right)^2 +  \alpha^2   }  \notag\\
 & +
 \Theta ( - \tilde{B}_0  )  \frac{   e P  }{  \left(  \tilde{B}_0 + 1 \right)^2 +  \alpha^2   } , 
 \label{effectivevoltage}
\end{align} 
where $ \tilde{B}_0 = (\gamma B_0) /  \omega^{\rm{ext}} $ is the dimensionless magnetic field and  $P = ( B^{\rm{ext}} )^2$. 
It is the quantity describing the power of electromagnetic field which we apply to derive the FMR.
$ \Theta ( \pm \tilde{B}_0 ) $  is the Heaviside step function.
We plot $\bar{V}^{\rm{SPI}}_{x}  ( \tilde{B}_0)$ in Fig. \ref{voltage}  by taking the electromagnetic-wave power $ P $ as a parameter while we fix the Gilbert-damping constant $\alpha$  to 0.15.
Here we plot $ \bar{V}^{\rm{SPI}}_{x}  $  for four different conditions; $ P = 0.0100, 0.0075, 0.0050$, and $ 0.0025$.
The full width of half maximum is equal to the Gilbert damping constant $\alpha.$ 
The most striking features of  $\bar{V}^{\rm{SPI}}_{x}   $ are (i) the emergence of two side peaks and (ii) the linear scaling of two peak values with respect to the power $ P $;
the two side peaks locate at  $  \tilde{B}_0  =\pm1. $ 
The values of two peaks  have the same absolute values ($=eP/\alpha^2 $) but the signs are opposite.
Let us now look at the experimental data \cite{ShiomiTISP,expFermiEdependencemagneticTI2}. 
First, in  \cite{ShiomiTISP} the (bulk insulating) TIs were chosen as Bi$_{1.5}$Sb$_{0.5}$Te$_{1.7}$Se$_{1.3}$ naming BSTS and Sn-doped Bi$_{2}$Te$_{2}$Se.
On the other hand, for the ferromagnetic material they choose Ni$_{81}$Fe$_{19}$.
Let us focus on Figs. 3(b) or 4(a) and 3(c). 
Fig. 3(b) is the experimental data of electric voltage for sample BSTS/Ni$_{81}$Fe$_{19}$ with the sample size of  BSTS is $4\times3\times0.1$ mm$^3$ for four different microwave-power conditions; 0.2mW, 0.15mW, 0.10mW, and 0.05 mW. 
 Fig 4(a) is the result of electric voltage for samples BSTS/Ni$_{81}$Fe$_{19}$ with three different sample sizes of BSTS; $4\times1\times0.1$ mm$^3$, $4\times3\times0.1$ mm$^3$, and $2\times1.5\times0.2$ mm$^3$.
They are plotted as functions of static magnetic field which corresponds to our $B_0$ (or $\tilde{B}_0$). 
Both of them show two side peaks as discussed in the previous description. 
Two peak spots appear symmetrically with respect to the origin of the magnetic-field axis and the two peak values have (almost) the same absolutes with opposite signs.
The similar result is reported in \cite{expFermiEdependencemagneticTI2}.
 In this work, the magnetic TI was engineered by creating the heterostructure of YIG and Cr-doped TI: YIG/Cr$_{0.08}$(Bi$_{0.37}$Sb$_{0.63}$)$_{1.92}$ Te$_{3}$.
 We will focus on Fig. 2(c) where the electric voltage is plotted as a function of magnetic field. It shows the similar features as the results shown in Fig. 3(b) or  Fig 4(a) in  \cite{ShiomiTISP}:
 the emergence of two side peaks having opposite signs. The difference between the electric voltage in  Fig. 3(b) or Fig 4(a) in  \cite{ShiomiTISP} and that in Fig. 2(c) in \cite{expFermiEdependencemagneticTI2}
 is that the signs of two peaks;  in  Fig. 3(b) the peak value at positive magnetic field is negative while it is positive in Fig. 2(c) it is positive.
 Such an opposite-sign behavior, however, is not essential for our analysis and we will not refer to its origin. 
 We note that in \cite{SPISHETINanoLett2015} the measurement of spin-pumping-induced voltage was performed using the bilayer systems of Bi$_{2}$Se$_{3}$ (TI) and CoFeB (ferromagnet). 
 In this experiment, it is considered that the dominant contribution to the spin-pumping-induced voltage is coming from the inverse spin Hall effect (bulk state) rather than the inverse Edelstein effect (surface state).
 Thus, although the measured spin-pumping-induced voltage shows similar characteristics (Figs. 2 and 3(a) and (b)) with $\bar{V}^{\rm{SPI}}_{x} $, we will not make a comparison with these experimental results.
 Next, let us take a look at Fig. 3(c) in  \cite{ShiomiTISP}.  
 It shows the microwave-power dependence of peak values for the BSTS sample size  $4\times1\times0.1$ mm$^3$.
Both the positive and negative peak values become larger as microwave power increases.
To summarize, from the above analysis we see that the physical behavior of our result  $\bar{V}^{\rm{SPI}}_{x}  $ represented by Eq.  \eqref{effectivevoltage} and Fig. \ref{voltage} match qualitatively
with these experimental results. 

\section{Conclusion}\label{CON}
In this paper, we have investigated the electric current on the surface of ferromagnetic TI induced by the spin pumping. 
First, we have presented the microscopic model of ferromagnetic TI surface and represented its time evolution. 
We have mathematically formulated how the system evolves from the thermal equilibrium state realized in the far past to the non-equilibrium steady state driven by the spin pumping (FMR).
Then we have used the Keldysh Green's function approach to analyze the generation of a spin-pumping-induced electric current.
We have calculated it by regarding the ac external magnetic field and the exchange interaction as perturbative terms.
In this way, we could clearly understand the way spin-pumping-induced electric current is generated by these two interactions. 
The mechanism is as follows. The FMR is triggered by the ac magnetic field and the magnon with the zero momentum emerges.
It is the fluctuation from the saturation magnetization. After then, the spin pumping is induced, and during such a process, the spin current flows from FM to TI carried by the zero-momentum magnon.
Through the exchange interaction, the zero-momentum magnon couples with the TI surface state and the spin is exchanged between them. 
Then owing to the spin-momentum locking, it is converted into momentum and effectively the TI surface state experiences this additional momentum as the applied electric field.
On the other hand, the TI surface state is affected by the non-magnetic impurity. 
As a result, at the non-equilibrium steady state these two effects commensurate and the static electric field, i.e., the spin-pumping-induced electric field is created leading to the generation of  
the spin-pumping-induced electric current.
It scales quadratically to the ac magnetic field (linear to the power of electromagnetic field) while it is linear to the strength of the exchange interaction.
The effective electric voltage $\bar{V}^{\rm{SPI}}_{x} $ in  Eq.  \eqref{effectivevoltage} is expressed by the spectrum of zero-momentum magnon
which clearly reflects that it is created by the spin pumping (FMR).  
The effective electric voltage $\bar{V}^{\rm{SPI}}_{x} $ shows two side peaks. 
They emerge when the absolute of external frequency of ac magnetic field becomes equivalent to the Zeeman gap of magnon.
 The absolutes of these two peak values are the same while they have the opposite signs. 
Further, the absolutes of two peaks are the increasing function of the microwave power.  
Such characteristics of our effective voltage $\bar{V}^{\rm{SPI}}_{x} $ show qualitatively the good matching with the experimental results of electric voltage reported in \cite{ShiomiTISP,expFermiEdependencemagneticTI2}. 
Consequently, the spin-pumping-induced electric current is the quantum  phenomena intrinsic in the hybrid quantum system of TI surface state and the zero-momentum magnon. 
It is the non-linear response to the ac magnetic field. 

Our microscopic theory based on the Keldysh Green's function approach makes not only the mechanism as well as the structure of spin-pumping-induced electric current (voltage) clear.
We believe that our theory can be extended in many other types of quantum phenomena occurring at the interface between the magnetic materials and TI.  
For instance, we would like to apply our Keldysh Green's function approach to analyze the heat current as well as the spin Seebeck effect and the spin-orbit torque in the future.       
In addition, we become able to extract more information on magnets and TI. 
For instance, by measuring the peak value of electric voltage we can estimate the exchange-coupling strength. 
Another important and interesting issue is the Fermi-energy dependence on spin-pumping-induced electric voltage.  
It is important to analyze whether the contribution to electric transport quantities (for instance, the electric voltage) is coming from the surface state or the bulk state \cite{SPISHETINanoLett2015,expFermiEdependencemagneticTI1,expFermiEdependencemagneticTI2}. 
The transport properties of Dirac electrons in solids are affected by many types of elements.
For instance, the Fermi-energy dependence of Dirac-electron conductivity in graphene differs whether the impurity potential is short-range (delta-function) type or long-range (Coulomb) type \cite{Nomuragraphene}.
For the TI surface state the characteristics of its conductivity is not only generated by the impurity effect but also by a scattering process due to a magnetic texture such as skyrmion \cite{ArakietalMTIPRB2017}.   
As our future work, we would like to explore the rich transport phenomena on the surface of a magnetic TI induced by the impurity potentials and the magnetic textures with many types
and analyze carefully the characteristics of electric voltage as well as the electrical conductivity as functions of the Fermi energy.

To discuss our result from the application point of view, the non-linear response in the magnetic field as well as the linear scaling in the exchange-coupling strength of the spin-pumping-induced electric current clearly
 indicates that the surface of ferromagnetic TI possesses the high-performing functionality of creating the electric charge current or voltage by the magnetic controlling. 
When we think of engineering spintronics devices, the merit of using the ferromagnetic TI 
comparing to the hybrid system of FM and metal like the NiFe/Pt is the lower energy consumption:
The joule heating is suppressed for the ferromagnetic TI because the bulk is insulating while it is unavoidable for the FM/metal hybrid system since the bulk is metallic.
By designing carefully the larger hybrid quantum systems based on the magnet and TI, 
we will become able to perform the coherent controlling of magnon dynamics and the quantum transport of TI surface state at the interface, and consequently,
make a high-efficient conversion of the spin and the electric charge current (coherent controlling of the magnetism and the electricity). 
Such investigations lead to an important progress on the realization of magnetic-TI-based spintronics devices.

\acknowledgements 
Y.~H thanks Kanta Asakawa for the having the discussion on the basics of FMR experiment, Yuki Shiomi for having fruitful discussion on Ref. \cite{ShiomiTISP},
 Minoru Kawamura for discussing  the physical interpretation on the non-linearity of spin-pumping-induced electric current,  Hiroyasu Yamahara for the 
 discussion on Refs. \cite{ShiomiTISP,SPISHETINanoLett2015} as well as the basics of FMR experiment.
This work was supported in part by the MEXT Grant-in-Aid for Scientific Research on Innovative Areas KAKENHI Grant Number JP15H05870 (Y.~H), and  
JSPS KAKENHI Grants Nos. JP15H0584 and JP17K05485, JST CREST Grant No. JPMJCR18T2, and JSPS KAKENHI Grant No. JP20H01830 (K.~N).

\appendix
\section{ Field Quantization,  Real-Time Green's Function, and Imaginary-time Green's function} \label{appendixA}
In this section, first we present the details of field quantization for the TI surface state. 
Then, we introduced the unperturbed real-time Green's function. 
Next we demonstrate the derivation of impurity-averaged  Green's function of TI surface state by using the imaginary-timeGreen's function formalism.
Further, we show the real-time Green's functions of magnon and  the retarded and advanced components of magnon Green's function including the Gilbert-damping effect.  

\subsection{ Field Quantization and Real-Time Green's Functions of TI Surface State} \label{AppA-1}
The spin-momentum-locking Hamiltonian of TI surface state in the momentum space is given by (see also Eq.  \eqref{TIhamiltoniandensity1})
\begin{align}
\mathcal{H}^{\rm{TI}}_{0,\alpha^\prime\alpha} (\boldsymbol{k})= \hbar v_{\rm{F}} \left(   k^x \sigma^y- k^y\sigma^x    \right )_{\alpha^\prime\alpha}, \label{3DTITIhamiltonian1} 
\end{align}
where $\alpha,\alpha^\prime=\uparrow,\downarrow$.
The eigenvalues of the above Hamiltonian are  $\pm \epsilon^{\rm{TI}} _{\boldsymbol{k}} = \pm   \hbar v_{\rm{F}} k $ with    $k=\sqrt{  (k^x)^2  +(k^y)^2    }. $
We denote the positive and negative-energy  plane-wave solutions as $u^{(+)}_{\boldsymbol{k} }(\boldsymbol{x}t)=u^{(+)}(\boldsymbol{k}) e^{ i ( \boldsymbol{k} \cdot \boldsymbol{x}- \omega^{\rm{TI}} (\boldsymbol{k})  t)}$
and $u^{(-)}_{\boldsymbol{k} }(\boldsymbol{x}t)=u^{(-)}(\boldsymbol{k}) e^{ i ( \boldsymbol{k} \cdot \boldsymbol{x}+ \omega^{\rm{TI}} (\boldsymbol{k})  t)}$, 
respectively.
The eigenfrequency  $\omega^{\rm{TI}} (\boldsymbol{k})$ is obtained from  $\epsilon^{\rm{TI}} _{\boldsymbol{k}}$ as $\omega^{\rm{TI}} (\boldsymbol{k})= \hbar^{-1} \epsilon^{\rm{TI}} (\boldsymbol{k})$.
 The vectors $u^{(+)}(\boldsymbol{k})= ( u_{\uparrow}(\boldsymbol{k}) , u_{\downarrow}(\boldsymbol{k})  )^{\rm{t}} $ and $u^{(-)}(\boldsymbol{k})= ( u^{(-)}_{\uparrow}(\boldsymbol{k}) , u^{(-)}_{\downarrow}(\boldsymbol{k})  )^{\rm{t}} $
  are two-column vectors with ``t" denoting the transpose. We take them as
   \begin{align}
& u^{(+)}(\boldsymbol{k}) 
=\frac{1}{\sqrt{2}} 
\left(
\begin{array}{c}
1  \\
\frac {ik^+}{k}   \\
\end{array}
\right), \quad
u^{(-)}( \boldsymbol{k}) 
=\frac{1}{\sqrt{2}} 
\left(
\begin{array}{c}
\frac {i k^-}{k}   \\
1  \\
\end{array}
\right), \label{eigenevector} 
\end{align} 
where $k^\pm = k^x\pm i k^y$. The two eigenvectors $u^{(+)}(\boldsymbol{k})$ and $u^{(-)}(\boldsymbol{k})$ satisfy the completeness relations
 \begin{align}
 & \sum_{\alpha=\uparrow,\downarrow} u^{(+)\dagger}_\alpha(\boldsymbol{k}) u^{(+)}_\alpha(\boldsymbol{k}) =  \sum_{\alpha=\uparrow,\downarrow} u^{(-)\dagger}_\alpha(\boldsymbol{k})  u^{(-)}_\alpha(\boldsymbol{k})=1, \notag\\
 & \sum_{\alpha=\uparrow,\downarrow} u^{(+)\dagger}_\alpha(\boldsymbol{k})  u^{(-)}_\alpha(\boldsymbol{k})  =  \sum_{\alpha=\uparrow,\downarrow} u^{(-)\dagger}_\alpha(\boldsymbol{k})  u^{(+)}_\alpha(\boldsymbol{k})=0.
 \label{eigenvectorrelations}
\end{align} 
With using the plane-wave solutions in Eq. \eqref{eigenevector}  and the completeness relations in Eq.  \eqref{eigenvectorrelations}, 
we construct the field operator of TI surface state. 
By denoting the field operator (annihilation operator) of TI surface state as $\psi_{\alpha}(\boldsymbol{x})$, it is given by
\begin{align}
\psi_\alpha(\boldsymbol{x})=\frac{1}{\sqrt{V}}\sum_{  \boldsymbol{k}, \lambda = \pm } \left(
 u^{(\lambda)}_\alpha(\boldsymbol{k})  e^{ i  \boldsymbol{k} \cdot \boldsymbol{x}}  
 \right) \cdot  c^{(\lambda)} (\boldsymbol{k}),
 \label{TIfieldquantization1}
\end{align} 
where $c^{(+)} (\boldsymbol{k}) $ and $c^{(-)} (\boldsymbol{k}) $ are annihilation operators of TI surface state whose energy and momentum are 
$( \epsilon^{\rm{TI}} (\boldsymbol{k}), \boldsymbol{k} )$ and $( -\epsilon^{\rm{TI}} (\boldsymbol{k}), \boldsymbol{k} )$ , respectively. $V$ is the area of TI surface.
For the ground state,  we choose the Dirac sea represented by
 \begin{align}
 |  0  \rangle  = \prod_{  \boldsymbol{k}  }  c^{(-)\dagger} (\boldsymbol{k})  |  \tilde{0}  \rangle,
 \label{Diracsea}
 \end{align}
 where $ |  \tilde{0}  \rangle$ is the Fock state which satisfies $ c^{(\pm)} (\boldsymbol{k}) |  \tilde{0}  \rangle=0$ for any $\boldsymbol{k}.$
 Correspondingly, we rewrite the field operator in Eq. \eqref{TIfieldquantization1} as
\begin{align}
\psi_\alpha(\boldsymbol{x})=\frac{1}{\sqrt{V}}\sum_{  \boldsymbol{k}} \left(
 u_\alpha(\boldsymbol{k})  e^{ i  \boldsymbol{k} \cdot \boldsymbol{x} } a (\boldsymbol{k}) +
 v_\alpha(\boldsymbol{k})  e^{ - i  \boldsymbol{k} \cdot \boldsymbol{x}} b^\dagger (\boldsymbol{k})
 \right),
 \label{TIfieldquantization2}
\end{align} 
where $a (\boldsymbol{k}) = c^{(+)} (\boldsymbol{k}), b^\dagger (-\boldsymbol{k}) = c^{(-)} (\boldsymbol{k})$,
$ u_\alpha(\boldsymbol{k})= u^{(+)}_\alpha(\boldsymbol{k})$, and  $ v_\alpha(\boldsymbol{k})= u^{(-)}_\alpha(-\boldsymbol{k})$.
The operator $a (\boldsymbol{k}) $ is annihilation operator of particle (electron) with the energy $+\epsilon^{\rm{TI}} (\boldsymbol{k})$ and momentum $\boldsymbol{k}$ while 
$b^\dagger (\boldsymbol{k}) $ is creation operator of anti-particle (hole) with the energy $+\hbar\omega^{\rm{TI}} (\boldsymbol{k})$ and momentum $\boldsymbol{k}$.
They satisfy the anti-commutation relations $\{     a (\boldsymbol{k}) ,  a^\dagger (\boldsymbol{k}^\prime)   \}  =\{     b (\boldsymbol{k}) ,  b^\dagger (\boldsymbol{k}^\prime)   \}
=\delta (\boldsymbol{k}-\boldsymbol{k}^\prime)$, and all the others are zero. 
From these anti-commutation relations and Eq.  \eqref{eigenvectorrelations}, we have 
$ \{ \psi_\alpha(\boldsymbol{x}), \psi^\dagger_{\alpha^\prime}(\boldsymbol{x}^\prime) \} = \delta (\boldsymbol{x} - \boldsymbol{x}^\prime)$ and 
 $ \{ \psi_\alpha(\boldsymbol{x}), \psi_{\alpha^\prime}(\boldsymbol{x}^\prime) \} = \{ \psi^\dagger_\alpha(\boldsymbol{x}), \psi^\dagger_{\alpha^\prime}(\boldsymbol{x}^\prime) \} = 0.$
By using the operator $\psi_\alpha(\boldsymbol{x})$ in Eq.  \eqref{TIfieldquantization2} and its Hermitian conjugate the free Hamiltonian,  momentum operator, and number operator are described as
 \begin{align}
 H^{\rm{TI}}_0 & =  \sum_{\boldsymbol{k}}  \epsilon^{\rm{TI}} (\boldsymbol{k}) \big{(}   a^\dagger (\boldsymbol{k})   a (\boldsymbol{k}) +  b^\dagger (\boldsymbol{k})   b (\boldsymbol{k})      \big{)}, \notag\\
 P^i  & = \sum_{\boldsymbol{k}}  \hbar k^i  \big{(}   a^\dagger (\boldsymbol{k})   a (\boldsymbol{k}) +  b^\dagger (\boldsymbol{k})   b (\boldsymbol{k})      \big{)}  \notag\\
 N^{\rm{TI}} & = \sum_{\boldsymbol{k}}    \big{(}    a^\dagger (\boldsymbol{k})   a (\boldsymbol{k}) -  b^\dagger (\boldsymbol{k})   b (\boldsymbol{k})      \big{)},    \label{HPNoperators1}
\end{align} 
where $i=x,y.$ We have neglected the constants in $H^{\rm{TI}}_0 $ and $N^{\rm{TI}}$ which are the contribution from the Dirac sea. 
Hereinafter, we just write $N^{\rm{TI}}$ as $N$.
Based on the previous argument, next we consider the Hamiltonian $ \bar{H}^{\rm{TI}}_{0}  = H^{\rm{TI}}_{0} -  \epsilon_{\rm{F}} N $ (see also Eq.  \eqref{TIhamiltonian1})
with $ \epsilon_{\rm{F}} $ ($>0$) the Fermi energy of TI. We have taken the chemical potential of TI surface state to be equal to $\epsilon_{\rm{F}}.$  
Correspondingly, we re-describe the field operator $ \psi_{ \alpha}  (\boldsymbol{x}) $ in Eq. \eqref{TIfieldquantization2}  by three operators  $a(\boldsymbol{k})$,  $b^{(+)}(\boldsymbol{k})$, and $b^{(-)}(\boldsymbol{k})$:
The operator $a(\boldsymbol{k})$ is the annihilation operator of momentum $\boldsymbol{k}$ with its energy higher than the Fermi energy $ \epsilon_{\rm{F}} $.
 On the other hand, the operator $b^{(+)}(\boldsymbol{k})$ is the annihilation operator of momentum $\boldsymbol{k}$ with a positive energy which is lower than $\epsilon_{\rm{F}}.$ 
  $b^{(-)}(\boldsymbol{k})$ is the annihilation operator of momentum $\boldsymbol{k}$ with a negative energy  \cite{GrapheneQFT2016}. 
The representation of operator $\psi_{ \alpha}  (\boldsymbol{x} )$  in terms of $a(\boldsymbol{k})$,  $b^{(+)}(\boldsymbol{k})$, and $b^{(-)}(\boldsymbol{k})$  is given as 
\begin{widetext}
\begin{align}
\psi_{\alpha}(\boldsymbol{x})
 & = \frac{1}{\sqrt{V}}\sum_{  \boldsymbol{k}} \left[
 \Theta (k - k_{\rm{F}}  )   u^a_\alpha(\boldsymbol{k})  e^{ i  \boldsymbol{k} \cdot \boldsymbol{x}  } a (\boldsymbol{k}) 
 + \left( \Theta ( k_{\rm{F}} - k )  v^{b^{(+)}}_\alpha(\boldsymbol{k})  b^{ (+) \dagger }(\boldsymbol{k}) 
 +  v^{b^{(-)}}_\alpha(\boldsymbol{k})   b^{ (-) \dagger }(\boldsymbol{k}) \right) e^{ - i  \boldsymbol{k} \cdot \boldsymbol{x}  } 
\right],  \label{TIfieldquantization3}
\end{align} 
\end{widetext}
where $ \Theta (k - k_{\rm{F}}  ) $ $ ( \Theta ( k_{\rm{F}} -k ) ) $ is the step function and $ k_{\rm{F}}  = ( \hbar v_{\rm{F}} )^{-1}   \epsilon_{\rm{F}}$. 
The operators $a(\boldsymbol{k})$,  $b^{(+)}(\boldsymbol{k})$, and $b^{(-)}(\boldsymbol{k})$ satisfy the anti-commutation relation
$\{     a (\boldsymbol{k}) ,  a^\dagger (\boldsymbol{k}^\prime)   \}  = \{     b^{(+)} (\boldsymbol{k}) ,  b^{(+)\dagger } (\boldsymbol{k}^\prime)   \}  = \{     b^{(-)} (\boldsymbol{k}) ,  b^{(-)\dagger } (\boldsymbol{k}^\prime)   \}
=\delta (\boldsymbol{k}-\boldsymbol{k}^\prime)$ and all the others are zero.
The two-column vectors $ u_\alpha(\boldsymbol{k}),$ $v^{b^{(+)}}_\alpha(\boldsymbol{k}),$ and $v^{b^{(-)}}_\alpha(\boldsymbol{k})$ are given by 
 \begin{widetext}
 \begin{align}
& u^{a}(\boldsymbol{k}) 
=\frac{1}{\sqrt{2}} 
\left(
\begin{array}{c}
1  \\
\frac {ik^+}{k}   \\
\end{array}
\right), \quad
  v^{b^{(+)}} (\boldsymbol{k}) 
=\frac{1}{\sqrt{2}} 
\left(
\begin{array}{c}
\frac {i k^-}{k}   \\
1  \\
\end{array}
\right), \quad
 v^{b^{(-)}} (\boldsymbol{k}) 
=\frac{1}{\sqrt{2}} 
\left(
\begin{array}{c}
-\frac {i k^-}{k}   \\
1  \\
\end{array}
\right).
\label{eigenevector2} 
\end{align} 
\end{widetext}
From Eqs. \eqref{TIfieldquantization3} and \eqref{eigenevector2}, the Hamiltonian $\bar{H}^{\rm{TI}}_{0}$ is expressed by the operators $a(\boldsymbol{k})$,  $b^{(+)}(\boldsymbol{k})$, and $b^{(-)}(\boldsymbol{k})$ as  
\begin{widetext}
\begin{align}
\bar{H}^{\rm{TI}}_{0}
  =  \sum_{  \boldsymbol{k}} \left[
 \Theta (k - k_{\rm{F}}  )   \xi^{\rm{TI}}_{a}  (\boldsymbol{k})  a^\dagger (\boldsymbol{k})  a (\boldsymbol{k}) 
 +  \Theta ( k_{\rm{F}} - k )   \xi^{\rm{TI}}_{b^{(+)}}  (\boldsymbol{k})  b^{ (+) \dagger }(\boldsymbol{k}) b^{ (+) } (\boldsymbol{k}) 
 +     \xi^{\rm{TI}}_{b^{(-)}}  (\boldsymbol{k}) b^{ (-) \dagger }(\boldsymbol{k})  b^{ (-)  }(\boldsymbol{k}) 
\right],  \label{TIsurfacebarhamiltonianapp}
\end{align} 
\end{widetext}
where  $ \xi^{\rm{TI}}_{a} (\boldsymbol{k}) = - \xi^{\rm{TI}}_{b^{(+)}}  (\boldsymbol{k}) = \epsilon^{\rm{TI}} (\boldsymbol{k}) -  \epsilon_{\rm{F}} $,     
and $\xi^{\rm{TI}}_{b^{(-)}}  (\boldsymbol{k}) = \epsilon^{\rm{TI}} (\boldsymbol{k}) +  \epsilon_{\rm{F}} $.
For deriving Eq. \eqref{TIsurfacebarhamiltonianapp}, we have used the anti-commutation relation 
$\{     b^{(+)} (\boldsymbol{k}) ,  b^{(+)\dagger } (\boldsymbol{k}^\prime)   \}  = \{     b^{(-)} (\boldsymbol{k}) ,  b^{(-)\dagger } (\boldsymbol{k}^\prime)   \} =\delta (\boldsymbol{k}-\boldsymbol{k}^\prime).$
Next, we introduce the field operator of TI surface state in the interaction picture with respect to the Hamiltonian 
Let us denote it as $\psi_{H_{0}  \alpha}  (\boldsymbol{x} t)$ which is defined by $ \psi_{H_{0}  \alpha}  (\boldsymbol{x}t)= e^{i \bar{H}^{\rm{TI}}_{0} t  /  \hbar }     \psi_{ \alpha}  (\boldsymbol{x})   e^{-i \bar{H}^{\rm{TI}}_{0} t  /  \hbar } $.
It is described by $a(\boldsymbol{k})$,  $b^{(+)}(\boldsymbol{k})$, and $b^{(-)}(\boldsymbol{k})$ as
\begin{widetext}
\begin{align}
\psi_{H_0\alpha}(\boldsymbol{x}t)
 & = \frac{1}{\sqrt{V}}\sum_{  \boldsymbol{k}} \Big{(}
 \Theta (k - k_{\rm{F}}  )   u^a_\alpha(\boldsymbol{k})  e^{ i  \left(\boldsymbol{k} \cdot \boldsymbol{x} -\omega^{\xi, \rm{TI}}_a (\boldsymbol{k})t \right)} a (\boldsymbol{k}) 
 + \Theta ( k_{\rm{F}} - k )  v^{b^{(+)}}_\alpha(\boldsymbol{k})  e^{ - i  \left(\boldsymbol{k} \cdot \boldsymbol{x}-\omega^{\xi, \rm{TI}}_{b^{(+)}} (\boldsymbol{k})t \right )  } b^{ (+) \dagger }(\boldsymbol{k}) \notag\\
& +  v^{b^{(-)}}_\alpha(\boldsymbol{k})  e^{ - i  \left(\boldsymbol{k} \cdot \boldsymbol{x}-\omega^{\xi, \rm{TI}}_{b^{(-)}} (\boldsymbol{k})t \right )  } b^{ (-) \dagger }(\boldsymbol{k})
 \Big{)},  \label{TIfieldquantization4}
\end{align} 
\end{widetext}
where    $ \omega^{\xi,\rm{TI}}_{a} (\boldsymbol{k})  = \hbar^{-1} \xi^{\rm{TI}}_{a}  (\boldsymbol{k}) ,    \omega^{\xi,\rm{TI}}_{b^{(+)}}  (\boldsymbol{k}) =  \hbar^{-1} \xi^{\rm{TI}}_{b^{(+)}}  (\boldsymbol{k}) $, 
and $\omega^{\xi,\rm{TI}}_{b^{(-)}}  (\boldsymbol{k}) = \hbar^{-1} \xi^{\rm{TI}}_{b^{(-)}}  (\boldsymbol{k})  $.
By using the field operator $\psi_{H_0\alpha}(\boldsymbol{x}t)$ in Eq. \eqref{TIfieldquantization3} and its Hermitian conjugate $\psi^\dagger_{H_0\alpha}(\boldsymbol{x}t)$, we introduce two-point real time Green's function in the interaction picture.
Let us denote the time-ordered, anti-time-ordered, retarded, and advanced Green's functions as
$g^{\rm{t}(0)}_{\alpha \alpha^\prime } (\boldsymbol{x}t;\boldsymbol{x}^\prime t^\prime)$,
$g^{\tilde{\rm{t}}(0)}_{\alpha \alpha^\prime } (\boldsymbol{x}t;\boldsymbol{x}^\prime t^\prime),$ $g^{\rm{r}(0)}_{\alpha \alpha^\prime } (\boldsymbol{x}t;\boldsymbol{x}^\prime t^\prime)$, and 
$g^{\rm{a}(0)}_{\alpha \alpha^\prime } (\boldsymbol{x}t;\boldsymbol{x}^\prime t^\prime),$
respectively.  They are given by
\begin{widetext}
\begin{align} 
 g^{\rm{t}(0)}_{\alpha \alpha^\prime } (\boldsymbol{x}t;\boldsymbol{x}^\prime t^\prime) & = -i T \langle  \psi_{H_0\alpha}(\boldsymbol{x}t)     \psi^\dagger_{H_0\alpha^\prime}(\boldsymbol{x}^\prime t^\prime)      \rangle_{\bar{0}} 
 = g^{\rm{t}(0)}_{\alpha \alpha^\prime } (\boldsymbol{x} - \boldsymbol{x}^\prime; t - t^\prime)  
  = \frac{1}{V}  \sum_{  \boldsymbol{k}} \int \frac{d\omega}{2\pi} e^{ i  \big{(}  \boldsymbol{k} \cdot ( \boldsymbol{x} - \boldsymbol{x}^\prime) -\omega (t-t^\prime)  \big{)} } 
  g^ {\rm{t} (0)} _{\alpha  \alpha^\prime } (\boldsymbol{k}\omega),
    \notag\\
  g^{\rm{t}(0)}_{\alpha \alpha^\prime } (\boldsymbol{k}\omega)  &= 
 \frac{ ( \boldsymbol{1}+\tilde{\mathcal{H}}_0 (\boldsymbol{k} ) )_{\alpha \alpha^\prime}  }{2}  
  \left[  
 \frac{ 1 - n_f (   \epsilon^{\rm{TI}}_{\boldsymbol{k}}  ) }{ \omega + \omega_{\rm{F}} -  \omega^{\rm{TI}}_{\boldsymbol{k}}   +  i \eta } +  
 \frac{ n_f (  \epsilon^{\rm{TI}}_{\boldsymbol{k}} ) }{\omega + \omega_{\rm{F}} -  \omega^{\rm{TI}}_{\boldsymbol{k}}  - i \eta } 
 \right ]    \notag\\
 &+  \frac{ (\boldsymbol{1} - \tilde{\mathcal{H}}_0 (\boldsymbol{k} )  )_{\alpha \alpha^\prime}  }{2}   
\left[  
 \frac{  \bar{n}_f (  \epsilon^{\rm{TI}}_{\boldsymbol{k}}) }{ \omega +  \omega_{\rm{F}}  +  \omega^{\rm{TI}}_{\boldsymbol{k}}   +i \eta } 
 +  \frac{  1 - \bar{n}_f (  \epsilon^{\rm{TI}}_{\boldsymbol{k}}  ) }{\omega +  \omega_{\rm{F}}  +  \omega^{\rm{TI}}_{\boldsymbol{k}}  - i \eta } 
 \right ], \notag\\ 
  g^{ \tilde{\rm{t}} (0)}_{\alpha \alpha^\prime } (\boldsymbol{x}t;\boldsymbol{x}^\prime t^\prime) & =   -i \tilde{T} \langle  \psi_{H_0\alpha}(\boldsymbol{x}t)     \psi^\dagger_{H_0\alpha^\prime}(\boldsymbol{x}^\prime t^\prime)  \rangle_{\bar{0}} 
  = g^{ \tilde{\rm{t}} (0)}_{\alpha \alpha^\prime } (\boldsymbol{x} - \boldsymbol{x}^\prime ; t - t^\prime) 
  = \frac{1}{V}  \sum_{  \boldsymbol{k}} \int \frac{d\omega}{2\pi} e^{ i  \big{(}  \boldsymbol{k} \cdot ( \boldsymbol{x} - \boldsymbol{x}^\prime) -\omega (t-t^\prime)  \big{)} }  g^ {  \tilde{\rm{t}} (0)} _{\alpha  \alpha^\prime } (\boldsymbol{k}\omega),
    \notag\\   
  g^{  \tilde{\rm{t}} (0)}_{\alpha \alpha^\prime } (\boldsymbol{k}\omega)  &= 
 \frac{ (\boldsymbol{1}+\tilde{\mathcal{H}}_0(\boldsymbol{k} )  )_{\alpha \alpha^\prime}  }{2}  
  \left[  
 \frac{  - n_f (   \epsilon^{\rm{TI}}_{\boldsymbol{k}}  ) }{ \omega + \omega_{\rm{F}} -  \omega^{\rm{TI}}_{\boldsymbol{k}}   +  i \eta } +  
 \frac{ n_f (  \epsilon^{\rm{TI}}_{\boldsymbol{k}} ) - 1 }{\omega + \omega_{\rm{F}} -  \omega^{\rm{TI}}_{\boldsymbol{k}}  - i \eta } 
 \right ]    \notag\\
 & +  \frac{ (\boldsymbol{1} - \tilde{\mathcal{H}}_0(\boldsymbol{k} )  )_{\alpha \alpha^\prime}  }{2}   
\left[  
 \frac{  \bar{n}_f (  \epsilon^{\rm{TI}}_{\boldsymbol{k}}) - 1 }{\omega +  \omega_{\rm{F}}  +  \omega^{\rm{TI}}_{\boldsymbol{k}}   +i \eta } 
 +  \frac{  - \bar{n}_f(  \epsilon^{\rm{TI}}_{\boldsymbol{k}}  ) }{\omega +  \omega_{\rm{F}}  +  \omega^{\rm{TI}}_{\boldsymbol{k}}  - i \eta } 
 \right ], \notag\\ 
 g^{\rm{r}(0)}_{\alpha \alpha^\prime } (\boldsymbol{x}t;\boldsymbol{x}^\prime t^\prime)  & =  -  i\theta(t - t^\prime )  \cdot \rho^{(0)}_{\alpha \alpha^\prime } (\boldsymbol{x}t;\boldsymbol{x}^\prime t^\prime)
 = g^{\rm{r}(0)}_{\alpha \alpha^\prime } (\boldsymbol{x} - \boldsymbol{x}^\prime; t -  t^\prime)   
 = \frac{1}{V}  \sum_{  \boldsymbol{k}} \int \frac{d\omega}{2\pi} e^{ i  \big{(}  \boldsymbol{k} \cdot ( \boldsymbol{x} - \boldsymbol{x}^\prime) -\omega (t-t^\prime)  \big{)} } g^ { \rm{r} (0)} _{\alpha  \alpha^\prime } (\boldsymbol{k}\omega), \notag\\
 g^{\rm{a}(0)}_{\alpha \alpha^\prime } (\boldsymbol{x}t;\boldsymbol{x}^\prime t^\prime) & =  i\theta(t^\prime - t )  \cdot \rho^{(0)}_{\alpha \alpha^\prime } (\boldsymbol{x}t;\boldsymbol{x}^\prime t^\prime)
= g^{\rm{a}(0)}_{\alpha \alpha^\prime } (\boldsymbol{x} - \boldsymbol{x}^\prime; t -  t^\prime) 
 = \frac{1}{V}  \sum_{  \boldsymbol{k}} \int \frac{d\omega}{2\pi} e^{ i  \big{(}  \boldsymbol{k} \cdot ( \boldsymbol{x} - \boldsymbol{x}^\prime) -\omega (t-t^\prime)  \big{)} }   g^ { \rm{a} (0)} _{\alpha  \alpha^\prime } (\boldsymbol{k}\omega), \notag\\
    g^{\rm{r}(0)}_{\alpha \alpha^\prime } (\boldsymbol{k}\omega)  &=   \int \frac{d \check{\omega} }{2\pi}  \frac{ \rho^{(0)} _{\alpha \alpha^\prime } (\boldsymbol{k}\check{\omega}) } {\omega-\check{\omega} + i\eta  }
 \quad
 g^{\rm{a}(0)}_{\alpha \alpha^\prime } (\boldsymbol{k}\omega) =  \int \frac{d \check{\omega} }{2\pi}  \frac{ \rho^{(0)} _{\alpha \alpha^\prime } (\boldsymbol{k}\check{\omega}) } {\omega-\check{\omega} - i\eta  },
 \label{realtimeGreen'sfunction1}
\end{align} 
\end{widetext}
where $ \langle \cdots \rangle_{\check{0} } $ denotes the thermal average taken by the density matrix $ \rho_{\rm{GC}}( \check{H}_0,\beta,\epsilon_{ \rm{F}   }   )$ 
with $ \check{H}_0 =  \bar{H}_0^{\rm{TI}}.$ $\eta$ is a positive infinitesimal. 
The symbols $T$ and $\tilde{T}$ are the time-ordering and anti-time-ordering operators, respectively.
For the step function $\theta(t - t^\prime )$ we have used $\theta(t - t^\prime ) = -\int \frac{d \omega }{2\pi i} \frac{e^{-i\omega(t- t^\prime)}}{\omega+i \eta}$.
The functions $n_f(\epsilon)$ and  $\bar{n}_f(\epsilon)$ are given by
$n_f(\epsilon) = \left(  1 + e^{\beta (\epsilon - \epsilon_{\rm{F}} )} \right)^{-1} $ and  $\bar{n}_f(\epsilon) =  \left(  1 + e^{\beta (\epsilon + \epsilon_{\rm{F}} )}  \right)^{-1} $.
 $n_f(\epsilon)$ represents the Fermi-Dirac distribution function for the energy $\epsilon$ with the chemical potential $\mu$ which we take to be equal to $ \epsilon_{\rm{F}} .$
   $\bar{n}_f(\epsilon)$ is the one with the chemical potential equal to $- \epsilon_{\rm{F}} $. 
   They are given by the thermal average of TI-surface-state field operators as 
   $n_f(  \epsilon^{\rm{TI}} (\boldsymbol{k}) ) =  \langle  a^\dagger (\boldsymbol{k})   a (\boldsymbol{k})  \rangle_{\bar{0} }     =  
   1-   \langle  b^{ (+) \dagger }(\boldsymbol{k})   b^{ (+)  }(\boldsymbol{k})  \rangle_{\bar{0} } $
   and $  \bar{n_f}(  \epsilon^{\rm{TI}} (\boldsymbol{k}) ) =  \langle  b^{ (-) \dagger }(\boldsymbol{k})   b^{ (-)  }(\boldsymbol{k})  \rangle_{\bar{0} }. $
  The matrix $ \tilde{\mathcal{H}}_0 $ is given by Eq.  \eqref{dimensionlessDiracHamiltonian} or  
   \begin{align*}
 \tilde{\mathcal{H}}_0 (\boldsymbol{k}) =  
 \left(
\begin{array}{cc}
0 &  - \frac{ i( k^x -i k^y) }{k} \\
 \frac{ i(k^x + i k^y) }{k}  & 0 \\
\end{array}
\right).   
\end{align*} 
The spectral functions $ \rho^{(0)}_{\alpha \alpha^\prime } (\boldsymbol{x}t;\boldsymbol{x}^\prime t^\prime)$ and $ \rho^{(0)}_{\alpha \alpha^\prime } (\boldsymbol{k}\omega)$  are 
\begin{widetext}
\begin{align} 
& \rho^{(0)}_{\alpha \alpha^\prime } (\boldsymbol{x}t;\boldsymbol{x}^\prime t^\prime) = \langle   \{  \psi_{H_0\alpha}(\boldsymbol{x}t)  ,   \psi^\dagger_{H_0\alpha^\prime}(\boldsymbol{x}^\prime t^\prime)   \}   \rangle_{\bar{0}} 
 = \rho^{(0)}_{\alpha \alpha^\prime } (\boldsymbol{x} - \boldsymbol{x}^\prime; t - t^\prime)
 =  \frac{1}{V}  \sum_{  \boldsymbol{k}} \int \frac{d\omega}{2\pi} e^{ i  \big{(}  \boldsymbol{k} \cdot ( \boldsymbol{x} - \boldsymbol{x}^\prime) -\omega (t-t^\prime)  \big{)} }   \rho^{(0)}_{\alpha \alpha^\prime } (\boldsymbol{k}\omega) ,   \notag\\
&  \rho^{(0)}_{\alpha \alpha^\prime } (\boldsymbol{k}\omega) = \pi \left[
 (\boldsymbol{1} + \tilde{\mathcal{H}}_0  (\boldsymbol{k}))_{\alpha \alpha^\prime}   \delta( \omega + \omega_{\rm{F}} -  \omega^{\rm{TI}}_{\boldsymbol{k}}   )
 +
  (\boldsymbol{1} - \tilde{\mathcal{H}}_0 (\boldsymbol{k}) )_{\alpha \alpha^\prime}    \delta( \omega + \omega_{\rm{F}} +  \omega^{\rm{TI}}_{\boldsymbol{k}}   )
\right],
   \label{realtimespectralfunction1}
\end{align} 
\end{widetext}
where $\{ \}$ in the above first equation denotes the anticommutator: $ \{ X,Y    \} = XY+YX$. \\ \noindent

Besides time-ordered, anti-time-ordered, retarded, and advanced components, there are lesser and greater components defined by 
\begin{widetext}
\begin{align} 
   g^{< (0)}_{\alpha \alpha^\prime } (\boldsymbol{x}t;\boldsymbol{x}^\prime t^\prime) & = i  \langle   \psi^\dagger_{H_0\alpha^\prime}(\boldsymbol{x}^\prime t^\prime)  \psi_{H_0\alpha}(\boldsymbol{x}t)    \rangle_{\bar{0}} 
 = g^{ < (0)}_{\alpha \alpha^\prime } (\boldsymbol{x} - \boldsymbol{x}^\prime; t - t^\prime)  
  = \frac{1}{V}  \sum_{  \boldsymbol{k}} \int \frac{d\omega}{2\pi} e^{ i  \big{(}  \boldsymbol{k} \cdot ( \boldsymbol{x} - \boldsymbol{x}^\prime) -\omega (t-t^\prime)  \big{)} } 
  g^ { < (0)} _{\alpha  \alpha^\prime } (\boldsymbol{k}\omega),
    \notag\\ 
  g^{ < (0)}_{\alpha \alpha^\prime } (\boldsymbol{k}\omega)  &=  i\pi  f (  \hbar \omega  )  \left[
 \delta ( \omega -  \omega^{\rm{TI}}_{\boldsymbol{k}}  + \omega_{\rm{F}}  )  (\boldsymbol{1}+\tilde{\mathcal{H}}_0 (\boldsymbol{k} ) )_{\alpha \alpha^\prime}
 +
  \delta ( \omega +  \omega^{\rm{TI}}_{\boldsymbol{k}} + \omega_{\rm{F}}  )  (\boldsymbol{1} - \tilde{\mathcal{H}}_0 (\boldsymbol{k} ) )_{\alpha \alpha^\prime}
  \right ], \notag\\  
  g^{>(0)}_{\alpha \alpha^\prime } (\boldsymbol{x}t;\boldsymbol{x}^\prime t^\prime) & = -i  \langle  \psi_{H_0\alpha}(\boldsymbol{x}t)     \psi^\dagger_{H_0\alpha^\prime}(\boldsymbol{x}^\prime t^\prime)      \rangle_{\bar{0}} 
 = g^{ > (0)}_{\alpha \alpha^\prime } (\boldsymbol{x} - \boldsymbol{x}^\prime; t - t^\prime)  
  = \frac{1}{V}  \sum_{  \boldsymbol{k}} \int \frac{d\omega}{2\pi} e^{ i  \big{(}  \boldsymbol{k} \cdot ( \boldsymbol{x} - \boldsymbol{x}^\prime) -\omega (t-t^\prime)  \big{)} } 
  g^ { > (0)} _{\alpha  \alpha^\prime } (\boldsymbol{k}\omega),
    \notag\\ 
  g^{ > (0)}_{\alpha \alpha^\prime } (\boldsymbol{k}\omega)  &=  -i\pi \big{(}1 - f (  \hbar \omega  ) \big{)} \left[
 \delta ( \omega -  \omega^{\rm{TI}}_{\boldsymbol{k}}  + \omega_{\rm{F}}  )  (\boldsymbol{1}+\tilde{\mathcal{H}}_0 (\boldsymbol{k} ) )_{\alpha \alpha^\prime}
 +
  \delta ( \omega +  \omega^{\rm{TI}}_{\boldsymbol{k}} + \omega_{\rm{F}}  )  (\boldsymbol{1} - \tilde{\mathcal{H}}_0 (\boldsymbol{k} ) )_{\alpha \alpha^\prime}
  \right ], 
   \label{realtimeGreen'sfunction2}
\end{align} 
\end{widetext}
where $f( \hbar \omega)=   \left(  1 + e^{ \beta   \hbar \omega  } \right)^{-1} $. 
The lesser and greater components of Green's functions are related to the time-ordered, anti-time-ordered, retarded, and advanced components through the relations \cite{NEQGreensfunctionRMPandtxtbook1,Mahantextbook}
\begin{widetext}
\begin{align} 
  g^{<(0)}_{\alpha \alpha^\prime } (\boldsymbol{x}t;\boldsymbol{x}^\prime t^\prime)  &= g^{\rm{t}(0)}_{\alpha \alpha^\prime } (\boldsymbol{x}t;\boldsymbol{x}^\prime t^\prime) - 
  g^{\rm{r}(0)}_{\alpha \alpha^\prime } (\boldsymbol{x}t;\boldsymbol{x}^\prime t^\prime)
  = g^{ \tilde{\rm{t}}(0)}_{\alpha \alpha^\prime } (\boldsymbol{x}t;\boldsymbol{x}^\prime t^\prime) + 
  g^{\rm{a}(0)}_{\alpha \alpha^\prime } (\boldsymbol{x}t;\boldsymbol{x}^\prime t^\prime), \notag\\
   g^{>(0)}_{\alpha \alpha^\prime } (\boldsymbol{x}t;\boldsymbol{x}^\prime t^\prime)  &= g^{\rm{t}(0)}_{\alpha \alpha^\prime } (\boldsymbol{x}t;\boldsymbol{x}^\prime t^\prime) -
  g^{\rm{a}(0)}_{\alpha \alpha^\prime } (\boldsymbol{x}t;\boldsymbol{x}^\prime t^\prime)
  = g^{ \tilde{\rm{t}}(0)}_{\alpha \alpha^\prime } (\boldsymbol{x}t;\boldsymbol{x}^\prime t^\prime) + 
  g^{\rm{r}(0)}_{\alpha \alpha^\prime } (\boldsymbol{x}t;\boldsymbol{x}^\prime t^\prime). 
\label{realtimeGreen'sfunctionrelations1}
\end{align} 
\end{widetext}
The above relation also holds for the Green's functions in the momentum-frequency representation. 
Further, from Eq.   \eqref{realtimeGreen'sfunction2} and the formula $\frac{1}{  z - z_0 \pm i\eta  }= {\rm{P}}  \left(  \frac{ 1}{   z - z_0}  \right) \mp i\pi \delta (z - z_0)$, we have
\begin{align} 
  g^{<(0)}_{\alpha \alpha^\prime } (\boldsymbol{k}\omega)  &= f(\hbar\omega) (
  g^{\rm{a}(0)}_{\alpha \alpha^\prime } ( \boldsymbol{k} \omega) -  g^{\rm{r}(0)}_{\alpha \alpha^\prime } ( \boldsymbol{k} \omega)), \notag\\
    g^{>(0)}_{\alpha \alpha^\prime } (\boldsymbol{k}\omega)  & = -(1- f(\hbar\omega)) (
  g^{\rm{a}(0)}_{\alpha \alpha^\prime } ( \boldsymbol{k} \omega) -  g^{\rm{r}(0)}_{\alpha \alpha^\prime } ( \boldsymbol{k} \omega) ).
\label{realtimeGreen'sfunctionrelations2}
\end{align} 

\subsection{Impurity-Averaged Imaginary-Time Green's Function}\label{IAIMGreen'sfunction}
We now include the impurity-potential effect and derive the impurity-averaged Green's function for the TI surface state. 

Let us denote a function described by the coordinates of impurities as $ F ( \boldsymbol{X}_1,  \ldots, \boldsymbol{X}_{ N_{\rm{imp}}   } ).$
The impurity average is defined by  
\begin{align} 
\langle  F ( \boldsymbol{X}_1, \ldots, \boldsymbol{X}_{ N_{\rm{imp}} } ) \rangle^{\rm{imp}}_{\rm{ave}} =
  \int \prod_{i=1}^{  N_{ \rm{imp}} } \frac{d  \boldsymbol{X}_i }{V} F ( \boldsymbol{X}_1, \ldots, \boldsymbol{X}_{ N_{\rm{imp}}    } ).
\label{defimpurityaverage}
\end{align} 
To perform this on the Green's functions of TI surface state, we derive the Dyson's equation for the imaginary-time (Matsubara) Green's functions with regarding the impurity potential as the perturbation which is given by Eq.  \eqref{impurityHamiltoniandensity}.
First, let us introduce the unperturbed  imaginary-time Green's function defined by \cite{FetterWaleckaQMPtxtbook,MBQtextbook,Mahantextbook}
\begin{widetext}
\begin{align} 
 \mathcal{G}^{(0)}_{\rm{M}, \alpha \alpha^\prime } (\boldsymbol{x}\tau_{\rm{M}};\boldsymbol{x}^\prime \tau^\prime_{\rm{M}}) &   =
 - \big{ \langle} T_{\rm{M}}  \psi_{H_0\alpha}(\boldsymbol{x} \tau_{\rm{M}})     \psi^\dagger_{H_0\alpha^\prime}(\boldsymbol{x}^\prime \tau^\prime_{\rm{M}})    \big{ \rangle } _{0,\rm{GC}}
 =  \mathcal{G}^{(0)}_{\rm{M}, \alpha \alpha^\prime }  (\boldsymbol{x} - \boldsymbol{x}^\prime; \tau_{\rm{M}} - \tau^\prime_{\rm{M}}),  \notag\\
 & = \frac{1}{\beta \hbar V}  \sum_{  \boldsymbol{k} m}  e^{ i  \big{(}  \boldsymbol{k} \cdot ( \boldsymbol{x} - \boldsymbol{x}^\prime) -i \omega_m ( \tau_{\rm{M}} - \tau^\prime_{\rm{M}})  \big{)} }  
   \mathcal{G}^{(0)}_{\rm{M}, \alpha \alpha^\prime } (\boldsymbol{k};i\omega_m) , \notag\\
  \mathcal{G}^{(0)}_{\rm{M}, \alpha \alpha^\prime } (\boldsymbol{k};i\omega_m) & = \int \frac{ d \omega^\prime}{2\pi} \frac {\rho^{(0)}_{\alpha \alpha^\prime } (\boldsymbol{k}\omega^\prime )} {i \omega_m - \omega^\prime}.
 \label{TISSMatsubaraGreensfunction1}
\end{align} 
\end{widetext}
Here $\tau_{\rm{M}},\tau^\prime_{\rm{M}}$ are the imaginary times and $T_{{\rm{M}}}$ represents the imaginary-time ordering. 
The fermionic imaginary-time Green's function $ \mathcal{G}^{(0)}_{\rm{M}, \alpha \alpha^\prime }  (\boldsymbol{x} - \boldsymbol{x}^\prime; \tau_{\rm{M}} - \tau^\prime_{\rm{M}})$ in the above equation is 
anti-periodic with respect to the the imaginary time: 
$  \mathcal{G}^{(0)}_{\rm{M}, \alpha \alpha^\prime }  (\boldsymbol{x} - \boldsymbol{x}^\prime; \tau_{\rm{M}} - \tau^\prime_{\rm{M}}) =
- \mathcal{G}^{(0)}_{\rm{M}, \alpha \alpha^\prime }  (\boldsymbol{x} - \boldsymbol{x}^\prime; \tau_{\rm{M}} - \tau^\prime_{\rm{M}}\pm \beta \hbar)$.
Correspondingly, the Matsubara frequency $\omega_m$ is given  by $\omega_m = (2m+1)\pi / (\beta \hbar) $ and the Fourier transform of the $ \mathcal{G}^{(0)}_{\rm{M}, \alpha \alpha^\prime }  (\boldsymbol{x} - \boldsymbol{x}^\prime; \tau_{\rm{M}} - \tau^\prime_{\rm{M}})$
for the temporal component becomes $ \mathcal{G}^{(0)}_{\rm{M}, \alpha \alpha^\prime }  (\boldsymbol{x} - \boldsymbol{x}^\prime; i\omega_m ) = \int_0^{\beta \hbar}  
e^{ i \omega_m ( \tau_{\rm{M}} - \tau^\prime_{\rm{M}})  }    \mathcal{G}^{(0)}_{\rm{M}, \alpha \alpha^\prime }  (\boldsymbol{x} - \boldsymbol{x}^\prime; \tau_{\rm{M}} - \tau^\prime_{\rm{M}}) d( \tau_{\rm{M}} - \tau^\prime_{\rm{M}} )   $.  
$\rho^{(0)}_{\alpha \alpha^\prime } (\boldsymbol{k}\omega)$  
is the spectral function defined in Eq.  \eqref{realtimespectralfunction1}. 
By performing the analytic continuation $i\omega_m \rightarrow \omega + i\eta$ on   $ \mathcal{G}^{(0)}_{\rm{M}, \alpha \alpha^\prime } (\boldsymbol{k}\omega)  $ in Eq.  \eqref{TISSMatsubaraGreensfunction1},
we obtain the retarded Green's function $g^{\rm{r}(0)}_{\alpha \alpha^\prime } (\boldsymbol{k}\omega) $ in Eq.  \eqref{realtimeGreen'sfunction1}.
On the other side, we have the advanced Green's function $g^{\rm{a}(0)}_{\alpha \alpha^\prime } (\boldsymbol{k}\omega) $ in Eq.  \eqref{realtimeGreen'sfunction1}
by $i\omega_m \rightarrow \omega - i\eta$.

Next, the Dyson equation owing to the impurity potential is given by \cite{MBQtextbook}
\begin{widetext}
\begin{align}
 \mathcal{G}_{\rm{M},\alpha \alpha^\prime}(\boldsymbol{x} \tau_{\rm{M}} ; \boldsymbol{x}^\prime \tau^\prime_{\rm{M}}) & 
  =  \mathcal{G}^0_{\rm{M},\alpha \alpha^\prime}(\boldsymbol{x} \tau_{\rm{M}} ; \boldsymbol{x}^\prime \tau^\prime_{\rm{M}})  \notag\\
 &  + \int_0^{ \beta \hbar } d \tau_{1}^{\rm{M}} \int d^2 x_1 \mathcal{G}^0_{\rm{M},\alpha \alpha^\prime_1}(\boldsymbol{x} \tau_{\rm{M}} ; \boldsymbol{x}_1 \tau_{1}^{\rm{M}}  )  
  \mathcal{H}^{\rm{imp}} _{\alpha^\prime_1 \alpha_1} (\boldsymbol{x}_1)   
 \mathcal{G}_{\rm{M},\alpha \alpha^\prime}(  \boldsymbol{x}_1 \tau_{1}^{\rm{M}}  ; \boldsymbol{x}^\prime \tau^\prime_{\rm{M}}). \label{imaginaryimpuritydysonequation1}
\end{align} 
\end{widetext}
Let us rewrite the right-hand side of Dyson Equation  \eqref{imaginaryimpuritydysonequation1} as  
$  \mathcal{G}_{\rm{M},\alpha \alpha^\prime}(\boldsymbol{x} \tau_{\rm{M}} ; \boldsymbol{x}^\prime \tau^\prime_{\rm{M}}) =
\sum_n   \mathcal{G}^{(n)}_{\rm{M},\alpha \alpha^\prime}(\boldsymbol{x} \tau_{\rm{M}} ; \boldsymbol{x}^\prime \tau^\prime_{\rm{M}})  $  
where $\mathcal{G}^{(n)}_{\rm{M},\alpha \alpha^\prime}(\boldsymbol{x} \tau_{\rm{M}} ; \boldsymbol{x}^\prime \tau^\prime_{\rm{M}})  $ is in the $n$-th order of impurity potential $ \mathcal{H}^{\rm{imp}} .$
Its form is represented as
\begin{widetext}
\begin{align} 
 \mathcal{G}^{(n)}_{\rm{M},\alpha \alpha^\prime}(\boldsymbol{x} \tau_{\rm{M}} ; \boldsymbol{x}^\prime \tau^\prime_{\rm{M}})  & = 
 \int_0^{ \beta \hbar } d \tau_{1}^{\rm{M}} \cdots  d \tau_{n}^{\rm{M}}    \int d^2 x_1 \cdots d^2 x_n 
  \mathcal{H}^{\rm{imp}} _{\alpha^\prime_1 \alpha_1} (\boldsymbol{x}_1) \cdots   \mathcal{H}^{\rm{imp}} _{\alpha^\prime_n \alpha_n} (\boldsymbol{x}_n)   \notag\\
& \times  \mathcal{G}^0_{\rm{M},\alpha \alpha^\prime_n}( \boldsymbol{x} - \boldsymbol{x}_n; \tau_{\rm{M}}  -  \tau_{n}^{ \rm{M}}  )  
 \mathcal{G}^0_{\rm{M},\alpha_n \alpha^\prime_{ n-1 } }( \boldsymbol{x}_n  - \boldsymbol{x}_{n-1}; \tau_{n}^{\rm{M}} - \tau_{n-1}^{\rm{M}}  )  \cdots
  \mathcal{G}^0_{\rm{M},\alpha_1 \alpha^\prime}(  \boldsymbol{x}_1  - \boldsymbol{x}^\prime;  \tau_{1}^{\rm{M}}  - \tau^\prime_{\rm{M}}),
 \label{imaginaryimpuritydysonequation2}
\end{align} 
\end{widetext}
where we have used $ \mathcal{G}^0_{\rm{M},\alpha \alpha^\prime }(\boldsymbol{x} \tau_{\rm{M}} ; \boldsymbol{x}^\prime \tau^\prime_{\rm{M}}  ) 
=  \mathcal{G}^0_{\rm{M},\alpha \alpha^\prime }(\boldsymbol{x} - \boldsymbol{x}^\prime; \tau_{\rm{M}} - \tau^\prime_{\rm{M}}  ) $.  
Let us perform the Fourier transformations on the Green's function $ \mathcal{G}^{(n)}_{\rm{M},\alpha \alpha^\prime}(\boldsymbol{x} \tau_{\rm{M}} ; \boldsymbol{x}^\prime \tau^\prime_{\rm{M}}) $.
 Since the impurity potential is time independent,   the Green's function $ \mathcal{G}^{(n)}_{\rm{M},\alpha \alpha^\prime}(\boldsymbol{x} \tau_{\rm{M}} ; \boldsymbol{x}^\prime \tau^\prime_{\rm{M}}) $ is described as
  $ \mathcal{G}^{(n)}_{\rm{M},\alpha \alpha^\prime}(\boldsymbol{x} \tau_{\rm{M}} ; \boldsymbol{x}^\prime \tau^\prime_{\rm{M}}) =  \mathcal{G}^{(n)}_{\rm{M},\alpha \alpha^\prime}(\boldsymbol{x}, \boldsymbol{x}^\prime;
  \tau_{\rm{M}} - \tau^\prime_{\rm{M}}) .$
Then the Fourier transformation is given as  
$  \mathcal{G}^{(n)}_{\rm{M},\alpha \alpha^\prime}( \boldsymbol{x} ,\boldsymbol{x}^\prime; \tau_{\rm{M}}  - \tau^\prime_{\rm{M}}) =
( \beta \hbar V )^{-1}  \sum_{ m \boldsymbol{k} \boldsymbol{k}^\prime } \mathcal{G}^{(n)}_{ \rm{M}, \alpha \alpha^\prime }  ( \boldsymbol{k},   \boldsymbol{k}^\prime  ; i\omega_m) 
e^{  i\left(  \boldsymbol{k} \cdot \boldsymbol{x} - \boldsymbol{k}^\prime \cdot \boldsymbol{x}^\prime - \omega_m ( \tau_{\rm{M}}  - \tau^\prime_{\rm{M}}  ) \right ) }.   $ 
 The formula of $ \mathcal{G}^{(n)}_{ \rm{M}, \alpha \alpha^\prime }  ( \boldsymbol{k},   \boldsymbol{k}^\prime  ; i\omega_m) $ is represented as
\begin{widetext}
\begin{align} 
  \mathcal{G}^{(n)}_{ \rm{M}, \alpha \alpha^\prime }  ( \boldsymbol{k},   \boldsymbol{k}^\prime  ; i\omega_m) &=  \frac{1}{V^{n }} \sum_{  \boldsymbol{k}_1,\ldots, \boldsymbol{k}_{n-1}      }
  \left[   \mathcal{G}^{(0)}_{\rm{M}, \alpha \alpha_{n} }  ( \boldsymbol{k}; i\omega_m)    v_{\rm{imp}}(  \boldsymbol{k} - \boldsymbol{k}_{n-1} )  \right]
   \left[   \mathcal{G}^{(0)}_{\rm{M}, \alpha_{n} \alpha_{n-1} }  ( \boldsymbol{k}_{n-1}; i\omega_m)    v_{\rm{imp}}(  \boldsymbol{k}_{n-1} - \boldsymbol{k}_{n-2} )  \right] \notag\\
   & \cdots  
     \left[   \mathcal{G}^{(0)}_{\rm{M}, \alpha_{3} \alpha_{2} }  ( \boldsymbol{k}_2; i\omega_m)    v_{\rm{imp}}(  \boldsymbol{k}_2 - \boldsymbol{k}_{1} )  \right]
   \left[   \mathcal{G}^{(0)}_{\rm{M}, \alpha_{2} \alpha_{2} }  ( \boldsymbol{k}_{1}; i\omega_m)    v_{\rm{imp}}(  \boldsymbol{k}_{1} - \boldsymbol{k}_{0} )  \right] 
   \mathcal{G}^{(0)}_{\rm{M}, \alpha_{1} \alpha^\prime }  ( \boldsymbol{k}^\prime; i\omega_m)  \notag\\
 & \times   \rho_{\rm{imp}} ( \boldsymbol{k} - \boldsymbol{k}_{n-1} )   \rho_{\rm{imp}} (  \boldsymbol{k}_{n-1} - \boldsymbol{k}_{n-2} ) \cdots
 \rho_{\rm{imp}} (  \boldsymbol{k}_{3} - \boldsymbol{k}_{2} )  \rho_{\rm{imp}} (  \boldsymbol{k}_{2} - \boldsymbol{k}_{1} )   \rho_{\rm{imp}} (  \boldsymbol{k}_{1} - \boldsymbol{k}^\prime),
 \label{imaginaryimpuritydysonequation3}
\end{align} 
\end{widetext}
where we have used $(\beta \hbar)^{-1} \int^{\beta \hbar} _0 d\tau \exp\big{(} i( \omega_m - \omega_{m^\prime} )\tau \big{)} = \delta_{m,m^\prime}$.
In the above equation, all the Matsubara frequencies of $n+1$ unperturbed Green's functions in the right-hand side are equivalent due to time independence of impurity potential. 
 We now perform the impurity average on Eq.  \eqref{imaginaryimpuritydysonequation3} by assuming that the total number of impurities $N_{\rm{imp}} $ is sufficiently large.
We use the formulas, for instance \cite{Mahantextbook},
 \begin{align} 
& \langle  \rho_{\rm{imp}} (  \boldsymbol{k} - \boldsymbol{k}^\prime )  \rangle^{\rm{imp}} _{\rm{ave}}  = N_{\rm{imp}}  \delta_{  \boldsymbol{k} , \boldsymbol{k}^\prime  }, \notag\\
& \langle  \rho_{\rm{imp}} (  \boldsymbol{k} )  \rho_{\rm{imp}}  ( \boldsymbol{k}^\prime )  \rangle^{\rm{imp}} _{\rm{ave}}  = N_{\rm{imp}}  \delta_{  \boldsymbol{k} + \boldsymbol{k}^\prime,   \boldsymbol{0}  }
+N^2_{\rm{imp}}  \delta_{  \boldsymbol{k},   \boldsymbol{0}  }  \delta_{   \boldsymbol{k}^\prime,   \boldsymbol{0}  }.
  \label{impurityaverageformulas1}
\end{align} 
For $ n = 0,1,2$, we have 
\begin{widetext}
 \begin{align} 
  \langle   \mathcal{G}^{(0)}_{\rm{M}, \alpha \alpha^\prime }  ( \boldsymbol{k},   \boldsymbol{k}^\prime  ; i\omega_m)   \rangle^{\rm{imp}} _{\rm{ave}} &=
 \delta_{ \boldsymbol{k},   \boldsymbol{k}^\prime }   \mathcal{G}^{(0)}_{\rm{M}, \alpha \alpha^\prime }  ( \boldsymbol{k}  ; i\omega_m), \notag\\
   \langle   \mathcal{G}^{(1)}_{\rm{M}, \alpha \alpha^\prime }  ( \boldsymbol{k},   \boldsymbol{k}^\prime  ; i\omega_m)   \rangle^{\rm{imp}} _{\rm{ave}} &=0, \notag\\
  \langle   \mathcal{G}^{(2)}_{\rm{M}, \alpha \alpha^\prime }  ( \boldsymbol{k},   \boldsymbol{k}^\prime  ; i\omega_m)   \rangle^{\rm{imp}} _{\rm{ave}} & =
  \delta_{ \boldsymbol{k},   \boldsymbol{k}^\prime }       \mathcal{G}^{(0)}_{\rm{M},    \alpha  \alpha _1 }  ( \boldsymbol{k}  ; i\omega_m)
   \left(  n_{\rm{imp}}     \int \frac{d^2q}{(2\pi)^2}   | v_{\rm{imp}}(  \boldsymbol{q}  - \boldsymbol{k} ) |^2
\mathcal{G}^{(0)}_{\rm{M}, \alpha_1 \alpha_0 }  ( \boldsymbol{q}  ; i\omega_m) \right)
 \mathcal{G}^{(0)}_{ \rm{M}, \alpha_0 \alpha^\prime  }  ( \boldsymbol{k}^\prime  ; i\omega_m) \notag\\
&  \equiv  \delta_{ \boldsymbol{k},   \boldsymbol{k}^\prime } \bar{ \mathcal{G}}^{(2)}_{\rm{M}, \alpha \alpha^\prime }  ( \boldsymbol{k}  ; i\omega_m),
  \label{impurityaverageformulas2}
\end{align}
\end{widetext}
where $ n_{\rm{imp}} =  N_{\rm{imp}}/V $ is the number density of impurities.
To derive Eq.    \eqref{impurityaverageformulas2},  we have taken the continuum limit $ V^{-1} \sum_{ \boldsymbol{q}} \to \int  d^2q/(2\pi)^2$.
Further, we used  $ v^\ast_{\rm{imp}}(  \boldsymbol{q}) = v_{\rm{imp}}( - \boldsymbol{q}) $ and  $v_{\rm{imp}}(  \boldsymbol{0}) = 0.$
The impurity-averaged Green's function   $   \langle   \mathcal{G}^{(n)}_{\rm{M}, \alpha \alpha^\prime }  ( \boldsymbol{k},   \boldsymbol{k}^\prime  ; i\omega_m)   \rangle^{\rm{imp}} _{\rm{ave}}  $  in  Eq. \eqref{impurityaverageformulas2} 
is diagonal with respect to momentum. Similarly, the impurity-averaged Green's function is diagonal in momentum  for $n \ge 3$ \cite{MBQtextbook,Mahantextbook}.
As a result, when the impurity average is taken on Green's function, it restores the translational symmetry.
Let us express $   \langle   \mathcal{G}_{\rm{M}, \alpha \alpha^\prime }  ( \boldsymbol{k},   \boldsymbol{k}^\prime  ; i\omega_m)   \rangle^{\rm{imp}} _{\rm{ave}}  $ 
and $   \langle   \mathcal{G}^{(n)}_{\rm{M}, \alpha \alpha^\prime }  ( \boldsymbol{k},   \boldsymbol{k}^\prime  ; i\omega_m)   \rangle^{\rm{imp}} _{\rm{ave}}  $ as
$\bar{  \mathcal{G}} _{\rm{M}, \alpha \alpha^\prime }  ( \boldsymbol{k}; i\omega_m)  \delta_{ \boldsymbol{k},   \boldsymbol{k}^\prime } $  and $\bar{  \mathcal{G}}  ^{(n)}_{\rm{M}, \alpha \alpha^\prime }  ( \boldsymbol{k}; i\omega_m)  \delta_{ \boldsymbol{k},   \boldsymbol{k}^\prime } $, respectively.
Their Fourier transforms are given by
$ \langle  \mathcal{G}_{\rm{M}, \alpha \alpha^\prime }  ( \boldsymbol{x},   \boldsymbol{x}^\prime   ; i\omega_m)  \rangle^{\rm{imp}} _{\rm{ave}} 
\equiv  \bar{  \mathcal{G}} _{\rm{M}, \alpha \alpha^\prime  }  ( \boldsymbol{x} - \boldsymbol{x}^\prime ; i\omega_m)
= V^{-1} \sum_{ \boldsymbol{k} }   e^{ i \boldsymbol{k} \cdot ( \boldsymbol{x} - \boldsymbol{x}^\prime ) } \bar{ \mathcal{G}}_{\rm{M}, \alpha \alpha^\prime  }  ( \boldsymbol{k}  ; i\omega_m) $
and
$ \langle  \mathcal{G}^{(n)}_{\rm{M}, \alpha \alpha^\prime }  ( \boldsymbol{x},   \boldsymbol{x}^\prime   ; i\omega_m)  \rangle^{\rm{imp}} _{\rm{ave}} 
\equiv  \bar{  \mathcal{G}}  ^{(n)}_{\rm{M}, \alpha \alpha^\prime  }  ( \boldsymbol{x} - \boldsymbol{x}^\prime ; i\omega_m)
= V^{-1} \sum_{ \boldsymbol{k} }   e^{ i \boldsymbol{k} \cdot ( \boldsymbol{x} - \boldsymbol{x}^\prime ) } \bar{ \mathcal{G}}^{(n)}_{\rm{M}, \alpha \alpha^\prime  }  ( \boldsymbol{k}  ; i\omega_m) $.

Next, we reorganize the perturbative expansion  $\sum_n  \bar{ \mathcal{G}}^{(n)}_{\rm{M}, \alpha \alpha^\prime  }  ( \boldsymbol{k}  ; i\omega_m) $  
by representing it as the sum of all irreducible Feynman diagrams. 
Let us denote the associated self-energy as $ \Sigma^{\rm{imp}} ( \boldsymbol{k},i\omega_m  )_{\alpha \alpha^\prime} .$
Then, the Dyson equation for the impurity-averaged Green's function  $  \bar{ \mathcal{G}}_{\rm{M}, \alpha \alpha^\prime  }  ( \boldsymbol{k}  ; i\omega_m)  $ is expressed as
 \begin{widetext}
\begin{align} 
  \bar{ \mathcal{G}}_{\rm{M}, \alpha \alpha^\prime  }  ( \boldsymbol{k}  ; i\omega_m) &=    \mathcal{G}^{ (0)  }_{\rm{M}, \alpha \alpha^\prime  }  ( \boldsymbol{k}  ; i\omega_m) 
 +     \mathcal{G}^{ (0) } _{\rm{M}, \alpha \alpha_2  }  ( \boldsymbol{k}  ; i\omega_m)   \Sigma^{\rm{imp}} ( \boldsymbol{k}; i\omega_m  )_{\alpha_2 \alpha_1}
 \bar{ \mathcal{G}}_{\rm{M}, \alpha_1 \alpha^\prime  }  ( \boldsymbol{k}  ; i\omega_m) .
 \label{imaginaryimpuritydysonequation4}
\end{align}
 \end{widetext}
To evaluate the self-energy $\Sigma^{\rm{imp}} ( \boldsymbol{k}; i\omega_m  )$, we take the impurity potential as $V_{\rm{imp}}(\boldsymbol{x}-\boldsymbol{X}^{\rm{imp}}_i) = v_0 \delta (\boldsymbol{x}-\boldsymbol{X}^{\rm{imp}}_i)$ with $v_0$ a constant and use the first-Born approximation.  Then, we have $ v_{\rm{imp}}(  \boldsymbol{q}) = v_0.$ 
Since the Fermi energy $ \epsilon_{\rm{F}} $ is positive,  the term in the unperturbed Green's function which contributes dominantly to this evaluation is   
$ 1/ ( i\omega_m + \omega_{\rm{F}} -  \omega^{\rm{TI}}_{\boldsymbol{k}} ) $. 
Therefore, when we perform the momentum integral for evaluating $\Sigma^{\rm{imp}} ( \boldsymbol{k},i\omega_m  )$, we just retain  $ 1/ ( i\omega_m + \omega_{\rm{F}} -  \omega^{\rm{TI}}_{\boldsymbol{k}} ) $.
Let us write the self-energy in the first-Born approximation as $\Sigma^{\rm{imp}}_{\rm{1BA}} ( \boldsymbol{k}; i\omega_m  )_{ \alpha \alpha^\prime   } $. It is evaluated as  
 \begin{widetext}
  \begin{align} 
  \Sigma {}^{\rm{imp}}_{\rm{1BA}} ( \boldsymbol{k}; i\omega_m  )_{ \alpha \alpha^\prime   } &=  \frac{ n_{\rm{imp}} }{\hbar} \int \frac{d^2q}{ (2\pi)^2}   | v_{\rm{imp}}( \boldsymbol{q}  -  \boldsymbol{k} ) |^2 
  \mathcal{G}^{(0)}_{\rm{M}, \alpha \alpha^\prime }  ( \boldsymbol{k}; i\omega_m) 
 =   \frac{ 1 }{2} n_{\rm{imp}} v_0^2 \int_{  - \epsilon_{\rm{F}}  }^\infty  d \xi  \tilde{N}(\xi)  \frac{1}{  i\epsilon_m -  \xi   } \notag\\
 & \approx  - \frac{ 1  }{2}   \tilde{N}(0) n_{\rm{imp}} v_0^2 \int_{  -\infty }^\infty  d \xi   \frac{\xi  +i \epsilon_m }{    \xi^2  + \epsilon_m^2 }   
 = - \frac{ i  }{2}   \pi \tilde{N}(0) n_{\rm{imp}} v_0^2   \rm{sgn} (\omega_m)
    \equiv   - \frac{i }{2\tau^{\rm{rel}}_{\rm{TI}} } \rm{sgn} (\omega_m) \delta_{ \alpha, \alpha^\prime },
    \label{1BAselfenergy}
\end{align} 
\end{widetext}
 where $ \epsilon_m = \hbar \omega_m$.
The quantity $\tilde{N}^{\rm{TI}} (\xi) $ is the density of states per volume of the TI surface state measured from the Fermi energy.
It is given by  $\tilde{N}^{\rm{TI}} (\xi) = (  \epsilon_{\rm{F}} +    \xi )/ \left( 2\pi (\hbar v_{\rm{F}})^2 \right).$   
 For going from the first to second line of Eq. \eqref{1BAselfenergy}, we have made an approximation such that the density of state $\tilde{N}^{\rm{TI}} (\xi) $ included in the integrand can be set with $\tilde{N}^{\rm{TI}} (0) $. 
This is because we can consider that the energy state in the vicinity of Fermi level dominantly contributes to the self energy $\Sigma^{\rm{imp}}_{\rm{1BA}} ( \boldsymbol{k}; i\omega_m  )_{ \alpha \alpha^\prime   } $: 
 $ | \boldsymbol{q} |, | \boldsymbol{k}  | \simeq k_{\rm{F}} $ and $ \tilde{N}(\xi) \simeq \tilde{N}(0)$. 
Further, we have replaced the lower limit $ -\epsilon_{\rm{F}}  $ with $-\infty$ since we consider that the number of electrons included in the surface with its area $V$ is large enough and the number density of TI surface $n^{\rm{TI}}_{2D}$
can be taken as large. The relation between $\epsilon_{\rm{F}} $ and $ n^{\rm{TI}}_{2D}$ is given by $  \epsilon_{\rm{F}} = \hbar v_{\rm{F}} \sqrt{ 4\pi n^{\rm{TI}}_{2D}}    $, and hence, we take $ \epsilon_{\rm{F}} \to \infty. $
The time $\tau^{\rm{rel}}_{\rm{TI}}$ is the relaxation time of TI surface state owing to the  impurity effect.
We now derive the impurity-averaged green's function $  \bar{ \mathcal{G}}_{\rm{M}, \alpha \alpha^\prime  }  ( \boldsymbol{k}  ; i\omega_m)$. 
From Eqs.  \eqref{imaginaryimpuritydysonequation4} and  \eqref{1BAselfenergy} we obtain
  \begin{widetext}
  \begin{align} 
  \bar{ \mathcal{G}}_{\rm{M}, \alpha \alpha^\prime  }  ( \boldsymbol{k}  ; i\omega_m) = \left[
 \left(   \mathcal{G}^{ (0)  }_{\rm{M} }  ( \boldsymbol{k}  ; i\omega_m) \right)^{-1} + \frac{i }{2\tau^{\rm{rel}}_{\rm{TI}} } \rm{sgn} (\omega_m) \cdot  \boldsymbol{1} 
  \right]^{-1}_{ \alpha \alpha^\prime   },
  \label{impurityaveragedimaginaryGreensfunction}
 \end{align} 
\end{widetext}
where 
  \begin{align} 
  \left(   \mathcal{G}^{ (0)  }_{\rm{M} }  ( \boldsymbol{k}  ; i\omega_m) \right)^{-1}_ { \alpha \alpha^\prime  }  =( \omega_m + \omega_{\rm{F}}   )\boldsymbol{1}_ { \alpha \alpha^\prime  } +
   \omega^{ \rm{TI} }_{ \boldsymbol{k} }    \tilde{\mathcal{H}}_{0,\alpha \alpha^\prime} .
  \label{inverseimaginaryGreensfunction}
 \end{align} 
By performing the analytic continuation $i \omega_m \to\omega + i {\rm{sgn}} (\omega_m) \eta  $, consequently,
we obtain the retarded and advanced components of impurity-averaged Green's functions  $\bar{g}^{\rm{r}} _{\alpha \alpha^\prime} (  \boldsymbol{k}, \omega   )$
and $\bar{g}^{\rm{a}} _{\alpha \alpha^\prime} (  \boldsymbol{k}, \omega   )$
in Eq. \eqref{unperturbedTIGreen'sfunctions}.  
The retarded component is obtained for  ${\rm{sgn}} (\omega_m)>0$ while we get the advanced component for ${\rm{sgn}} (\omega_m)<0$.

\subsection{Magnon Green's Functions}\label{MGGreen'sfunction}
We present the real-time magnon Green's functions. 
First let us show them without the damping effect.  
Like given in Eq. \eqref{realtimeGreen'sfunction1}, the time-ordered, anti-time-ordered, retarded, and advanced components of 
magnon Green's functions in the interaction picture are  
\begin{widetext}
\begin{align} 
 D^{\rm{t}} (\boldsymbol{p}t;\boldsymbol{p}^\prime t^\prime) &  = -i T \langle a_{H_0}(\boldsymbol{p}t)    a^\dagger_{H_0}(\boldsymbol{p}^\prime t^\prime)      \rangle_{0} 
 =  \delta_{\boldsymbol{p}, \boldsymbol{p}^\prime} D^{\rm{t}} (\boldsymbol{p}, t-  t^\prime)
 = \int \frac{d\omega}{2\pi} e^{ - i \omega (t-t^\prime)}   \delta_{\boldsymbol{p}, \boldsymbol{p}^\prime}  D^{\rm{t}} (\boldsymbol{p}\omega), \notag\\ 
  D^{\tilde{\rm{t}}} (\boldsymbol{p}t;\boldsymbol{p}^\prime t^\prime) &  =  -i \tilde{T} \langle a_{H_0}(\boldsymbol{p}t)    a^\dagger_{H_0}(\boldsymbol{p}^\prime t^\prime)      \rangle_{0} 
  =  \delta_{\boldsymbol{p}, \boldsymbol{p}^\prime}  D^{\tilde{\rm{t}}} (\boldsymbol{p},  t-  t^\prime) 
 = \int \frac{d\omega}{2\pi} e^{ - i \omega (t-t^\prime)}   \delta_{\boldsymbol{p}, \boldsymbol{p}^\prime}  D^{\tilde{\rm{t}}} (\boldsymbol{p}\omega), \notag\\ 
 D^{\rm{t}} (\boldsymbol{p}\omega)  & = 
 \frac{ 1 +  n_b (   \epsilon^{\rm{FM}}_{\boldsymbol{p}}  ) }{ \omega -  \omega^{\rm{FM}}_{\boldsymbol{p}}   +  i \eta } -  
 \frac{ n_b (  \epsilon^{\rm{FM}}_{\boldsymbol{p}} ) }{\omega  -  \omega^{\rm{FM}}_{\boldsymbol{p}}  - i \eta } , \quad
  D^{\tilde{\rm{t}}} (\boldsymbol{p}\omega)  = 
 \frac{   n_b (   \epsilon^{\rm{FM}}_{\boldsymbol{p}}  ) }{ \omega -  \omega^{\rm{FM}}_{\boldsymbol{p}}   +  i \eta } -  
 \frac{ 1 + n_b (  \epsilon^{\rm{FM}}_{\boldsymbol{p}} ) }{\omega  -  \omega^{\rm{FM}}_{\boldsymbol{p}}  - i \eta } , \notag\\ 
 D^{\rm{r}} (\boldsymbol{p}t;\boldsymbol{p}^\prime t^\prime) & = -  i\theta(t - t^\prime )  \cdot \rho^{(0)}(\boldsymbol{p}t;\boldsymbol{p}^\prime t^\prime) =
  \delta_{\boldsymbol{p}, \boldsymbol{p}^\prime} D^{\rm{r}} (\boldsymbol{p}, t -  t^\prime) 
 =  \delta_{\boldsymbol{p}, \boldsymbol{p}^\prime}  \int \frac{d\omega}{2\pi}   e^{ - i \omega (t - t^\prime) }  D^{\rm{r}} (\boldsymbol{p}\omega),  \notag\\
 D^{\rm{a}} (\boldsymbol{p}t;\boldsymbol{p}^\prime t^\prime) & =   i\theta( t^\prime - t )  \cdot \rho^{(0)}(\boldsymbol{p}t;\boldsymbol{p}^\prime t^\prime) =
  \delta_{\boldsymbol{p}, \boldsymbol{p}^\prime} D^{\rm{a}} (\boldsymbol{p}, t -  t^\prime) 
 =  \delta_{\boldsymbol{p}, \boldsymbol{p}^\prime}  \int \frac{d\omega}{2\pi}   e^{ - i \omega (t - t^\prime) }  D^{\rm{a}} (\boldsymbol{p}\omega), \notag\\
 \rho^{(0)}(\boldsymbol{p}t;\boldsymbol{p}^\prime t^\prime)  & =  \delta_{\boldsymbol{p}, \boldsymbol{p}^\prime}  e^{ - i \omega^{\rm{FM}}_{\boldsymbol{p}} (t - t^\prime) }  , \quad
 D^{\rm{r}} (\boldsymbol{p}\omega)  = 
 \frac{ 1  }{ \omega -  \omega^{\rm{FM}}_{\boldsymbol{p}}   +  i \eta } , \quad 
 D^{\rm{a}} (\boldsymbol{p}\omega)  = 
 \frac{ 1  }{ \omega -  \omega^{\rm{FM}}_{\boldsymbol{p}}   -  i \eta } , 
 \label{magnonrealtimeGreen'sfunction1}
\end{align} 
\end{widetext}
where $a_{H_0}(\boldsymbol{p}t) = e^{iH_0t/\hbar} a(\boldsymbol{p}) e^{-iH_0t/\hbar}$ and $a^\dagger_{H_0}(\boldsymbol{p}t) = e^{iH_0t/\hbar} a^\dagger(\boldsymbol{p}) e^{-iH_0t/\hbar}$.
$n_b (\epsilon) = \left( e^{\beta \epsilon  }  - 1 \right)^{-1}  $ is the Bose-Einstein distribution function.
The lesser and greater Green's functions are given by
\begin{widetext}
\begin{align} 
  D^{ > } (\boldsymbol{p}t;\boldsymbol{p}^\prime t^\prime) &  = - i  \langle a_{H_0}(\boldsymbol{p}t)    a^\dagger_{H_0}(\boldsymbol{p}^\prime t^\prime)      \rangle_{0} 
 =  \delta_{\boldsymbol{p}, \boldsymbol{p}^\prime} D^{ > } (\boldsymbol{p}, t-  t^\prime)
 = \int \frac{d\omega}{2\pi} e^{ - i \omega (t-t^\prime)}   \delta_{\boldsymbol{p}, \boldsymbol{p}^\prime}  D^{ > } (\boldsymbol{p}\omega), \notag\\
 D^{ < } (\boldsymbol{p}t;\boldsymbol{p}^\prime t^\prime) &  = - i  \langle    a^\dagger_{H_0}(\boldsymbol{p}^\prime t^\prime)   a_{H_0}(\boldsymbol{p}t)    \rangle_{0} 
 =  \delta_{\boldsymbol{p}, \boldsymbol{p}^\prime} D^{ < } (\boldsymbol{p}, t-  t^\prime)
 = \int \frac{d\omega}{2\pi} e^{ - i \omega (t-t^\prime)}   \delta_{\boldsymbol{p}, \boldsymbol{p}^\prime}  D^{ < } (\boldsymbol{p}\omega), \notag\\
 D^{ > } (\boldsymbol{p}\omega) & = -2\pi i  \delta ( \omega -  \omega^{\rm{FM}}_{\boldsymbol{p}}   )  \big{(} 1 + n_b (  \epsilon^{\rm{FM}}_{\boldsymbol{p}} ) \big{)}, \quad
 D^{ < } (\boldsymbol{p}\omega)  = -2\pi i  \delta ( \omega -  \omega^{\rm{FM}}_{\boldsymbol{p}}   )   n_b (  \epsilon^{\rm{FM}}_{\boldsymbol{p}} ).
 \label{realtimeGreen'sfunction3}
\end{align} 
\end{widetext}
The magnon Green's functions presented above satisfy exactly the same relations given in Eq. \eqref{realtimeGreen'sfunctionrelations1}.  
Further, from Eqs.  \eqref{magnonrealtimeGreen'sfunction1} and  \eqref{realtimeGreen'sfunction3}, we can verify that with similar to Eq. \eqref{realtimeGreen'sfunctionrelations2} these components satisfy the relations 
\begin{align} 
 D^{ < } (\boldsymbol{p}\omega)  & =  - n_b (\hbar \omega)   \left(    D^{\rm{a}} (\boldsymbol{p}\omega) - D^{\rm{r}} (\boldsymbol{p}\omega)      \right), \notag\\
 D^{ > } (\boldsymbol{p}\omega)  & = - \big{ (}  1 + n_b (\hbar \omega) \big{)}  \left(    D^{\rm{a}} (\boldsymbol{p}\omega) - D^{\rm{r}} (\boldsymbol{p}\omega)      \right).
\end{align} 
Next, we show the magnon Green's functions including the damping effect.  
This is obtained by solving the Landau-Lifshitz-Gilbert equation 
  \begin{align}
 \frac{d \boldsymbol{S}_i  }{ dt } =  \gamma  \left( \boldsymbol{B}_0 + \boldsymbol{B}^{\rm{ext}} (t) \right)  \times
  \boldsymbol{S}_i  - \frac{\alpha}{S_0} \left( \boldsymbol{S}_i   \times  \frac{d \boldsymbol{S}_i  }{ dt } \right), 
\label{LLGEq}
\end{align} 
where $\boldsymbol{B}_0 = (0, B_0, 0) $  with $B_0$ a constant. 
For $\boldsymbol{B}^{\rm{ext}} (t)$, we take the same magnetic-field configuration as we did 
in Sec. \ref{microscopictheory}:  $\boldsymbol{B}^{\rm{ext}}(t)= 
B^{\rm{ext}}\left(\sin \big{(}  { \rm{sgn} }(B_0) \cdot  \omega^{\rm{ext}}t  \big{)}  ,0, \cos \big{(}  { \rm{sgn} }(B_0) \cdot  \omega^{\rm{ext}}t  \big{)}   \right)$. The solution is given in the form  $S^y_i= - S_0  { \rm{sgn} }(B_0)$  and
 $S^-_i = S^z_i - i S^x_i = \bar{S}^-_i  e^{ - i  { \rm{sgn} }(B_0) \omega^{\rm{ext}} t} $  where   $\bar{S}^-_i$ is a complex constant. 
The demagnetizing coefficient is going to be excluded.
By introducing $ B^{ \rm{ext},- } = B^{ \rm{ext}, z }  - i B^{ \rm{ext}, x } $ and the magnetic susceptibility as $\chi^{\rm{mag}} $,   for $ {\rm{sgn}}(B_0) > 0$
we re-express $S^-_i $ as $S^-_i =  \chi^{\rm{mag}} B^{ \rm{ext},- }$ \cite{spintronicsRMP2005}. As a result, we obtain  
 \begin{align}
 \chi^{\rm{mag}} = \frac{ \gamma S_0}{ \omega^{\rm{ext}} -   \omega^{\rm{FM}}_{\boldsymbol{0}}  + i   \alpha  \omega^{\rm{ext}}  },
\label{magneticsusceptibility}
\end{align} 
where $ \omega^{\rm{FM}}_{\boldsymbol{0}} =   \gamma |B_0|$, which is the Zeeman gap of magnon. 
With eliminating the factor $ \gamma S_0$ in Eq. \eqref{magneticsusceptibility}, we identify the magnetic susceptibility $\chi^{\rm{mag}} $
with the retarded  Green's function  $\bar{D}^{\rm{r}}  (  \boldsymbol{0}, \omega   ) $  in Eq. \eqref{unperturbedmagnonGreen'sfunctions}.
On the other hand, for $ {\rm{sgn}}(B_0) < 0$ we identify the retarded function with the response (magnetic susceptibility) of $S^+_i = S^z_i + i S^x_i $ to $B^+_i = B^z_i + i B^x_i $
and this is equal to $ \chi^{\rm{mag}} $ in Eq. \eqref{magneticsusceptibility}.
The advanced component is given by the complex conjugate of retarded component.  

\section{ Keldysh Green's Functions}\label{appendixB}
In this section,  first we present the formalism for the Keldysh Green's function.
Next, we show how the Keldysh Green's function is related to the real-time Green's function via the real-time projection.
Further, with presenting some useful formulas obtained by the the real-time projection, 
 we demonstrate the derivation of impurity-averaged real-time Green's functions. 
\subsection{Real-Time Projection}\label{real-timeP}

As discussed in subsec.  \ref{PNeqGreensfunc}, our starting point is  the full lesser Green's function of TI surface state in Eq.  \eqref{TIECCexpv2}.
We rewrite this with the Keldysh Green's function given by Eq. \eqref{KGreen'sfunctiondef} or
\begin{align*}
iG_{C,\alpha\alpha^\prime}(\boldsymbol{x}\tau;\boldsymbol{x}^\prime \tau^\prime)
=  \Big{\langle}  T_{C}\big{[} \mathcal{U}^{\rm{exc}}_{C} \mathcal{U}^{\rm{ext}}_{C}     
\psi^\dagger_{ H_0 \alpha^\prime} (\boldsymbol{x}^\prime \tau^\prime) \psi_{ H_0 \alpha} (\boldsymbol{x} \tau)  \big{]}
 \Big{\rangle}_0.   
\end{align*}
The time-evolution operators  $  \mathcal{U}^{\rm{exc}}_C$ and $ \mathcal{U}^{\rm{ext}}_{C}   $ in the above equation are defined by Eq. \eqref{Uoperators} or 
 \begin{align*}
& \mathcal{U}^{\rm{exc}}_C =\exp\left(-\frac{i}{\hbar}\int_C d \check{\tau} V^{\rm{exc}}_{H_0}( \check{\tau}) \right), \notag\\ 
& \mathcal{U}^{\rm{ext}}_C=\exp\left(-\frac{i}{\hbar}\int_C d \tilde{\tau}  H^{\rm{ext}}_{H_0}( \tilde{\tau}) \right).
\end{align*}
The perturbative calculation is performed by expanding $ \mathcal{U}^{\rm{exc}}_{C}$ and $ \mathcal{U}^{\rm{ext}}_{C}   $ with respect to  $V^{\rm{exc}}_{H_0}(\check{\tau})$ and $H^{\rm{ext}}_{H_0}( \tilde{\tau})$, respectively.
Then by taking thermal average on them, these perturbative expansions are described by the unperturbed Keldysh Green's functions given by Eqs. \eqref{TICGreensfunction1} and  \eqref{magnonCGreensfunction1} or
\begin{align*}
&i  \mathcal{G}^0_{C,\alpha \alpha^\prime}(\boldsymbol{x} \tau ; \boldsymbol{x}^\prime \tau^\prime)= \Big{\langle}  T_{C}\big{[}   
\psi_{H_0\alpha} (\boldsymbol{x} \tau) \psi^\dagger_{H_0  \alpha^\prime} (\boldsymbol{x}^\prime \tau^\prime) \big{]}
\Big{\rangle}_0,   \\
&i \mathcal{D}^0_{C}(\boldsymbol{q} \tau ; \boldsymbol{q}^\prime \tau^\prime)= \Big{\langle}  T_{C}\big{[}   
a_{H_0} (\boldsymbol{q} \tau) a^\dagger_{H_0 } (\boldsymbol{q}^\prime\tau^\prime)  \big{]}
\Big{\rangle}_0.
\end{align*}
We perform the real-time projection to the above Keldysh Green's functions in order to calculate the physical observables.  
We do this by classifying whether the contour time $\tau$ belongs to the path $C_-$  or $C_+$ while $\tau^\prime$ to $C_-$  or $C_+$ (see Fig. \ref{SchwingerKeldyshcontour}).
We have four different configurations.
To represent this situation clearly, let us introduce a two-by-two-matrix Green's function (Schwinger-Keldysh Green's function) \cite{NEQGreensfunctionRMPandtxtbook1}
\begin{align}
\hat{\mathcal{G}}_{\alpha \alpha^\prime }(\boldsymbol{x}\tau; \boldsymbol{x}^\prime \tau^\prime)=\left(
\begin{array}{ccc}
\hat{\mathcal{G}}^{--}_{\alpha \alpha^\prime }(\boldsymbol{x}t^-; \boldsymbol{x}^\prime t^{\prime-}) &  \hat{\mathcal{G}}^{-+}_{\alpha \alpha^\prime }(\boldsymbol{x}t^-; \boldsymbol{x}^\prime t^{\prime+}) \\
\hat{\mathcal{G}}^{+-}_{\alpha \alpha^\prime }(\boldsymbol{x}t^+; \boldsymbol{x}^\prime t^{\prime-})& \hat{\mathcal{G}}^{++}_{\alpha \alpha^\prime }(\boldsymbol{x}t^+; \boldsymbol{x}^\prime t^{\prime+}) \\
\end{array}
\right), \label{Schwinger-KeldyshGreensfunction1}
\end{align} 
where $t^{\pm}$ and $t^{\prime\pm}$ are both real times. 
The component $\hat{\mathcal{G}}^{\mu \mu^\prime }_{\alpha \alpha^\prime }(\boldsymbol{x} t^\mu; \boldsymbol{x}^\prime t^{\prime\mu^\prime})$ $(\mu,\mu^\prime=\mp)$ is representing that 
the contour time $\tau = t^\mu $ belongs to the contour  $C_\mu$ while $\tau^\prime = t^{\prime \mu^\prime}$ belongs to $C_{\mu^\prime}$.  
The components $ \hat{\mathcal{G}}^{--}_{\alpha \alpha^\prime }(\boldsymbol{x}t^-; \boldsymbol{x}^\prime t^{\prime-}),\hat{\mathcal{G}}^{-+}_{\alpha \alpha^\prime }(\boldsymbol{x}t^-; \boldsymbol{x}^\prime t^{\prime+}),
 \hat{\mathcal{G}}^{+-}_{\alpha \alpha^\prime }(\boldsymbol{x}t^+; \boldsymbol{x}^\prime t^{\prime-}),$ and  $\hat{\mathcal{G}}^{++}_{\alpha \alpha^\prime }(\boldsymbol{x}t^+; \boldsymbol{x}^\prime t^{\prime+})$ are
equivalent to time-ordered, lesser, greater, and anti-time-ordered components, respectively.

Similarly, we introduce the Schwinger-Keldysh Green's function of magnon given by 
\begin{align}
\hat{\mathcal{D}} (\boldsymbol{q}\tau; \boldsymbol{q}^\prime \tau^\prime)=\left(
\begin{array}{ccc}
\hat{\mathcal{D}}^{--} (\boldsymbol{q}t^-; \boldsymbol{q}^\prime t^{\prime-}) &  \hat{\mathcal{D}}^{-+}  (\boldsymbol{q}t^-; \boldsymbol{q}^\prime t^{\prime+}) \\
\hat{\mathcal{D}}^{+-} (\boldsymbol{q}t^+; \boldsymbol{q}^\prime t^{\prime-})& \hat{\mathcal{D}}^{++}  (\boldsymbol{q}t^+; \boldsymbol{q}^\prime t^{\prime+}) \\
\end{array}
\right), \label{Schwinger-KeldyshGreensfunction2}
\end{align} 
where the components $ \hat{\mathcal{D}}^{--} (\boldsymbol{q}t^-; \boldsymbol{q}^\prime t^{\prime-}),  \hat{\mathcal{D}}^{-+} (\boldsymbol{q}t^-; \boldsymbol{q}^\prime t^{\prime+})$,
$\hat{\mathcal{D}}^{+-} (\boldsymbol{q}t^+; \boldsymbol{q}^\prime t^{\prime-}), $ and $ \hat{\mathcal{D}}^{++} (\boldsymbol{q}t^+; \boldsymbol{q}^\prime t^{\prime+}) $
are equivalent to time-ordered, lesser, greater, and anti-time-ordered Green's functions, respectively.

In the following, let us show some examples of calculation for the real-time projection on the Keldysh Green's functions.  
We will just write the time arguments of functions and omit the arguments of spatial coordinate or momentum
since what we want to demonstrate here is the calculation for real-time projection and integrals of real-time variables. 
We perform the integral along the contour $C$ by decomposing it into $C_-$ and $C_+$ and rewrite them by the real-time variables.  

For practice, first let us show the simplest example of integral along the contour $C$ given by a single contour-time variable $\tau_1$. It has a form 
\begin{align}
 f (\tau , \tau^\prime) & =
 \int_{C } d\tau_1 g(\tau, \tau_1)  h(\tau_1, \tau^\prime) \notag\\
& = \int_{-\infty} ^{+\infty } \tau_z^{\mu_1 \mu_1 } d t^{\mu_1} _1
 f(t, t^{\mu_1} _1 )  g(t^{\mu_1}_1, t^\prime),
\label{CtoRtimeintegral1}
\end{align}
where 
 \begin{align}
\tau_z^{\mu_1 \mu^\prime_1}=\left(
\begin{array}{cc}
\tau_z^{--} & \tau_z^{-+} \\
\tau_z^{+-} & \tau_z^{++}  \\
\end{array}
\right)= \left(
\begin{array}{cc}
1& 0 \\
0 & -1 \\
\end{array}
\right),
\label{Schwinger-Keldysh-spacematrix1}
\end{align} 
and $t^{\mu_1}_1$ is the real-time variable. 
Via the real-time projection, let us rewrite the function $ f (\tau , \tau^\prime)$ as $ f^{ \nu_{\mu \mu^\prime}  } (t ,t^\prime)$.
 Here $t$ and $t^\prime$ are real-time variables corresponding to the real-time projection of the contour times $\tau$ and  $\tau^\prime$, respectively.
The superscript $\nu_{\mu \mu^\prime} = { \rm{t} },<,>, { \tilde{\rm{t}} }$ with $\mu,\mu^\prime=\pm$. 
It describes the situation such that $\tau  \in C_\mu $ while  $\tau^\prime  \in C_{\mu^\prime} $. 
For instance, when  $\tau  \in C_- $ while  $\tau^\prime  \in C_{+} $ the function $ f (\tau , \tau^\prime)$ becomes $ f^{ <  } (t ,t^\prime)$.
In the following,  we list the four cases of  $ f (\tau , \tau^\prime)$ given as
\begin{widetext}
\begin{align}
 f^{ {\rm{t}} } (t ,t^\prime) & =
 \int_{-\infty} ^{+\infty }d t_1  \big{(}
 g^{ \rm{t} }(t, t_1)  h^{   {\rm{t}} }(t_1, t^\prime)    -  g^{ < }(t, t_1)  h^{ > }(t_1, t^\prime)  \big{)} , \notag\\
 f^{ < } (t ,t^\prime) & =
 \int_{-\infty} ^{+\infty }d t_1  \big{(}
 g^{ \rm{t} }(t, t_1)  h^{<}(t_1, t^\prime)    -  g^{ < }(t, t_1)  h^{ \tilde{ \rm{t}} }(t_1, t^\prime)  \big{)} , \notag\\
 f^{ > } (t ,t^\prime) & =
 \int_{-\infty} ^{+\infty }d t_1  \big{(}
 g^{ > }(t, t_1)  h^{ {\rm{t}} }(t_1, t^\prime)    -  g^{  \tilde{ \rm{t}}    }(t, t_1)  h^{ > }(t_1, t^\prime)  \big{)} , \notag\\
  f^{ \tilde{ \rm{t}}  } (t ,t^\prime) & =
 \int_{-\infty} ^{+\infty }d t_1  \big{(}
 g^{ > }(t, t_1)  h^{ <    }(t_1, t^\prime)    -  g^{ \tilde{ \rm{t}} }(t, t_1)  h^{ \tilde{ \rm{t}} }(t_1, t^\prime)  \big{)} .
 \label{CtoRtimeintegral2}
\end{align}
\end{widetext}
By using the relations in Eq. \eqref{realtimeGreen'sfunctionrelations1}, Eq. \eqref{CtoRtimeintegral2} can be redescribed by the lesser, greater, retarded, and advanced components as
\begin{widetext}
\begin{align}
f^{ < } (t ,t^\prime) & =
 \int_{-\infty} ^{+\infty }d t_1  \big{(}
     g^{ < }(t, t_1)  h^{  \rm{a} }(t_1, t^\prime) +g^{ \rm{r} }(t, t_1)  h^{<}(t_1, t^\prime)   \big{)} , \notag\\
     f^{ > } (t ,t^\prime) & =
 \int_{-\infty} ^{+\infty }d t_1  \big{(}
     g^{ > }(t, t_1)  h^{  \rm{a} }(t_1, t^\prime) +g^{ \rm{r} }(t, t_1)  h^{  > }(t_1, t^\prime)   \big{)} , \notag\\
 f^{ {\rm{r}} } (t ,t^\prime)  & =
 \int_{-\infty} ^{+\infty }d t_1 
 g^{ \rm{r} }(t, t_1)  h^{   {\rm{r}} }(t_1, t^\prime)     , \qquad
  f^{  \rm{a} } (t ,t^\prime)  =
 \int_{-\infty} ^{+\infty }d t_1 g^{ \rm{a} }(t, t_1)  h^{   {\rm{a}} }(t_1, t^\prime)  .
 \label{ReCtoRtimeintegral2}
\end{align}
\end{widetext} 
Next, we present an example of temporal function represented by two contour-time variables $ \tau_1$ and  $ \tau_2$ given by
\begin{align}
f (\tau , \tau^\prime) =
 \int_{C } d\tau_1 d\tau_2 g(\tau, \tau_2)  h(\tau_2, \tau_1) l(\tau_1, \tau^\prime). 
\label{CtoRtimeintegral3}
\end{align}
Like we did in Eq.  \eqref{CtoRtimeintegral2},  we perform the real-time projection on $\tau_1$ and $\tau_2$ and rewrite them as $t_1$ and $t_2$, respectively.
As a result, we have
\begin{widetext}
\begin{align}
f^{ <} (t ,t^\prime)  & =
 \int_{-\infty} ^{+\infty }d t_1  d t_2
 \left(
  g^{ < }(t, t_2) h^{  \rm{a}   }(t_2, t_1) l^{ \rm{a}  }(t_1, t^\prime) +
 g^{  \rm{r} }(t, t_2)  h^{ <   }(t_2, t_1) l^{ \rm{a}  }(t_1, t^\prime) +
  g^{  \rm{r} }(t, t_2)  h^{  \rm{r}  }(t_2, t_1) l^{ < }(t_1, t^\prime)
    \right), \notag\\ 
  f^{ <} (t ,t^\prime)  & =
 \int_{-\infty} ^{+\infty }d t_1  d t_2
 \left(
  g^{ > }(t, t_2) h^{  \rm{a}   }(t_2, t_1) l^{ \rm{a}  }(t_1, t^\prime) +
 g^{  \rm{r} }(t, t_2)  h^{ >   }(t_2, t_1) l^{ \rm{a}  }(t_1, t^\prime) +
  g^{  \rm{r} }(t, t_2)  h^{  \rm{r}  }(t_2, t_1) l^{ > }(t_1, t^\prime)
    \right), \notag\\  
f^{ \rm{r} } (t ,t^\prime)  & =
 \int_{-\infty} ^{+\infty }d t_1  d t_2 \left(
  g^{  \rm{r}  }(t, t_2)   h^{  \rm{r}  }(t_2, t_1) l^{  \rm{r}  }(t_1, t^\prime) \right), \qquad 
 f^{ \rm{a} } (t ,t^\prime)   =
 \int_{-\infty} ^{+\infty }d t_1  d t_2 \left(
  g^{  \rm{a}  }(t, t_2)   h^{  \rm{a}  }(t_2, t_1) l^{  \rm{a}  }(t_1, t^\prime) \right). 
   \label{CtoRtimeintegral4}
\end{align}
\end{widetext}
Eqs.   \eqref{ReCtoRtimeintegral2} and \eqref{CtoRtimeintegral4} are called Langreth rules \cite{NEQGreensfunctionRMPandtxtbook1,NEQGreensfunctiontxtbook2}. 

As a last example,  we demonstrate a calculation represented by three contour-time variables $ \tilde{\tau}_1,  \tilde{\tau}_2,$ and  $ \tau_1$.
The integral which we calculate is  
\begin{align}
f(\tau,\tau^\prime) =  \int_{C } d \tilde{\tau}_1 d \tilde{\tau}_2  d \tau_1  l( \tilde{\tau}_1, \tau_1 ) m( \tau_1,  \tilde{\tau}_2 ) n( \tau, \tau_1 ) o( \tau_1,  \tau^\prime ) .
\label{CtoRtimeintegral5}
\end{align}
With using the real-time variables  $ \tilde{t}_1 $, $ \tilde{t}_2$, and $ t_1$ corresponding to $ \tilde{\tau}_1,  \tilde{\tau}_2,$ and  $ \tau_1$, respectively, the right-hand side of Eq. \eqref{CtoRtimeintegral5} is rewritten as 
\begin{widetext}
\begin{align}
 f^{<}(t, t^\prime) & =
 \int_{-\infty} ^{+\infty }  d \tilde{t}_1 d \tilde{t}_2 d t_1
  l^{ \rm{a} }( \tilde{t}_1, t_1)  m^{  \rm{r} }( t_1, \tilde{t}_2)  
   \left(
    n^{ < }(t, t_1)  o^{ \rm{a} }(t_1, t^\prime) + n^{ \rm{r} }(t, t_1)  o^{<}(t_1, t^\prime)
  \right), \notag\\
  f^{>}(t, t^\prime) & =
 \int_{-\infty} ^{+\infty }  d \tilde{t}_1 d \tilde{t}_2 d t_1
  l^{ \rm{a} }( \tilde{t}_1, t_1)  m^{  \rm{r} }( t_1, \tilde{t}_2)  
   \left(
 n^{ > }(t, t_1)  o^{ \rm{a} }(t_1, t^\prime) + n^{ \rm{r} }(t, t_1)  o^{>}(t_1, t^\prime)
  \right), \notag\\
   f^{  { \rm{r} }    }(t, t^\prime) & =
 \int_{-\infty} ^{+\infty }  d \tilde{t}_1 d \tilde{t}_2 d t_1
  l^{ \rm{a} }( \tilde{t}_1, t_1)  m^{  \rm{r} }( t_1, \tilde{t}_2)  
 n^{ \rm{r} }(t, t_1)  o^{ \rm{r}  }(t_1, t^\prime) , \notag\\
  f^{  \rm{a}   }(t, t^\prime) & =
 \int_{-\infty} ^{+\infty }  d \tilde{t}_1 d \tilde{t}_2 d t_1
  l^{ \rm{a} }( \tilde{t}_1, t_1)  m^{  \rm{r} }( t_1, \tilde{t}_2)  
  n^{  \rm{a} }(t, t_1)  o^{  \rm{a} }(t_1, t^\prime) .   \label{CtoRtimeintegral6}
\end{align}
\end{widetext}

\subsection{Impurity-Averaged Real-Time Green's Function}\label{IARTGreen'sfunction}
Let us apply the Keldysh Green's function formalism to derive the retarded, advanced, lesser, and greater components of impurity-averaged real-time Green's functions.    

The Dyson equation for the Keldysh Green's function of TI surface state due to the non-magnetic impurity effect is given by \cite{spintronicsPR2008,NEQGreensfunctionRMPandtxtbook1,NEQGreensfunctiontxtbook2}
\begin{widetext}
\begin{align} 
 \mathcal{G}_{C,\alpha \alpha^\prime}(\boldsymbol{x} \tau ; \boldsymbol{x}^\prime \tau^\prime) & =  \mathcal{G}^0_{C,\alpha \alpha^\prime}(\boldsymbol{x} \tau ; \boldsymbol{x}^\prime \tau^\prime) 
 + \int_C d \tau_1 \int d^2 x_1  \mathcal{G}^0_{C,\alpha \alpha^\prime_1}(\boldsymbol{x} \tau ; \boldsymbol{x}_1 \tau_1)    \mathcal{H}^{\rm{imp}} _{\alpha^\prime_1 \alpha_1} (\boldsymbol{x}_1)   
 \mathcal{G}_{C,\alpha_1 \alpha^\prime}(\boldsymbol{x}_1 \tau_1 ; \boldsymbol{x}^\prime \tau^\prime), \notag\\
 & =  \mathcal{G}^0_{C,\alpha \alpha^\prime}(\boldsymbol{x} \tau ; \boldsymbol{x}^\prime \tau^\prime)
 + \int_C d \tau_1 \int d^2 x_1  \mathcal{G}_{C,\alpha \alpha^\prime_1}(\boldsymbol{x} \tau ; \boldsymbol{x}_1 \tau_1)    \mathcal{H}^{\rm{imp}} _{\alpha^\prime_1 \alpha_1} (\boldsymbol{x}_1)   
 \mathcal{G}^0_{C,\alpha_1 \alpha^\prime}(\boldsymbol{x}_1 \tau_1 ; \boldsymbol{x}^\prime \tau^\prime).  \label{Keldyshimpurity1}
\end{align}
\end{widetext}
We use the formulas given in Eq.  \eqref{ReCtoRtimeintegral2}   and perform the real-time projection on the contour times $ \tau ,  \tau ^\prime$ and $\tau_1$ in Eq.  \eqref{Keldyshimpurity1}.
Then, we obtain the Dyson equations  for retarded, advanced, lesser, and greater Green's functions given by
\begin{widetext}
\begin{align}
 g^{ { \rm{r} } }_{\alpha \alpha^\prime}(\boldsymbol{x} t ; \boldsymbol{x}^\prime t^\prime) & = g^{ { \rm{r} } (0) }_{\alpha \alpha^\prime}(\boldsymbol{x} t ; \boldsymbol{x}^\prime t^\prime) 
 + \int d t_1 d^2 x_1   \mathcal{H}^{\rm{imp}} _{\alpha^\prime_1 \alpha_1}  (\boldsymbol{x}_1)  
g^{ {\rm{r}}  (0)}_{\alpha \alpha^\prime_1}(\boldsymbol{x} t ; \boldsymbol{x}_1 t_1)      g^{ {\rm{r}}  }  _{\alpha_1 \alpha^\prime}(\boldsymbol{x}_1 t_1 ; \boldsymbol{x}^\prime t^\prime) \notag\\
 & = g^{ { \rm{r} } (0) }_{\alpha \alpha^\prime}(\boldsymbol{x} t ; \boldsymbol{x}^\prime t^\prime) 
 + \int d t_1 d^2 x_1   \mathcal{H}^{\rm{imp}} _{\alpha^\prime_1 \alpha_1}  (\boldsymbol{x}_1)  
g^{ {\rm{r}} }_{\alpha \alpha^\prime_1}(\boldsymbol{x} t ; \boldsymbol{x}_1 t_1)      g^{ {\rm{r}} (0) }  _{\alpha_1 \alpha^\prime}(\boldsymbol{x}_1 t_1 ; \boldsymbol{x}^\prime t^\prime) \notag\\
 g^{ { \rm{a} } }_{\alpha \alpha^\prime}(\boldsymbol{x} t ; \boldsymbol{x}^\prime t^\prime) & = g^{ { \rm{a} } (0) }_{\alpha \alpha^\prime}(\boldsymbol{x} t ; \boldsymbol{x}^\prime t^\prime) 
 + \int d t_1 d^2 x_1   \mathcal{H}^{\rm{imp}} _{\alpha^\prime_1 \alpha_1}  (\boldsymbol{x}_1)  
g^{ {\rm{a}}  (0)}_{\alpha \alpha^\prime_1}(\boldsymbol{x} t ; \boldsymbol{x}_1 t_1)      g^{ {\rm{a}}   }  _{\alpha_1 \alpha^\prime}(\boldsymbol{x}_1 t_1 ; \boldsymbol{x}^\prime t^\prime) \notag\\
 & = g^{ { \rm{a} } (0) }_{\alpha \alpha^\prime}(\boldsymbol{x} t ; \boldsymbol{x}^\prime t^\prime) 
 + \int d t_1 d^2 x_1   \mathcal{H}^{\rm{imp}} _{\alpha^\prime_1 \alpha_1}  (\boldsymbol{x}_1)  
g^{ {\rm{a}}  }_{\alpha \alpha^\prime_1}(\boldsymbol{x} t ; \boldsymbol{x}_1 t_1)      g^{ {\rm{a}}  (0) }  _{\alpha_1 \alpha^\prime}(\boldsymbol{x}_1 t_1 ; \boldsymbol{x}^\prime t^\prime) \notag\\
g^{<}_{\alpha \alpha^\prime}(\boldsymbol{x} t ; \boldsymbol{x}^\prime t^\prime) & =  g^{<(0)}_{\alpha \alpha^\prime}(\boldsymbol{x} t ; \boldsymbol{x}^\prime t^\prime) 
 + \int d t_1 d^2 x_1   \mathcal{H}^{\rm{imp}} _{\alpha^\prime_1 \alpha_1}  (\boldsymbol{x}_1)   \left(
 g^{ <  (0) }_{\alpha \alpha^\prime_1}(\boldsymbol{x} t ; \boldsymbol{x}_1 t_1)   g^{ {\rm{a}}   }  _{\alpha_1 \alpha^\prime}(\boldsymbol{x}_1 t_1 ; \boldsymbol{x}^\prime t^\prime)
 +
 g^{ {\rm{r}}  (0)}_{\alpha \alpha^\prime_1}(\boldsymbol{x} t ; \boldsymbol{x}_1 t_1)   g^{ <    }_{\alpha_1 \alpha^\prime}(\boldsymbol{x}_1 t_1 ; \boldsymbol{x}^\prime t^\prime)
 \right) \notag\\ 
 & =  g^{<(0)}_{\alpha \alpha^\prime}(\boldsymbol{x} t ; \boldsymbol{x}^\prime t^\prime)
 + \int d t_1 d^2 x_1   \mathcal{H}^{\rm{imp}} _{\alpha^\prime_1 \alpha_1}  (\boldsymbol{x}_1)   \left(
 g^{ <   }_{\alpha \alpha^\prime_1}(\boldsymbol{x} t ; \boldsymbol{x}_1 t_1)   g^{ {\rm{a}}  (0) }  _{\alpha_1 \alpha^\prime}(\boldsymbol{x}_1 t_1 ; \boldsymbol{x}^\prime t^\prime)
 +
 g^{ {\rm{r}}  }_{\alpha \alpha^\prime_1}(\boldsymbol{x} t ; \boldsymbol{x}_1 t_1)   g^{ <  (0)  }_{\alpha_1 \alpha^\prime}(\boldsymbol{x}_1 t_1 ; \boldsymbol{x}^\prime t^\prime)
 \right) \notag\\  
 g^{>}_{\alpha \alpha^\prime}(\boldsymbol{x} t ; \boldsymbol{x}^\prime t^\prime) & =   g^{ > (0)}_{\alpha \alpha^\prime}(\boldsymbol{x} t ; \boldsymbol{x}^\prime t^\prime) 
 + \int d t_1 d^2 x_1   \mathcal{H}^{\rm{imp}} _{\alpha^\prime_1 \alpha_1}  (\boldsymbol{x}_1)   \left(
 g^{ >  (0) }_{\alpha \alpha^\prime_1}(\boldsymbol{x} t ; \boldsymbol{x}_1 t_1)   g^{ {\rm{a}}   }  _{\alpha_1 \alpha^\prime}(\boldsymbol{x}_1 t_1 ; \boldsymbol{x}^\prime t^\prime)
 +
 g^{ {\rm{r}}  (0)}_{\alpha \alpha^\prime_1}(\boldsymbol{x} t ; \boldsymbol{x}_1 t_1)   g^{ >    }_{\alpha_1 \alpha^\prime}(\boldsymbol{x}_1 t_1 ; \boldsymbol{x}^\prime t^\prime)
 \right) \notag\\ 
 & =  g^{ > (0)}_{\alpha \alpha^\prime}(\boldsymbol{x} t ; \boldsymbol{x}^\prime t^\prime)
 + \int d t_1 d^2 x_1   \mathcal{H}^{\rm{imp}} _{\alpha^\prime_1 \alpha_1}  (\boldsymbol{x}_1)   \left(
 g^{ >   }_{\alpha \alpha^\prime_1}(\boldsymbol{x} t ; \boldsymbol{x}_1 t_1)   g^{ {\rm{a}}  (0) }  _{\alpha_1 \alpha^\prime}(\boldsymbol{x}_1 t_1 ; \boldsymbol{x}^\prime t^\prime)
 +
 g^{ {\rm{r}}  }_{\alpha \alpha^\prime_1}(\boldsymbol{x} t ; \boldsymbol{x}_1 t_1)   g^{ >  (0)  }_{\alpha_1 \alpha^\prime}(\boldsymbol{x}_1 t_1 ; \boldsymbol{x}^\prime t^\prime)
 \right) , \notag\\
 \label{Keldyshimpurity2}
\end{align} 
\end{widetext}
where we have used the relations $ g^{ {\rm{t}}  } =  g^{ {\rm{r}}  } + g^{ <  } = g^{ {\rm{a}}  } + g^{ >  }$ 
and  $ g^{ \tilde{\rm{t}}  } =  -g^{ {\rm{r}}  } + g^{ >  } = -g^{ {\rm{a}}  } + g^{ <  }$  as given in Eq. \eqref{realtimeGreen'sfunctionrelations1}.
Based on Eq. \eqref{Keldyshimpurity2},  we derive the impurity-averaged Green's functions for all of these four components.
Basically, we can obtain them by adopting the same argument given in subSec. \ref{IAIMGreen'sfunction}.  
At first, let us focus on the retarded component. 
We express  $g^{ { \rm{r} } }_{\alpha \alpha^\prime}(\boldsymbol{x} t ; \boldsymbol{x}^\prime t^\prime) $ in the perturbative expansion form as 
$ g^{ { \rm{r} } }_{\alpha \alpha^\prime}(\boldsymbol{x} t ; \boldsymbol{x}^\prime t^\prime) = \sum_n g^{  { \rm{r} } (n) }_{\alpha \alpha^\prime}(\boldsymbol{x} t ; \boldsymbol{x}^\prime t^\prime)$ 
where $g^{  { \rm{r} }  (n)}_{\alpha \alpha^\prime}(\boldsymbol{x} t ; \boldsymbol{x}^\prime t^\prime)$  is in the $n$-th order of $\mathcal{H}^{\rm{imp}}$.
Then we perform the Fourier transformations  
\begin{widetext}
\begin{align} 
& g^{ { \rm{r} } }_{\alpha \alpha^\prime}(\boldsymbol{x} t ; \boldsymbol{x}^\prime t^\prime) = \frac{1}{V} \sum_{ \boldsymbol{k} \boldsymbol{k}^\prime  }\int  \frac{d\omega d\omega^\prime}{(2\pi)^2}  
e^{ i \left( ( \boldsymbol{k}  \boldsymbol{x} - \boldsymbol{k}^\prime \boldsymbol{x}^\prime ) - ( \omega t - \omega^\prime t^\prime )  \right)     } 
g^{ { \rm{r} } } _{\alpha \alpha^\prime} (  \boldsymbol{k} \omega;  \boldsymbol{k}^\prime \omega^\prime) , \notag\\
& g^{  { \rm{r} } (n) }_{\alpha \alpha^\prime}(\boldsymbol{x} t ; \boldsymbol{x}^\prime t^\prime) =  \frac{1}{V}  \sum_{ \boldsymbol{k} \boldsymbol{k}^\prime  }\int \frac{d\omega d\omega^\prime}{(2\pi)^2}  
e^{ i \left( ( \boldsymbol{k}  \boldsymbol{x} - \boldsymbol{k}^\prime \boldsymbol{x}^\prime ) - ( \omega t - \omega^\prime t^\prime )  \right)   }
 g^{  { \rm{r} } (n) } _{\alpha \alpha^\prime} (  \boldsymbol{k} \omega;  \boldsymbol{k}^\prime \omega^\prime) .
\end{align} 
\end{widetext}
We take the impurity average on $g^{ { \rm{r} } } _{\alpha \alpha^\prime} (  \boldsymbol{k} \omega;  \boldsymbol{k}^\prime \omega^\prime) $ 
as well as $g^{ { \rm{r} } (n) } _{\alpha \alpha^\prime} (  \boldsymbol{k} \omega;  \boldsymbol{k}^\prime \omega^\prime) $.
Let us denote them as $  \langle g^{  { \rm{r} } } _{\alpha \alpha^\prime} (  \boldsymbol{k} \omega;  \boldsymbol{k}^\prime \omega^\prime)  \rangle^{\rm{imp}} _{\rm{ave}}   $
and  $  \langle g^{  { \rm{r} } (n)} _{\alpha \alpha^\prime} (  \boldsymbol{k} \omega;  \boldsymbol{k}^\prime \omega^\prime)  \rangle^{\rm{imp}} _{\rm{ave}}   $, respectively.
As a result, both of them become diagonal in momentum and frequency represented as
\begin{align} 
& \langle g^{  { \rm{r} }  } _{\alpha \alpha^\prime} (  \boldsymbol{k} \omega;  \boldsymbol{k}^\prime \omega^\prime)  \rangle^{\rm{imp}} _{\rm{ave}}   
=  \bar{g}^{ { \rm{r} } }_{\alpha \alpha^\prime}( \boldsymbol{k} \omega ) \delta_{  \boldsymbol{k}, \boldsymbol{k}^\prime  }  \cdot
\left(   2\pi   \delta(   \omega - \omega^\prime  )    \right), \notag\\
& \langle g^{  { \rm{r} } (n) } _{\alpha \alpha^\prime} (  \boldsymbol{k} \omega;  \boldsymbol{k}^\prime \omega^\prime)  \rangle^{\rm{imp}} _{\rm{ave}}   
=  \bar{g}^{ { \rm{r} } (n) }_{\alpha \alpha^\prime}( \boldsymbol{k} \omega ) \delta_{  \boldsymbol{k}, \boldsymbol{k}^\prime  }  \cdot
\left(   2\pi   \delta(   \omega - \omega^\prime  )    \right). \label{KeldyshFretardedTinv}
\end{align} 
Consequently, the impurity average of $g^{ { \rm{r} } }_{\alpha \alpha^\prime}(\boldsymbol{x} t ; \boldsymbol{x}^\prime t^\prime)$ is represented in the translational invariant form: 
$ \langle g^{  { \rm{r} }  } _{\alpha \alpha^\prime} (  \boldsymbol{x} t;  \boldsymbol{x}^\prime t^\prime)  \rangle^{\rm{imp}} _{\rm{ave}}  = 
\bar{g}^{  { \rm{r} }  } _{\alpha \alpha^\prime} (  \boldsymbol{x} -  \boldsymbol{x}^\prime; t- t^\prime)$. 
The Fourier transform of impurity-averaged Green's function $ \bar{g}^{ { \rm{r} } }_{\alpha \alpha^\prime}( \boldsymbol{k} \omega ) $ is obtained from the two-point Green's function
 $ \bar{g}^{  { \rm{r} }  } _{\alpha \alpha^\prime} (  \boldsymbol{x} -  \boldsymbol{x}^\prime; t- t^\prime)$  as
\begin{widetext}
\begin{align} 
 \bar{g}^{  { \rm{r} }  } _{\alpha \alpha^\prime} (  \boldsymbol{x} -  \boldsymbol{x}^\prime; t- t^\prime)
 = \frac{1}{V} \sum_{ \boldsymbol{k}  } \int  \frac{d\omega }{2\pi}  
e^{ i \left(  \boldsymbol{k} ( \boldsymbol{x} - \boldsymbol{x}^\prime  ) - \omega ( t - t^\prime )  \right)     } 
\bar{g}^{ { \rm{r} } }_{\alpha \alpha^\prime}( \boldsymbol{k} \omega )   .
\end{align}
 \end{widetext}
Next, we re-sum the perturbative expansion $  \bar{g}^{ { \rm{r} } }_{\alpha \alpha^\prime}( \boldsymbol{k} \omega ) = \sum_n  \bar{g}^{ { \rm{r} } (n) }_{\alpha \alpha^\prime}( \boldsymbol{k} \omega )  $ 
and express it in terms of the self-energy. Here we take the first-Born approximation to evaluate this as we did in subSec. \ref{IAIMGreen'sfunction}.  
As a result,  we obtain
\begin{align}
\bar{g}^{ { \rm{r} } }_{\alpha \alpha^\prime}( \boldsymbol{k} \omega ) & = g^{ {\rm{r}} (0)}_{\alpha \alpha^\prime} (\boldsymbol{k} \omega) + 
g^{ {\rm{r}}  (0)}_{\alpha \alpha_2} (\boldsymbol{k} \omega)  
\Sigma^{ {\rm{r}}  (0)}_{ \alpha_2  \alpha_1 }  (\boldsymbol{k} \omega)   \bar{g}^{ {\rm{r}}   }_{\alpha_1 \alpha^\prime} (\boldsymbol{k} \omega) \notag\\
 & = g^{ {\rm{r}} (0)}_{\alpha \alpha^\prime} (\boldsymbol{k} \omega) + \bar{g}^{ {\rm{r}}  }_{\alpha \alpha_2} (\boldsymbol{k} \omega)  
\Sigma^{ {\rm{r}}  (0)}_{ \alpha_2  \alpha_1 }  (\boldsymbol{k} \omega)   g^{ {\rm{r}}  (0) }_{\alpha_1 \alpha^\prime} (\boldsymbol{k} \omega), 
 \label{Keldyshimpurity3}
\end{align}
where the self-energy $\Sigma^{ {\rm{r}}  (0)}_{ \alpha_2  \alpha_1 }  (\boldsymbol{k} \omega)$ is given by 
  \begin{align} 
  \Sigma^{ { \rm{r  }  } (0) }_{ \alpha_2  \alpha_1   } ( \boldsymbol{k} \omega  ) &=  \frac{ n_{\rm{imp}} }{ \hbar V} \sum_{\boldsymbol{q}}   | v_{\rm{imp}}( \boldsymbol{q}  -  \boldsymbol{k} ) |^2 
  g^{ { \rm{r} } (0) }_{     \alpha_2  \alpha_1 }  ( \boldsymbol{q} \omega). 
    \label{realtime1BAselfenergy}
\end{align} 
Formally, we can solve Eq.  \eqref{Keldyshimpurity3} for  $\bar{g}^{ { \rm{r} } }_{\alpha \alpha^\prime}( \boldsymbol{k} \omega )$ as  
\begin{align}
\bar{g}^{ { \rm{r} } }_{ \alpha \alpha^\prime}( \boldsymbol{k} \omega )  & =  
\left[  
\left(  g^{ {\rm{r}} (0)} (\boldsymbol{k} \omega)  \right)^{-1}  -  \Sigma^{ { \rm{r }  } (0) } ( \boldsymbol{k} \omega  )
 \right]^{-1}_{   \alpha \alpha_1 }  \notag\\
 \left(  g^{ {\rm{r}} (0)} (\boldsymbol{k} \omega)  \right)^{-1}  _{  \alpha   \alpha^\prime}  & =
  ( \omega    + \omega_{\rm{F}}  + i\eta )  \boldsymbol{1}_{\alpha \alpha^\prime} -      \omega^{\rm{TI}}_{\boldsymbol{k}} \tilde{\mathcal{H}}_{0,\alpha \alpha^\prime }.
    \label{impaveragedretardedgreeensfunction}
\end{align}
Next, let us evaluate the self-energy $\Sigma^{ { \rm{r  }  } (0) }_{ \alpha_2  \alpha_1   } ( \boldsymbol{k} \omega  ) $ in Eq.  \eqref{realtime1BAselfenergy}.
In order to do this, we only retain the term proportional to $\omega +  \omega_{\rm{F}}  -  \omega^{\rm{TI}}_{\boldsymbol{q}}   +i \eta$ in $ g^{ { \rm{r} } (0) }_{     \alpha_2  \alpha_1 }$. 
As a result, we have
\begin{widetext}
 \begin{align} 
 \Sigma^{ {\rm{r}}  (0)}_{  \alpha_2  \alpha_1 }  (\boldsymbol{k} \omega)    & = \frac{ 1 }{2} n_{\rm{imp}} v_0^2 \int_0^\infty d \epsilon N(\epsilon)  \frac{1}{  \epsilon_{\rm{F}}  -  \epsilon  +i \eta } 
 = \frac{ 1 }{2} n_{\rm{imp}} v_0^2 \int_{  - \epsilon_{\rm{F}}  }^\infty  d \xi  \tilde{N}(\xi)  \frac{1}{   -  \xi  +i \eta } \notag\\
& \approx  - \frac{ 1  }{2}   \tilde{N}(0) n_{\rm{imp}} v_0^2 \int_{  -\infty }^\infty  d \xi   \frac{\xi  +i \eta }{    \xi^2  + \eta^2 }  
 = - \frac{1  }{2}   \pi \tilde{N}(0) n_{\rm{imp}} v_0^2 
= - \frac{i   }{2 \tau^{\rm{rel}}_{\rm{TI}} } \delta_{ \alpha_2  \alpha_1  },
    \label{retarded1BAselfenergy}
\end{align} 
\end{widetext}
where $N(\epsilon) =  \epsilon / \left( 2\pi (\hbar v_{\rm{F}})^2 \right)$ is the density of states per volume of TI surface state  with $\epsilon = \hbar \omega$. 
$  \xi =  \epsilon - \epsilon_{\rm{F}} $ and $  \tilde{N}(\xi) = ( \epsilon_{\rm{F}} + \xi )/ \left( 2\pi (\hbar v_{\rm{F}})^2 \right).$
The time $\tau^{\rm{rel}}_{\rm{TI}}$ in the above equation is  the same quantity appearing in Eq.  \eqref{1BAselfenergy}.  
As we derived Eq.  \eqref{1BAselfenergy}, in the above analysis we considered that the quantum tranport phenomena is induced by the electrons in the energy state in the vicinity of Fermi surface and the number density of TI surface to be sufficiently large.
Thus, we assume  $ | \boldsymbol{q} |, | \boldsymbol{k}  | \simeq k_{\rm{F}}, \epsilon \ll \epsilon_{\rm{F}} $ and set  $ \tilde{N}(\xi)= \tilde{N}(0)$ with taking $\epsilon_{\rm{F}} \to \infty$ for the lower limit 
in the first line of above equation.
From Eqs.   \eqref{Keldyshimpurity3} and \eqref{retarded1BAselfenergy}, we have the formula of $ \bar{g}^{ { \rm{r} } }_{\alpha \alpha^\prime}( \boldsymbol{k} \omega ) $
and it is  the same as the one given in Eq.  \eqref{unperturbedTIGreen'sfunctions}.
By applying the similar argument,  we can derive the advanced components of impurtiy-averaged Green's function  $\bar{g}^{ { \rm{a} } }_{\alpha \alpha^\prime}( \boldsymbol{k} \omega )$.
Like Eq.  \eqref{Keldyshimpurity3}, we have 
\begin{align}
\bar{g}^{ { \rm{a} } }_{\alpha \alpha^\prime}( \boldsymbol{k} \omega ) & = g^{ {\rm{a}} (0)}_{\alpha \alpha^\prime} (\boldsymbol{k} \omega) + 
g^{ {\rm{a}}  (0)}_{\alpha \alpha_2} (\boldsymbol{k} \omega)  
\Sigma^{ {\rm{a}}  (0)}_{ \alpha_2  \alpha_1 }  (\boldsymbol{k} \omega)   \bar{g}^{ {\rm{a}}   }_{\alpha_1 \alpha^\prime} (\boldsymbol{k} \omega) \notag\\
 & = g^{ {\rm{a}} (0)}_{\alpha \alpha^\prime} (\boldsymbol{k} \omega) + \bar{g}^{ {\rm{a}}  }_{\alpha \alpha_2} (\boldsymbol{k} \omega)  
\Sigma^{ {\rm{a}}  (0)}_{ \alpha_2  \alpha_1 }  (\boldsymbol{k} \omega)   g^{ {\rm{a}}  (0) }_{\alpha_1 \alpha^\prime} (\boldsymbol{k} \omega), 
 \label{Keldyshimpurity4}
\end{align}
where 
  \begin{align} 
  \Sigma^{ { \rm{a  }  } (0) }_{ \alpha_2  \alpha_1   } ( \boldsymbol{k} \omega  ) &=  \frac{ n_{\rm{imp}} }{\hbar V} \sum_{\boldsymbol{q}}   | v_{\rm{imp}}( \boldsymbol{q}  -  \boldsymbol{k} ) |^2 
  g^{ { \rm{a} } (0) }_{     \alpha_2  \alpha_1 }  ( \boldsymbol{q} \omega). 
    \label{realtime1BAselfenergy2}
\end{align} 
The self-energy $  \Sigma^{ { \rm{a  }  } (0) }_{ \alpha_2  \alpha_1   } ( \boldsymbol{k} \omega  ) $ in Eq. \eqref{realtime1BAselfenergy2} can be evaluated by  the same analysis which we did exactly for deriving Eq. \eqref{retarded1BAselfenergy}.
We have $  \Sigma^{ { \rm{a  }  } (0) }_{ \alpha_2  \alpha_1   } ( \boldsymbol{k} \omega  ) = -   \Sigma^{ { \rm{r  }  } (0) }_{ \alpha_2  \alpha_1   } ( \boldsymbol{k} \omega  ).$ 
We solve Eq.  \eqref{Keldyshimpurity4} for $\bar{g}^{ { \rm{a} } }_{\alpha \alpha^\prime}( \boldsymbol{k} \omega ) $ and obtain
\begin{align}
\bar{g}^{ { \rm{a} } }_{ \alpha \alpha^\prime}( \boldsymbol{k} \omega )  & =  
\left[  
\left(  g^{ {\rm{a }} (0)} (\boldsymbol{k} \omega)  \right)^{-1}  -  \Sigma^{ { \rm{a }  } (0) } ( \boldsymbol{k} \omega  )
 \right]^{-1}_{   \alpha \alpha_1 }  \notag\\
 \left(  g^{ {\rm{a}} (0)} (\boldsymbol{k} \omega)  \right)^{-1}  _{  \alpha   \alpha^\prime}  & =
  ( \omega    + \omega_{\rm{F}}  -  i\eta )  \boldsymbol{1}_{\alpha \alpha^\prime} -      \omega^{\rm{TI}}_{\boldsymbol{k}} \tilde{\mathcal{H}}_{0,\alpha \alpha^\prime },
    \label{impaveragedadvancedgreeensfunction}
\end{align}
which is equal to the advanced Green's function in Eq. \eqref{unperturbedTIGreen'sfunctions}.
Lastly, we derive the impurity-averaged lesser Green's function  referring to the analysis in \cite{spintronicsPR2008}.  
By the similar analysis used for deriving Eqs.  \eqref{Keldyshimpurity3} or  \eqref{Keldyshimpurity4}, we obtain
\begin{widetext}
\begin{align} 
 \bar{g}^{<}_{\alpha \alpha^\prime} (\boldsymbol{k} \omega) & =   g^{<(0)}_{\alpha \alpha^\prime} (\boldsymbol{k} \omega) + g^{ < (0)}_{\alpha \alpha^\prime_2} (\boldsymbol{k} \omega)  \Sigma^{ {\rm{a}}  (0) }_{  \alpha^\prime_2 \alpha_1 } (\boldsymbol{k} \omega)   \bar{g}^{ {\rm{a}} }  _{\alpha_1 \alpha^\prime} (\boldsymbol{k} \omega) \notag\\
 & + g^{ {\rm{r}}  (0)}_{\alpha \alpha^\prime_2} (\boldsymbol{k} \omega)   
  \left[
\Sigma^{ {\rm{r}}  (0)}_{ \alpha^\prime_2  \alpha_1 }  (\boldsymbol{k} \omega)   \bar{g}^{ <   }_{\alpha_1 \alpha^\prime} (\boldsymbol{k} \omega)
 +
 \Sigma^{ <  (0) }_{  \alpha^\prime_2 \alpha_1 } (\boldsymbol{k} \omega)   \bar{g}^{ {\rm{a}}  }  _{\alpha_1 \alpha^\prime} (\boldsymbol{k} \omega)
\right],  \notag\\
 \label{Keldyshimpurity5}
\end{align}
\end{widetext}
where
\begin{align} 
  \Sigma^{ <  (0) }_{ \alpha_2  \alpha_1   } ( \boldsymbol{k} \omega  ) &=  \frac{ n_{\rm{imp}} }{\hbar V} \sum_{\boldsymbol{q}}   | v_{\rm{imp}}( \boldsymbol{q}  -  \boldsymbol{k} ) |^2 
  g^{ < (0) }_{     \alpha_2  \alpha_1 }  ( \boldsymbol{q} \omega). 
    \label{realtime1BAselfenergy3}
\end{align} 
From Eqs.  \eqref{impaveragedretardedgreeensfunction},  \eqref{impaveragedadvancedgreeensfunction}, and  \eqref{realtime1BAselfenergy3}, Eq.  \eqref{Keldyshimpurity5} is rewritten as
\begin{widetext}
\begin{align} 
 \bar{g}^{<}_{\alpha \alpha^\prime} (\boldsymbol{k} \omega) & = \bar{g}^{ {\rm{r}} }  _{\alpha \alpha_2} (\boldsymbol{k} \omega)  \Sigma^{ <  (0) }_{ \alpha_2  \alpha_1   } ( \boldsymbol{k} \omega  ) 
 \bar{g}^{ {\rm{a}} }  _{\alpha_1 \alpha^\prime} (\boldsymbol{k} \omega) \notag\\
& + \bar{g}^{ {\rm{r}} }  _{\alpha \alpha_2} (\boldsymbol{k} \omega) 
\left[
\left(   g^{ {\rm{r}}  (0)} (\boldsymbol{k} \omega)     \right)^{-1} \cdot   g^{<(0)} (\boldsymbol{k} \omega)  \cdot
 \left(   g^{ {\rm{a}}  (0)} (\boldsymbol{k} \omega)     \right)^{-1}
\right]_{  \alpha_2 \alpha_1 }
\bar{g}^{ {\rm{a}} }  _{\alpha_1 \alpha^\prime} (\boldsymbol{k} \omega)  .
 \label{Keldyshimpurity6}
\end{align}
\end{widetext}
First, let us evaluate the second term in Eq.  \eqref{Keldyshimpurity6}. 
From Eqs. \eqref{realtimeGreen'sfunctionrelations2},  \eqref{impaveragedretardedgreeensfunction}, and \eqref{impaveragedadvancedgreeensfunction} we have
\begin{widetext}
\begin{align} 
 \left[
\left(   g^{ {\rm{r}}  (0)} (\boldsymbol{k} \omega)     \right)^{-1} \cdot   g^{<(0)} (\boldsymbol{k} \omega)  \cdot
 \left(   g^{ {\rm{a}}  (0)} (\boldsymbol{k} \omega)     \right)^{-1}
\right]_{  \alpha_2 \alpha_1 }  
&= f( \hbar \omega  )
\left[
\left(   g^{ {\rm{r}}  (0)} (\boldsymbol{k} \omega)     \right)^{-1} -  \left(   g^{ {\rm{a}}  (0)} (\boldsymbol{k} \omega)     \right)^{-1}
\right]_{  \alpha_2 \alpha_1 } \notag\\
& =  2i \eta  f( \hbar \omega  ) \boldsymbol{ 1 }_{ \alpha_2 \alpha_1}.
\end{align}
\end{widetext}
By taking the limit $\eta \to 0+$, we see that the second term in Eq.  \eqref{Keldyshimpurity6} vanishes. 
Next, let us evaluate the first term in Eq. \eqref{Keldyshimpurity6} using Eqs.  \eqref{unperturbedTIGreen'sfunctions} and \eqref{realtimeGreen'sfunctionrelations2}. 
This becomes
\begin{widetext}
\begin{align} 
 \bar{g}^{ {\rm{r}} }  _{\alpha \alpha_2} (\boldsymbol{k} \omega)  \Sigma^{ <  (0) }_{ \alpha_2  \alpha_1   } ( \boldsymbol{k} \omega  ) 
 \bar{g}^{ {\rm{a}} }  _{\alpha_1 \alpha^\prime} (\boldsymbol{k} \omega)  =   f( \hbar \omega  ) 
  \left(  \bar{g}^{ {\rm{a}} }  _{\alpha \alpha^\prime} (\boldsymbol{k} \omega)   -  \bar{g}^{ {\rm{r}} }  _{\alpha \alpha^\prime} (\boldsymbol{k} \omega) 
   \right).
\end{align} 
\end{widetext}
By doing exactly the same analysis for the greater component starting from the Dyson equation  \eqref{Keldyshimpurity2},
as a result,   the lesser and greater components of impurity-averaged Green's functions are represented by the retarded and advanced components as    
\begin{align} 
  \bar{g}^{<}_{\alpha \alpha^\prime } (\boldsymbol{k}\omega)  &= f(\hbar\omega) (
   \bar{g}^{\rm{a}}_{\alpha \alpha^\prime } ( \boldsymbol{k} \omega) -   \bar{g}^{\rm{r}}_{\alpha \alpha^\prime } ( \boldsymbol{k} \omega)), \notag\\
     \bar{g}^{>}_{\alpha \alpha^\prime } (\boldsymbol{k}\omega)  & = -(1- f(\hbar\omega)) (
   \bar{g}^{\rm{a}}_{\alpha \alpha^\prime } ( \boldsymbol{k} \omega) -   \bar{g}^{\rm{r}}_{\alpha \alpha^\prime } ( \boldsymbol{k} \omega) ).
\label{impurityaveragedGreen'sfunctionrelations1}
\end{align} 
Consequently, the impurity-averaged Green's functions satisfy exactly the same relations with the ones for non-impurity-averaged Green's functions presented in Eq. \eqref{realtimeGreen'sfunctionrelations2}.

\end{document}